\documentclass[review]{elsarticle}
\makeatletter
\def\ps@pprintTitle{%
 \let\@oddhead\@empty
 \let\@evenhead\@empty
 \def\@oddfoot{\centerline{\thepage}}%
 \let\@evenfoot\@oddfoot}
\makeatother

\pdfoutput=1
\usepackage{graphicx}
\usepackage{amsfonts, amsmath, amssymb}
\usepackage{mathabx}
\usepackage{booktabs,siunitx}
\usepackage[svgnames,table]{xcolor}
\usepackage[tableposition=above]{caption}
\usepackage{pifont}
\usepackage{bm}
\usepackage{cancel}
\usepackage{caption}
\usepackage{color}
\usepackage{enumerate}
\usepackage{float}
\usepackage{soul}
\usepackage{ulem}
\usepackage{hyperref}
\usepackage[english]{babel}
\usepackage[margin=1in]{geometry}
\usepackage{mathtools}
\usepackage{multirow}
\usepackage{setspace}
\usepackage{subfigure}
\usepackage{lineno}
\DeclareUnicodeCharacter{00B3}{\textsuperscript{3}}
\usepackage[ruled,linesnumbered]{algorithm2e}
\RestyleAlgo{ruled}
\SetKwComment{Comment}{/* }{ */}

\SetCommentSty{mycommfont}
\DeclareMathAlphabet{\mathpzc}{OT1}{pzc}{m}{it}

\renewcommand \d[2]{\frac{{\rm d} #1}{{\rm d} #2}}
\renewcommand \d[1]{\rm{d} #1}
\newcommand \D[2]{\frac{\partial #1}{\partial #2}}

\newcommand \DDD [2]{\frac{{\rm D} #1}{{\rm D} #2}}
\renewcommand{\vec}[1]{\bm{\mathrm{#1}}}
\newcommand{\V}[1]{\bm{\mathrm{#1}}}

\def \div{\nabla \cdot \mbox{}}
\def \grad{\nabla}

\def \x{\vec{x}}

\def \n{\vec{n}}
\def \r{\vec{r}}
\def \u{\vec{u}}

\def \I{\vec{I}}

\def \U{\vec{U}}
\def \L{\vec{L}}

\def \cM{\mathcal{M}}

\def \Sb{S_\text{b}}

\def \A{\vec{A}}
\def \b{\vec{b}}
\def \bu{\vec{b_u}}
\def \bp{\vec{b_p}}

\def \C{\vec{C}}
\def \vD{\vec{D}}
\def \cF{\vec{\mathcal{F}}}

\def \g{\vec{g}}
\def \G{\vec{G}}
\def \HG{H_\text{G}}
\def \I{\vec{I}}

\def \Lmu{\vec{L_{\mu}}}
\def \vrho{\vec{\rho}}
\def \vrhochi { \vrho_{\V \chi} }

\def \vmu{\vec{\mu}}
\def \Lrho{\vec{L_{\rho}}}
\def \Lrhochi{\vec{L_{\rho_\chi}}}
\def \M{\vec{M}}

\def \Nx{N_x}
\def \Ny{N_y}

\def \Omegag{\Omega_{\text{g}}}

\def \pexact{p_\text{exact}}

\def \U{\vec{U}}

\def \uexact{u_\text{exact}}

\def \W{\vec{W}}

\def \X{\vec{X}}
\def \Xcom{\X_{\text{com}}}

\def \f{\vec{f}}
\def \fu{\vec{f_u}}

\def \half{\frac{1}{2}}
\def \3half{\frac{3}{2}}
\def \5half{\frac{5}{2}}

\def \mul{\mu^{\text{L}}}
\def \mus{\mu^{\text{S}}}
\def \mug{\mu^{\text{G}}}
\def \n{\vec{n}}

\def \ncells{n_{\text{cells}}}
\def \ncycles{n_{\text{cycles}}}

\def \rhol{\rho^{\text{L}}}
\def \rhos{\rho^{\text{S}}}
\def \rhog{\rho^{\text{G}}}
\def \kl{k^{\text{L}}}
\def \ks{k^{\text{S}}}
\def \kg{k^{\text{G}}}

\def \S{\vec{S}}
\def \sgn{\textrm{sgn}}
\def \u{\vec{u}}

\def \vexact{v_\text{exact}}

\def \vvarphi{\vec{\phi}}

\def \x{\vec{x}}
\def \xu{\vec{x_u}}
\def \xp{\vec{x_p}}

\def \div{\nabla \cdot \mbox{}}
\def \grad{\nabla}

\def \dt{\Delta t}
\def \dx{\Delta x}
\def \dy{\Delta y}
\def \dz{\Delta z}

\def \hen {\mathpzc{h}}

\def \Tsol{T^{\rm sol}}
\def \Tliq{T^{\rm liq}}
\def \hsol{\hen^{\rm sol}}
\def \hliq{\hen^{\rm liq}}
\def \cps{C^{\rm S}}
\def \cpl{C^{\rm L}}
\def \cpg{C^{\rm G}}

\newcommand{\upperRomannumeral}[1]{\uppercase\expandafter{\romannumeral#1}}

\begin{document}
\let\today\relax

\begin{frontmatter}
	
\title{An effective preconditioning strategy for volume penalized incompressible/low Mach multiphase flow solvers}
\author[SDSU]{Ramakrishnan Thirumalaisamy}
\author[SDSU]{Kaustubh Khedkar}
\author[LBNL]{Pieter Ghysels}
\author[SDSU]{Amneet Pal Singh Bhalla\corref{mycorrespondingauthor}}
\ead{asbhalla@sdsu.edu}

\address[SDSU]{Department of Mechanical Engineering, San Diego State University, San Diego, CA}
\address[LBNL]{Scalable Solvers Group, Lawrence Berkeley National Laboratory, Berkeley, CA}
\cortext[mycorrespondingauthor]{Corresponding author}

\begin{abstract}
The volume penalization (VP) or the Brinkman penalization (BP) method is a diffuse interface method for simulating multiphase fluid-structure interaction (FSI) problems in ocean engineering and/or phase change problems in thermal sciences and engineering. The method relies on a penalty factor (which is inversely related to body's permeability $\kappa$) that must be large to enforce rigid body velocity in the solid domain. When the penalty factor is large, the discrete system of equations becomes stiff and difficult to solve numerically. In this paper, we propose a projection method-based preconditioning strategy for solving volume penalized (VP) incompressible and low-Mach Navier-Stokes  equations. The projection preconditioner enables the monolithic solution of the coupled velocity-pressure system in both single phase (uniform density and viscosity) and multiphase (variable density and viscosity) flow settings.  In this approach, the penalty force is treated implicitly, which is allowed to take arbitrary large values without affecting the solver's convergence rate or causing numerical stiffness. It is made possible by including the penalty term in the pressure Poisson equation (PPE), which was not included in previous works that solved VP incompressible Navier-Stokes equations using the projection method. We show how and where the Brinkman penalty term enters the PPE by re-deriving the projection algorithm for the VP method.  Solver scalability under grid refinement is demonstrated, i.e., convergence is achieved with the same number of iterations regardless of the problem size. A manufactured solution in a single phase setting is  used to determine the spatial accuracy of the penalized solution. Various values of body's permeability $\kappa$ are considered. Second-order pointwise accuracy is achieved for both velocity and pressure solutions for reasonably small values of $\kappa$. Error saturation occurs when $\kappa$ is extremely small, but the convergence rate of the solver does not degrade. The solver converges faster as $\kappa$ decreases, contrary to prior experience. Two multiphase fluid-structure interaction (FSI) problems from the ocean engineering literature are also simulated to evaluate the solver's robustness and performance (in terms of its number of iterations). The proposed solver also allows us to investigate the effect of $\kappa$ on the motion of the contact line over the surface of the immersed body. It also allows us to investigate the dynamics of the free surface of a solidifying metal.  
   
\end{abstract}


\end{frontmatter}

\section{Introduction}

The volume penalization (VP) method, also known as the Brinkman penalization method belongs to the family of fictitious domain methods---a field pioneered by Glowinski and colleagues~\cite{glowinski1994fictitious,glowinski1999distributed,Patankar2000}---that aim to reduce the numerical complexity of solving partial differential equations (PDEs) defined over irregular domains by embedding them inside  larger, regular domains. In the fictitious domain approach, the governing PDE is \textit{extended} to the larger domain and appropriate (volumetric) forcing functions are used to account for the boundary conditions over the irregular boundary. 

The VP method was originally developed by Arquis and Caltagirone~\cite{Arquis1984} to simulate isothermal obstacles in incompressible single-phase flows.  In 1999, Angot et al.~\cite{Angot1999} provided convergence proofs and error estimates of the penalized solution in terms of the \textit{penalty} parameter. Inspired by Brinkman's work~\cite{Brinkman1949}, the VP technique treats solids embedded in a fluid as porous media with extremely low permeabilities $\kappa \ll 1$. The velocity boundary condition on the fluid-solid interface, which is of Dirichlet type, is imposed through a volumetric feedback force that is inversely proportional to the body's permeability $\kappa$. The feedback force is commonly referred to as the Brinkman penalty force. Due to the fact that both fluids and solids are represented on a single Eulerian grid, the VP method can be implemented in parallel codes much more easily~\cite{BhallaBP2019,Rossinelli2010}.  A number of extensions and improvements have been made to the VP method over the years due to its simplicity, robustness, ease of implementation, and ability to handle complex geometries. Thirumalaisamy et al.~\cite{Thirumalaisamy2021, Thirumalaisamy2022} have recently extended the VP method to allow spatially-varying Neumann and Robin boundary conditions over complex interfaces for advection-diffusion PDEs. In their VP method~\cite{Thirumalaisamy2021, Thirumalaisamy2022}, the parameter $\kappa$ has no physical relationship to the solid's permeability, but rather is a numerical diffusion parameter that ensures flux continuity across the interface. To achieve high-order (up to fourth-order) spatial accuracy of the penalized solution, Kou et al.~\cite{Kou2022} used a combination of VP and high-order flux reconstruction techniques in a discontinuous Galerkin framework. VP is also used to model phase change problems, but under a different name, the Carman-Kozeny drag model~\cite{Carman1937, Voller1987}. In melting/solidification problems, a volumetric penalty/drag force is applied to retard solid phase motion~\cite{Voller1987, Huang2022}. In literature, volume penalization is most commonly used to model fluid-structure interaction (FSI). The technique has been extensively used to model FSI in incompressible single phase flows (uniform density and viscosity)~\cite{Gazzola2011b, Bergmann2011, engels2015numerical}. The VP technique has also recently gained popularity in modeling multiphase FSI, such as solid motion in incompressible gas-liquid flows. In Bhalla et al.~\cite{BhallaBP2019}, water entry/exit problems were simulated and hydrodynamic loads were calculated for solid bodies slamming into air-water interfaces. Khedkar et al.~\cite{Khedkar2020,khedkar2022model} modeled FSI and optimal control of wave energy converters. Sharaborin et al.~\cite{Sharaborin2021} combined the VP method with a volume of fluid approach to model the prescribed motion of rigid bodies in gas-liquid flows. Bergmann used the VP method to investigate the hydrodynamics of a dolphin jumping out of water~\cite{bergmann2022numerical}. In addition, there are some alternative methods for modeling multiphase FSI that are worth mentioning. In recent years, sharp interface cut cell methods have been proposed to simulate solid-liquid-gas flows~\cite{xie2020three,van2023two}. They have the advantage of conserving mass and resolving three-phase triple points over VP methods. These methods also employ discontinuous interface tracking methods, such as the volume of fluid method to truncate the air-water interface at the solid surface. This is different from a fictitious domain/VP method that employs a continuous representation of the air-water interface, whose implications are discussed in Sec.~\ref{sec_kappa_effect_wedge} of this article. The implementation of cut cell methods requires more work, especially when considering three spatial dimensions, and special care must be taken when addressing non-prescribed motions of solid bodies and the intersection of multiple phases~\cite{van2023two}.  

In spite of the fact that the VP method has been extensively studied to understand the accuracy of the penalized solution, there is no study (to our knowledge) that proposes efficient solvers for the volume penalized Navier-Stokes equations, particularly when the system becomes stiff as $\kappa \rightarrow 0$. Several studies have treated the Brinkman penalty term explicitly~\cite{Kolomenskiy2009, Sakurai2019} (i.e., using the prior time value of the penalty force) or via an operator-splitting approach~\cite{Gazzola2011,Gazzola2011b,Beaugendre2018,Rossinelli2010} (i.e., accounting the penalty force separately in a substep), but this limits the time step size $\Delta t$ and/or $\kappa$ values in order to avoid numerical stiffness/instability. An interesting linearization-based technique was recently proposed by Kou et al.~\cite{Kou2022} to treat the stiff penalty term in an implicit-explicit manner in the context of volume penalized compressible flows, which allowed the authors to take $\kappa$ one to two orders of magnitude lower than needed in an explicit treatment. As we aim to develop solver technology for arbitrary small $\kappa$ values (e.g., four orders lower compared to that required in an explicit treatment) for incompressible flows in this paper, we do not discuss the explicit and operator-splitting approaches further. This is needed in situations like modeling phase change problems, where it is necessary to experiment with increasingly large values of $\kappa^{-1}$ in order to find out what is large enough to obtain physically correct phase change dynamic. Similarly, for multiphase FSI problems $\kappa$ controls the contact line motion over the immersed surface, as demonstrated in this work. To alleviate the aforementioned issues related to numerical stiffness and instability, the stiff penalty term can be treated implicitly, and this is commonly accomplished by solving the penalized system using projection solvers in literature. Bergmann and Iolla~\cite{Bergmann2011} describe the projection method to solve the volume penalized single-phase incompressible Navier-Stokes (INS) system, and more recently, Sharaborin et al.~\cite{Sharaborin2021} have extended and described the projection approach for volume penalized multiphase flows. Prior works~\cite{Bergmann2011,Sharaborin2021} have employed an incorrect projection algorithm to solve the volume penalized INS equations, which is corrected in this study.

An alternative to the projection solver is the monolithic velocity-pressure solver, which does not split velocity and pressure degrees of freedom. Monolithic flow solvers are believed to be computationally inefficient compared to velocity-pressure split solvers such as the projection solver. Griffith~\cite{Griffith2009} and Cai et al.~\cite{Cai2014} point out that this is a misconception, and that the coupled velocity-pressure system can be solved as efficiently as the projection method. Moreover, the monolithic approach does not suffer from order of accuracy reductions caused by operator-splitting or artificial boundary conditions that are required in the projection algorithm. In 2009, Griffith proposed to use an \textit{inexact} projection solver as a preconditioner for the GMRES/FGMRES solver to solve the single phase INS equations monolithically and efficiently. Cai et al.~\cite{Cai2014} extended the projection preconditioner approach to the variable density and viscosity INS system in 2014. The authors considered periodic, free-slip, and velocity boundary conditions for the multiphase solver in their work~\cite{Cai2014}.  Later in 2019, Nangia et al.~\cite{Nangia2019MF} demonstrated that the multiphase projection preconditioner can also handle spatially and temporally varying traction boundary conditions. Based on the success of using projection method as a preconditioner for the coupled velocity-pressure system, this work extends the technique to volume penalized single and multiphase INS systems. The same preconditioner can also be used to solve volume penalized low Mach Navier-Stokes equations. Low Mach volume penalized systems, for instance, are used to model the melting/solidification of phase change materials undergoing volume changes (i.e., velocity is not divergence-free) during the phase change process.  An example of this can be found in the motivating Sec.~\ref{sec_motivation}. It is interesting to note that some phase change problems can also be modeled under isothermal conditions; see for example Ahlkrona and Elfverson~\cite{ahlkrona2021cut} and L{\"o}fgren~\cite{lofgren2022increasing},  who simulate glacier melting due to the shear thinning of ice. Unlike Sec.~\ref{sec_motivation}, these works assume divergence-free velocity.

An obvious choice for the projection preconditioner (to solve the volume penalized INS system) is \textit{inexact} versions of the projection solvers of Bergmann and Iolla~\cite{Bergmann2011} and Sharaborin et al.~\cite{Sharaborin2021}. An inexact solver solves the system of equations only approximately, such as by using few iterations or setting a loose convergence tolerance. Empirical testing shows, however, that the projection algorithms of~\cite{Bergmann2011, Sharaborin2021} do not lead to robust convergence of the monolithic solver, particularly when $\kappa$ is small. This is because prior projection algorithms have not taken into account the Brinkman penalty term in the pressure Poisson equation (PPE). By re-deriving the projection algorithm for the VP method, we show how and where the Brinkman penalty term enters the PPE. Using the correct projection algorithm, we are able to achieve robust convergence of the monolithic multiphase flow solver, even when the value of $\kappa$ is very small. In addition, the solver is demonstrated to remain scalable under grid refinement, i.e., the number of iterations required to converge remains essentially the same regardless of the problem size. To test the spatial accuracy of the penalized solution, we consider a manufactured solution for a uniform density and viscosity flow. A wide range of $\kappa$ values is considered in the test. For both velocity and pressure solutions, second-order pointwise accuracy is achieved for reasonably small values of $\kappa$. In the case of extremely small $\kappa$, error saturation occurs, but the convergence rate of the solver does not deteriorate. In fact, as $\kappa$ values decrease, the proposed solver converges faster, contrary to prior experience~\cite{Kolomenskiy2009, Sakurai2019,Rossinelli2010,Gazzola2011,Gazzola2011b}, where the system becomes stiff (difficult to solve) at small values of $\kappa$. An additional test evaluates the solver's performance (in terms of iteration count) for a multiphase FSI problem with two- to three-order differences in density and viscosity. The convergence rates remain robust in the multiphase case as well. Finally,  the proposed preconditioner also allows us to study the effect of $\kappa$ on the contact line motion over the immersed surface.   

 \section{Motivation behind the proposed solver and preconditioner} \label{sec_motivation}
 
 \begin{figure}
  \centering
    \includegraphics[scale = 0.3]{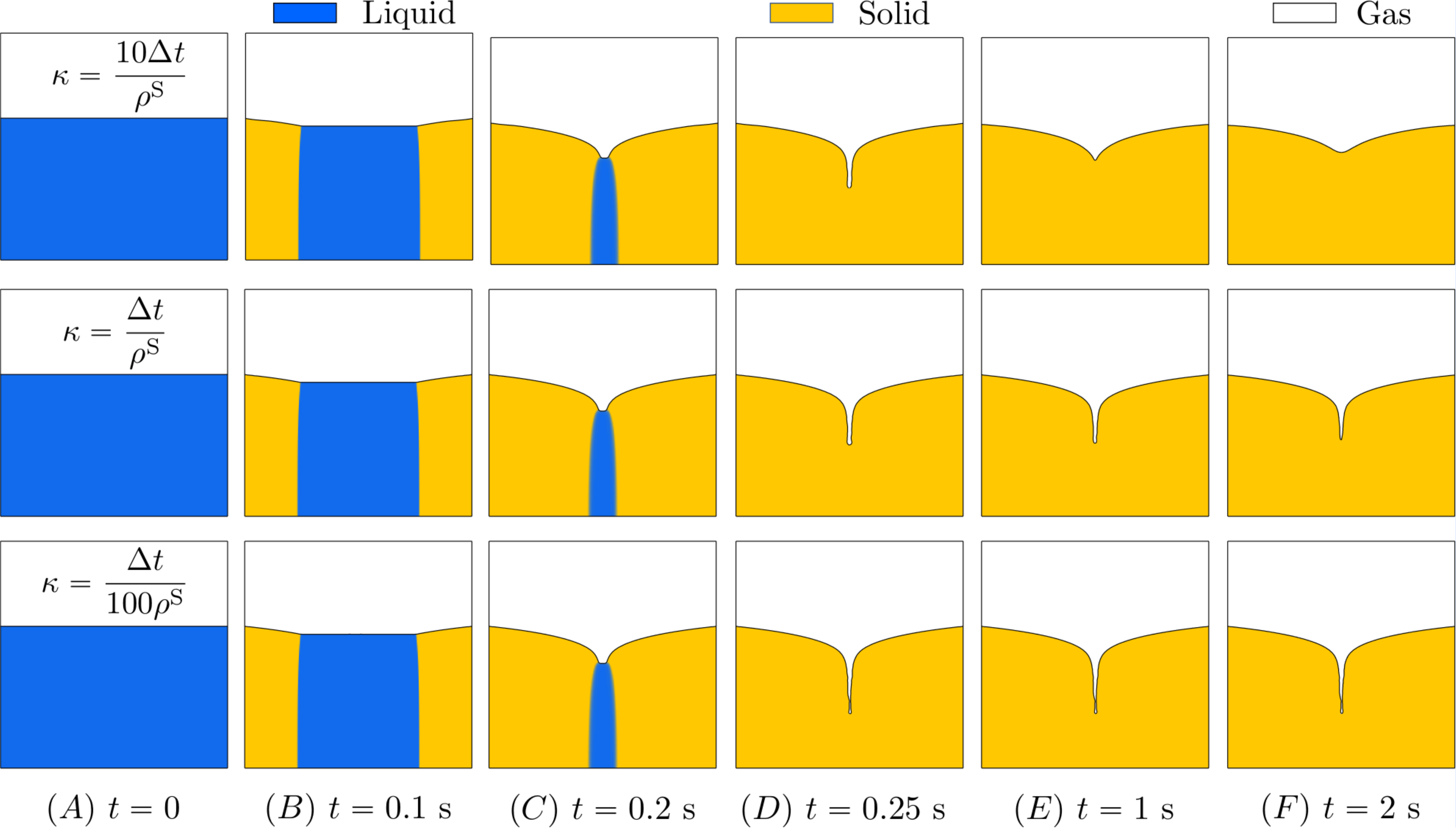} 
  \caption{Solidification of liquid aluminum in presence of gas phase. As the liquid aluminum solidifies a pipe defect appears due to the volume shrinkage effect. The effect of the permeability parameter $\kappa = \epsilon \Delta t/(\rhos)$ on the solidification dynamics is shown. Based on the results of this simulation  $\kappa =  \Delta t/(100 \, \rhos)$ or lower is suggested for this class of problems. } 
    \label{fig_phase_change}
\end{figure}
 
The motivation behind the development of the preconditioner proposed in this work comes from our efforts to model phase change problems involving melting/solidification of phase change materials (PCMs). Consider, for example, liquid aluminum/PCM (blue color) solidifying in a cast in the presence of air (white color) as illustrated in Fig.~\ref{fig_phase_change}. The computational domain is taken to be a square of extents $\Omega \in [0, 8 \times 10^{-3}]^2$, which is discretized into $N\times N = 256 \times 256$ uniform cells.Temperature $T$ is fixed on all boundaries at $0.5 T_m$ ($T_m = 933.6$ K is aluminum's solidification temperature), except at the bottom wall, where zero-heat flux is imposed. Initially, liquid aluminum has a temperature of $2T_m$, whereas gas has a temperature of $0.5T_m$. The imposed boundary conditions cause solidification to begin on the right and left sides of the domain. Solidification is not affected by the top open boundary (which is exposed to the atmosphere) since air has a low thermal conductivity $k$ (gas, liquid, and solid conductivities are taken to be $\kg = 6.1 \times 10^{-2}$, $\kl = 91$, and $\ks = 211$ W/m$\cdot$K, respectively). Solid and liquid aluminum and gas have densities of $\rhos = 2700$, $\rhol= 2475$, and  $\rhog = 0.4$ kg/m$^3$, respectively, specific heats of $\cps = 910$, $\cpl = 1042.4$, and $\cpg = 1100$ J/kg$\cdot$K, respectively, and viscosities of $\mul =  \mus =  1.4 \times 10^{-3}$,  and $\mug = 4.5 \times 10^{-5}$ Pa$\cdot$s, respectively. Viscosity in the solid phase is fictitious and does not affect numerical results. The surface tension coefficient between liquid aluminum and gas  is taken to be $\gamma = 0.87$ N/m. The latent heat of fusion/melting of aluminum is $L = 383840$ J/kg.
 
The time evolution of the non-isothermal phase changing gas-liquid-solid system is governed by the equation of state, and conservation of mass, momentum and energy equations that read as
\begin{subequations} 
\begin{alignat}{2}
&\text{Equation of state (EOS):} \qquad && \rho = \rhog +( \rhos- \rhog) \beta + ( \rhol - \rhos)\beta \varphi, \label{eq_EOS_lm} \\
&\text{Indicator advection:} \qquad && \DDD{\beta}{t} =  \frac{ \rm{d} \beta}{\rm{d} \sigma} \left(\D{\sigma}{t} + \u \cdot \grad \sigma \right) = \beta^{'} \DDD{\sigma}{t} = 0,  \label{eq_beta_lm} \\
&\text{Mass/Low-Mach:} \qquad && \div \u  = -\frac{1}{\rho}\DDD{\rho}{t} =  \frac{(\rhos-\rhol)}{\rho} \beta \DDD{\varphi}{t}, \label{eq_lm} \\
& \text{Momentum:}  \qquad  && \D{\left(\rho \u\right)}{t}+\div{\left(\rho \u \otimes \u\right)}  = -\grad{p}+\div\left[{\mu\left(\grad{\u}+\grad{\u}^\intercal \right)}\right] +\rho \g - A_d \u,  \label{eq_mom_lm}  \\
& \text{Energy/Enthalpy:} \qquad  && \D{\left(\rho \hen \right)}{t}+\div{\left(\rho \u \hen \right)} = \div\left({ k \grad{T}}\right) + Q_{\rm src}.  \label{eq_enthalpy_lm}
\end{alignat}
\end{subequations} 
The EOS (Eq.~\eqref{eq_EOS_lm}) defines density in terms of: (i) an indicator function $\beta$ that is defined to be 1 in the solid-liquid PCM region and 0 in the gas region; and (ii) a liquid fraction variable $\varphi$ that is defined to be 1 in the liquid, 0 in the solid, and between 0 and 1 in the ``mushy" zone. $\beta$ is transported with the local velocity $\u$ (Eq.~\eqref{eq_beta_lm}), whereas $\varphi$ is defined to be an explicit function of enthalpy and evolves with it. The indicator function $\beta$ is defined in terms of a signed distance/level set function $\sigma$ that satisfies the same linear advection equation as $\beta$; see Eq.~\eqref{eq_beta_lm}. The $\varphi$-$\hen$ relation is derived in the Appendix sec.~\ref{sec_enthalpy}. More details on the novel low Mach formulation of the enthalpy method, its numerical implementation, and its validation with an analytical solution to a two-phase Stefan problem involving jumps in density, kinetic energy, and specific heat can be found in our recent work~\cite{thirumalaisamy2023low}.
We note that the enthalpy $\hen$ and temperature $T$ of the system evolves due to the boundary conditions and/or heat source/sink term $Q_{\rm src}$; see Eq.~\eqref{eq_enthalpy_lm}. $Q_{\rm src}$ is taken to be zero for this case. 

It is also instructive to provide a physical rationale for re-formulating the original enthalpy method \footnote{The original enthalpy method considered matched density of fluid and solid phases. Barring the (temperature) advective term, the heat transfer equation becomes decoupled from the fluid solver. As a consequence, the original enthalpy method focused primarily on the energy/enthalpy equation and ignored its coupling (via density) to the momentum equation.} of Voller and colleagues~\cite{Voller1987,voller1991eral}) as a low Mach technique. The two phases---solid and liquid---that undergo phase change (melting and solidification) are assumed to be incompressible. This means that the characteristic sound speed is infinite in both of these media. It also implies that in the bulk of both phases, the Mach number of the flow is zero. The mushy region between all solid and liquid phases is a very narrow area that is of the order of a few atomic/molecular diameters. Consequently, the characteristic sound speed in the mushy region is expected not to deviate significantly from the bulk solid and liquid phases, and it remains close to infinity. This \textit{ansatz} allows us to employ a low Mach model to express density as a function of liquid fraction that varies with enthalpy. Low Mach models also imply that variations in density do not affect the thermodynamic pressure $\tilde{p}$. Additionally, the pressure variable $p$ which appears in the momentum equation is mechanical in origin. It serves as a Lagrange multiplier that enforces the kinematic constraint on the velocity field as written in Eq.~\eqref{eq_lm}. Furthermore, for problems that have an open boundary, such as the one considered in this section, the spatially-uniform thermodynamic pressure remains temporally constant as well. For closed systems there is an additional term of the form $\partial \tilde{p}/\partial t$ that appears in the right-hand side of Eq.~\eqref{eq_enthalpy_lm} that is zero in this case. We remark that although we call the new enthalpy method  a ``low Mach" method, it is actually a zero Mach method. This is the common name for the class of models described by equations such as~\eqref{eq_EOS_lm} and~\eqref{eq_lm}. It is similar to how ``low Reynolds number" is most commonly used to mean ``zero Reynolds number."

\begin{figure}
  \centering
    \includegraphics[scale = 0.08]{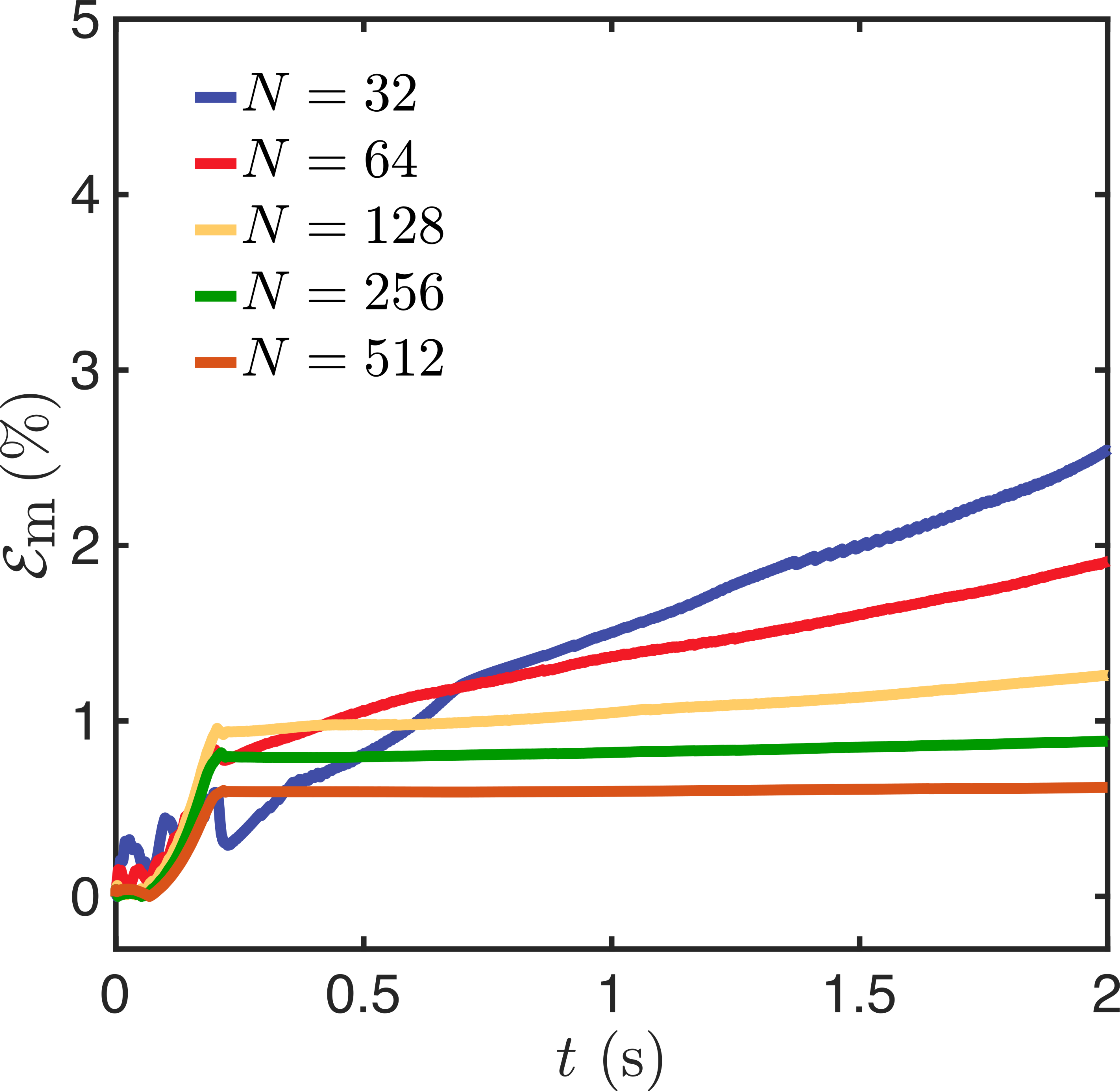} 
  \caption{Percentage change $\mathcal{E}_\textrm{m}$ in aluminum's mass as it solidifies over time for different grid sizes. A uniform time step size $\Delta t$ is used in the simulations. For the coarsest grid $N^2 = 32^2$, a uniform time step size of $\Delta t=8\times10^{-5}$ is employed, and for each successively refined grid $\Delta t$ is halved.}
    \label{fig_pipe_shrinkage_mass_error}
\end{figure}

The momentum Eq.~\eqref{eq_mom_lm} contains a volume penalization term $A_d \u$ that retards any fluid motion in the solid region. Here, $A_d = C_d \frac{\displaystyle {\varphi_{\rm{S}}}^2}{ \displaystyle (1-\varphi_{\rm{S}})^3+\epsilon}$ is the Carman-Kozeny drag coefficient, $\varphi_{\rm S} = \beta(1-\varphi)$ is solid fraction of the grid cell, and $\epsilon$ is a tunable parameter that controls the strength of  
 the permeability parameter $\kappa = \epsilon/C_d = \epsilon \Delta t/(\rhos)$ in the solid region;  small values of $\epsilon$ (or $\kappa$) increase the drag force and retard the motion of solid. This particular choice of $C_d = \rhos/ \Delta t$ is based on an inertial scale and will be explained later. Here, $\Delta t$ is the time step size. $\kappa$ is usually chosen through numerical experiments where one can start with $\epsilon \sim \mathcal{O}(10)$ or $\mathcal{O}(1)$ and gradually reduce its value until no further changes are discernible in the solution or phase change dynamics.  As an example, Fig.~\ref{fig_phase_change}  shows the solidification dynamics of liquid aluminum using $\epsilon = {10, 1, 10^{-2}}$ in the permeability parameter $\kappa$. Additionally, we also simulated the solidification dynamics with $\epsilon = 10^{-3}$; the solidification dynamics remained qualitatively the same as in the $\epsilon = 10^{-2}$ case. In the case of $\epsilon = 10$ and 1, the solidification dynamics are incorrect---a lower value of drag force is not sufficient to prevent the gas-solid interface from moving upon complete solidification, which occurs around $t = 0.25$ s in the simulation. However, lowering $\epsilon$ to $10^{-2}$ or below gives the correct dynamics, which has the solidified metal remaining stationary for $t > 0.25$ s.  In addition, the low Mach Eq.~\eqref{eq_lm} captures the volume shrinkage/pipe defect due to the density difference between solid and liquid metal ($\rhos > \rhol$). We also compute the percentage change in aluminum's mass as it solidifies over time for different grid sizes.  It is computed as $\mathcal{E}_\textrm{m} = \frac{|m(t)-m_0|}{m_0} \times 100$,  in which $m(t) = \int_{\Omega} \beta [\rhos(1-\varphi) + \rhol\varphi] \, \d \Omega$ is aluminum's mass at time $t$ and $m_0 = m(t = 0)$ is its initial mass when it is all liquid. As can be observed from Fig.~\ref{fig_pipe_shrinkage_mass_error},   $\mathcal{E}_\textrm{m}$ decreases significantly under grid refinement. For the finest grid $N^2 = 512^2$, the percentage mass change is approximately 0.62\%.
 
At first, we were unable to go below $\epsilon = 1$ without breaking the monolithic velocity-pressure solver; the linear solver would take a large number of iterations (and  long time) to converge, particularly at high mesh resolutions. This was because our initial implementation of the preconditioner employed the projection algorithm suggested by Bergmann and Iolla~\cite{Bergmann2011} for solving volume penalized Navier-Stokes equations. Later in this work, we demonstrate that the prior projection algorithm does not consider the volume penalized term in the pressure Poisson equation (PPE), which leads to poor performance of the solver\footnote{This foresight came after much struggle and time.}. For incompressible or low Mach systems, one can also use segregated velocity-pressure solvers instead of monolithic ones (where issues related to incorrect/inaccurate PPE would remain ``hidden"), but we advocate using monolithic solvers since they have several advantages. This is discussed in Sec.~\ref{sec_proj_pc}. Monolithic solvers require robust preconditioners, and the proposed preconditioner is an effective strategy.


In what follows, we explain the solution strategy for solving volume penalized equations by considering an isothermal (no phase change) and incompressible multiphase system. It avoids the additional complexity associated with heat transfer and phase change. There is no change to the coupled velocity and pressure linear system for low Mach and incompressible Navier-Stokes equations, except for the non-zero right hand side of Eq.~\eqref{eq_lm}.

\section{Equations of motion}
\subsection{The continuous isothermal multiphase equations}\label{sec_continuous_eqns}

\begin{figure}
  \centering
  \subfigure[Continuous domain]{
    \includegraphics[scale = 0.45]{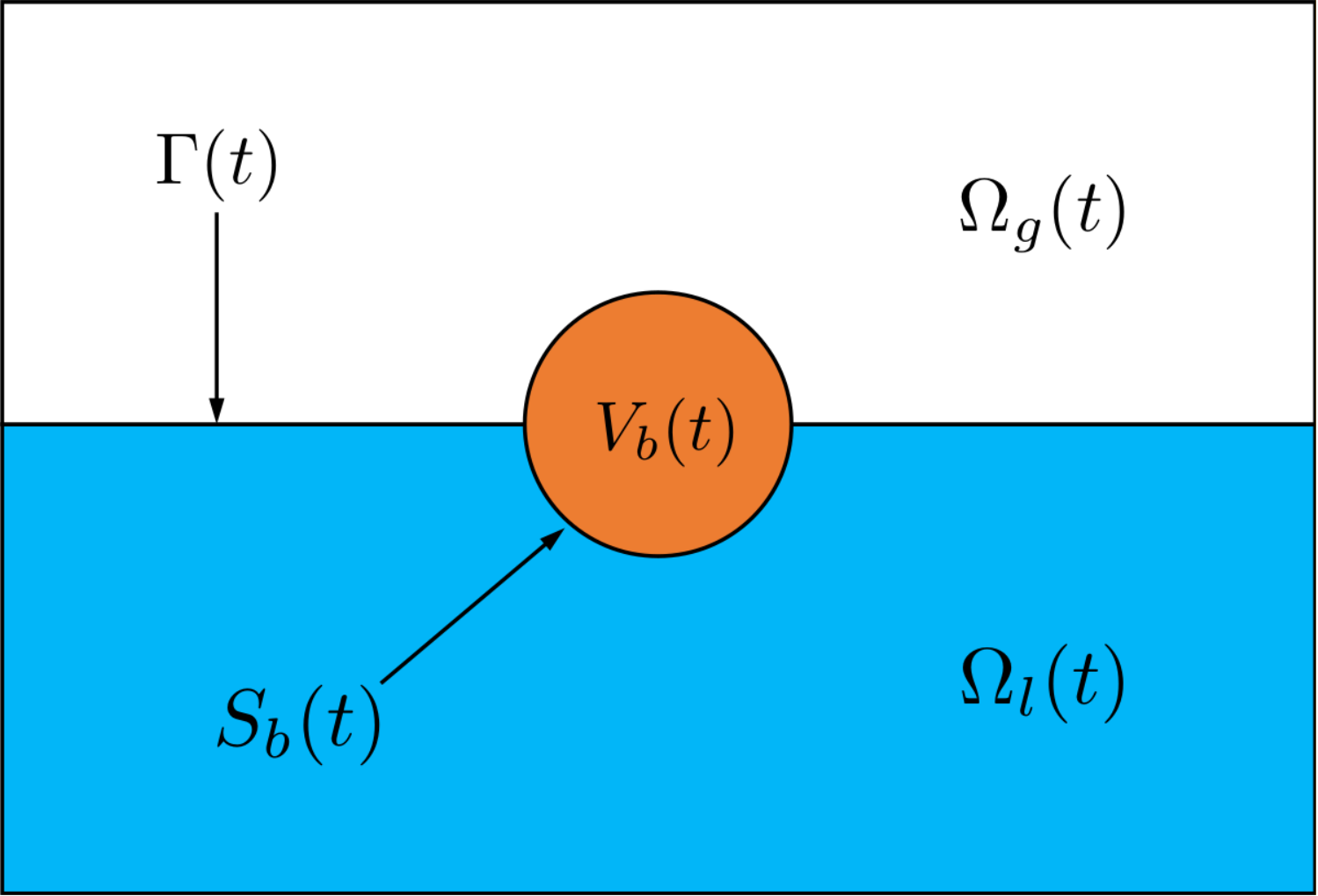} 
    \label{fig_cont_domain}
  }
   \subfigure[Discretized domain]{
    \includegraphics[scale = 0.45]{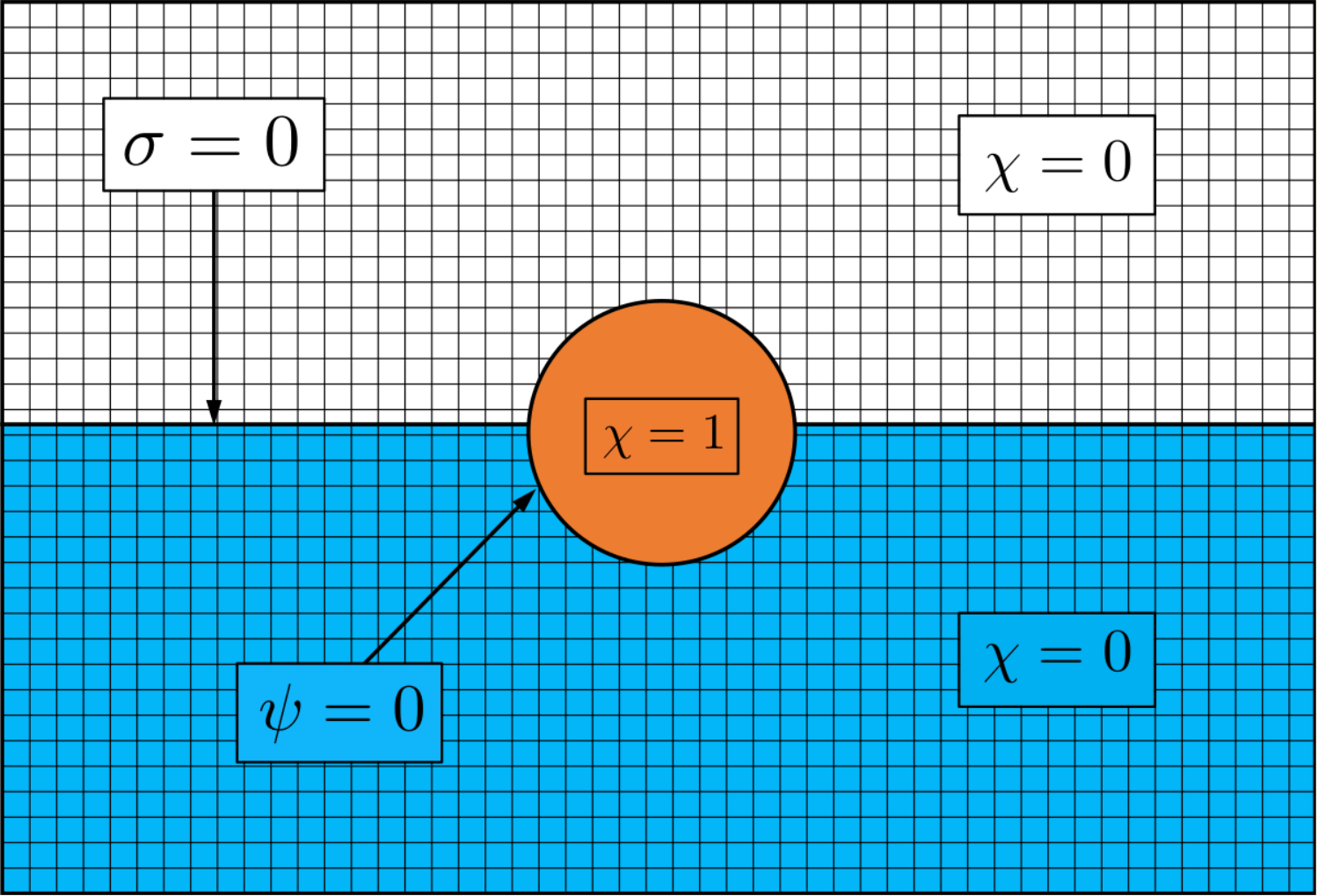}
    \label{fig_discrete_domain}
  }
  \caption{\subref{fig_cont_domain} Two-dimensional computational domain $\Omega$ showing a body immersed in a fluid and interacting with two fluids.~\subref{fig_discrete_domain} Discretization of the domain $\Omega$ on a Cartesian mesh and values of the indicator function $\chi(\x,t)$ used to differentiate the fluid and solid regions in the fictitious domain volume penalization method. Here, $\chi(\x,t)= 1$ inside the solid domain and $\chi(\x,t) = 0$ in liquid and gas domains. The liquid-gas interface $\Gamma(t)$ is tracked by the zero-contour of $\sigma(\x,t)$, while the zero-contour of $\psi(\x,t)$ tracks the solid-fluid interface $\Sb(t)$.
  }
\label{fig_cfd_domains}
\end{figure}

Let $\Omega \subset \mathbb{R}^d$ represent a fixed region of space in spatial dimensions $d = 2 \text { or } 3$. The volume penalized  incompressible Navier-Stokes (INS) equations governing the dynamics of the coupled multiphase fluid-structure system are:
\begin{align}
  \D{\rho \u(\x,t)}{t} + \div (\rho \, \u(\x,t) \otimes \u(\x,t)) &= -\grad p(\x,t) + \div \left[\mu \left(\grad \u(\x,t) + \grad \u(\x,t)^\intercal\right) \right] + \f  \nonumber \\ 
   & \qquad + \frac{\chi(\x,t)}{\kappa}\left(\u_b(\x,t) - \u(\x,t)\right), \label{eqn_momentum}\\
  \div \u(\x,t) &= 0, \label{eqn_continuity} 
\end{align}
which describe the momentum and incompressibility of a fluid with velocity $\u(\x,t)$ and pressure $p(\x,t)$ in an Eulerian coordinate system $\x \in \mathbb{R}^d$.  Eqs.~\eqref{eqn_momentum} and~\eqref{eqn_continuity} are written for the entire computational domain $\Omega$. The domain $\Omega$ is further decomposed into two non-overlapping regions, one occupied by the fluid---liquid and gas---$\Omega_f(t) = \Omega_l(t) \cup \Omega_g(t) \subset \Omega$ and the other by an immersed body $\Omega_b(t) \subset \Omega$, so that $\Omega = \Omega_f(t) \cup \Omega_b(t)$. Fig.~\ref{fig_cfd_domains} shows the schematic representation of the domain occupied by the three (air, water, solid) phases.

The right-hand side of Eq.~\eqref{eqn_momentum} involves the Brinkman penalty force 
\begin{align}  
	\V {\lambda} (\x,t)  = \frac{\chi(\x,t)}{\kappa}\left(\u_b(\x,t) - \u(\x,t)\right)
\label{eqn_brinkman_force}
\end{align}
 that imposes the structural velocity $\u_b(\x,t)$ onto the fictitious fluid contained within $\Omega_b(t)$. In this work, we consider $\u_b$ to be a rigid body velocity. The immersed body is treated as a porous region with vanishing permeability $\kappa \ll 1$, and is tracked using an indicator function $\chi(\x,t)$ that is defined to be one inside $\Omega_b(t)$ and zero outside. In the limit $\kappa \rightarrow 0$, the Brinkman penalty coefficient $\chi/\kappa \rightarrow \infty$, and $\V {\lambda}$ becomes an unknown Lagrange multiplier that needs to be solved for. The Lagrange multiplier formulation is not considered here and we refer the readers to Kallemov et al.~\cite{Kallemov16} and Usabiaga et al.~\cite{Usabiaga17} for a solution strategy to this problem. Here, we examine the case of finite, but small values of $\kappa$ that can make the system of Eqs.~\eqref{eqn_momentum}-\eqref{eqn_continuity} stiff if treated explicitly or via operator-splitting.  The density and viscosity fields vary spatiotemporally and are denoted $\rho(\x,t)$ and $\mu(\x,t)$, respectively. In Eq.~\eqref{eqn_momentum}, $\f$ represents an additional body force term, such as gravity.  The rigid body velocity $\u_b(\x,t)$ in the solid region $\Omega_b(t)$ can either be prescribed or determined by the combined actions of the hydrodynamic and external forces (e.g., gravity).  
 
When describing multiphase flows, it is useful to introduce additional scalar fields, such as the level set/signed distance function (SDF) whose zero-contour defines the two-phase interface implicitly~\cite{Osher1988,Sussman1994}. 
To describe three phase solid-liquid-gas flows, two level set functions are required: $\sigma(\x,t)$ and $\psi(\x,t)$.  
The level set function $\sigma(\x,t)$ is used to demarcate the liquid (e.g., water) and gas (e.g., air) regions, $\Omega_l \subset \Omega$ and $\Omega_g \subset \Omega$, respectively, in the computational domain. The zero-contour of $\sigma$ defines the two fluid interface $\Gamma(t) = \Omega_l \cap \Omegag$. Similarly, the surface of the immersed body $\Sb(t) = \partial V_b(t)$ is tracked using the zero-contour of the level set function $\psi(\x,t)$; see Fig.~\ref{fig_discrete_domain}. The indicator function $\chi(\x,t)$ for the solid domain is computed based on the level set function $\psi$. The two SDFs are advected using the local fluid velocity:
\begin{align}
	\D{\sigma}{t} + \u \cdot \grad \sigma &= 0, \label{eq_ls_fluid_advection} \\
	\D{\psi}{t} +  \u \cdot \grad \psi  &= 0. \label{eq_ls_solid_advection}
\end{align}
The density and viscosity in the entire computational domain is expressed as a function of $\sigma(\x,t)$ and $\psi(\x,t)$ using the signed distance property: 
\begin{align}
\rho (\x,t) &= \rho(\sigma(\x,t), \psi(\x,t)), \label{eq_rho_ls} \\
\mu (\x,t) &= \mu(\sigma(\x,t), \psi(\x,t)). \label{eq_mu_ls}
\end{align}
To maintain their signed distance property, both level set functions need to be reinitialized after each time step. To reinitialize $\sigma$, the relaxation approach of Sussman et al.~\cite{Sussman1994} is used to compute the steady state solution to the Hamilton-Jacobi equation.  This is explained in Sec.~\ref{sec_volume_loss}.  For simple solid geometries (e.g., cylinder, sphere, wedge) $\psi$ can be reinitialized analytically by using constructive solid geometry operators (i.e., the min/max operator) on primitive shapes~\cite{Zhang2019}.  

\subsection{Spatial discretization} 
\label{sec_spatial_discretization}

\begin{figure}
  \centering
  \subfigure[2D staggered Cartesian grid]{
    \includegraphics[scale = 0.45]{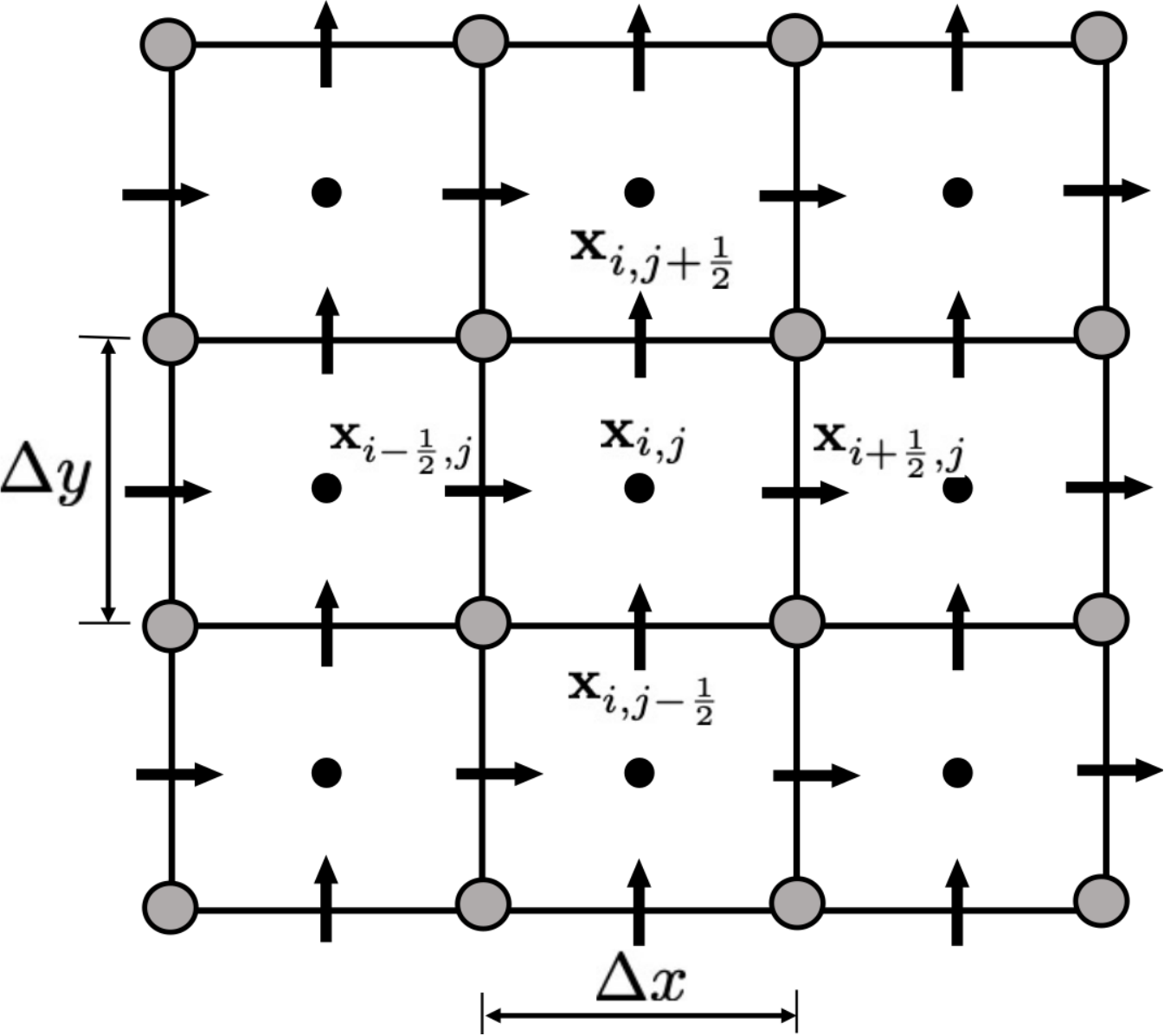} 
    \label{fig_discretized_staggered_grid}
  }
   \subfigure[Single grid cell]{
    \includegraphics[scale = 0.35]{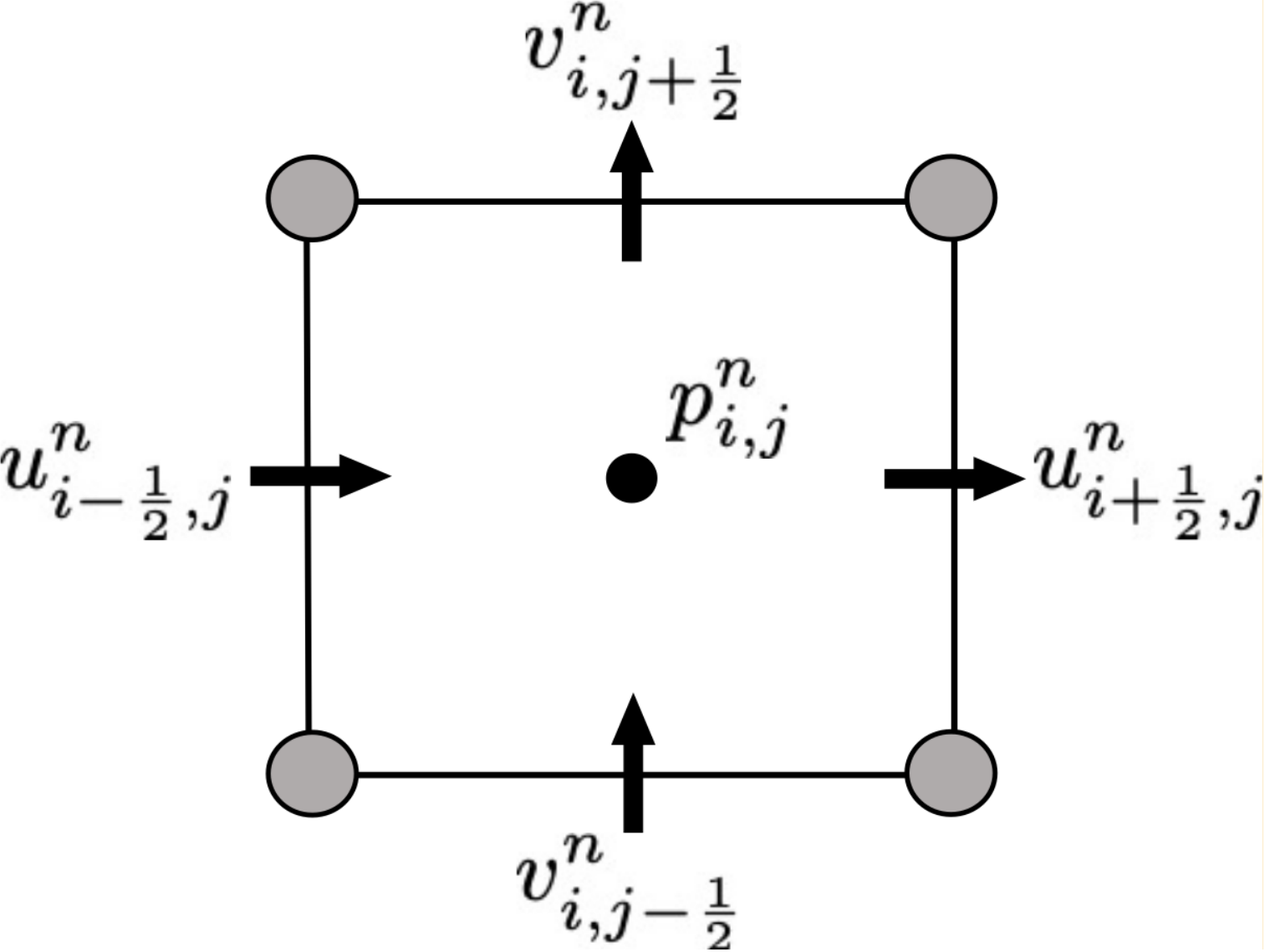}
    \label{fig_single_cell}
  }
  \caption{Schematic representation of a 2D staggered Cartesian grid. \subref{fig_discretized_staggered_grid} shows the coordinate system for the staggered grid. \subref{fig_single_cell} shows a single grid cell with velocity components $u$ and $v$ approximated at the cell faces (${\bf{\rightarrow}}$) and scalar variable pressure $p$ approximated at the cell center ($\bullet$) at $n^\text{th}$ time step.}
\label{fig_cfd_grid}
\end{figure}

The continuous equations of motion given by Eqs.~\eqref{eqn_momentum} and~\eqref{eqn_continuity} are 
discretized on a staggered Cartesian grid.  Without loss of generality, we explain 
the spatial discretization in $d = 2$ spatial dimensions. Extension to three spatial dimensions is analogous.  
A discrete $\Nx \times \Ny$ Cartesian grid covers the physical domain 
$\Omega$ with mesh spacing $\dx$ and $\dy$ in each direction.
The bottom left corner of the domain is situated
at the origin $(0,0)$. The position of each grid cell center is then given by 
$\x_{i,j} = \left((i + \half)\dx,(j + \half)\dy\right)$, where $i = 0, \ldots, \Nx - 1$ and $j = 0, \ldots, \Ny - 1$. 
For a given cell center, $\x_{i-\half,j} = \left(i\dx,(j + \half)\dy\right)$ where
$i = 0, \ldots, \Nx$ and $j = 0, \ldots, \Ny - 1$ denotes the physical location of the cell face 
that is half a grid space away from $\x_{i,j}$ in the negative $x$-direction. 
Similarly $\x_{i,j-\half} =\left((i + \half)\dx,j\dy\right)$, where 
$i = 0, \ldots, \Nx - 1$ and $j = 0, \ldots, \Ny$ denotes the physical location of the cell 
face that is half a grid cell away from  $\x_{i,j}$ in the negative $y$-direction.
The pressure is defined at cell centers of the staggered grid and are denoted by
$p_{i,j}^{n} = p\left(\x_{i,j},t^{n}\right)$, where $t^n$ is the time at time step $n$.
Velocity components are defined at cell faces: $u_{i-\half,j}^{n} = u\left(\x_{i-\half,j}, t^{n}\right)$ and 
$v_{i,j-\half}^{n} = v\left(\x_{i,j-\half}, t^{n}\right)$. The components of the body force $\f = (f_1, f_2)$
are also defined at $x$- and $y$-faces of the staggered grid cells, respectively. The density and viscosity
are defined at cell centers of the staggered grid and are denoted by $\rho_{i,j}^{n} = \rho\left(\x_{i,j},t^{n}\right)$
and  $\mu_{i,j}^{n} = \mu\left(\x_{i,j},t^{n}\right)$, and are interpolated onto the required degrees of freedom as
needed. Similarly, the phase interface is tracked via the level set functions, which are also defined
at cell centers and denoted by $\sigma_{i,j}^{n} = \sigma\left(\x_{ij}, t^n\right)$ and $\psi_{i,j}^{n} = \psi\left(\x_{ij}, t^n\right)$.
Fig.~\ref{fig_cfd_grid} shows the staggered-grid discretization of $\Omega$.

Standard second-order finite differences are used to approximate the 
spatial differential operators~\cite{Griffith2009,Cai2014,Harlow1965,Guermond2006}. 
These are briefly described here to facilitate the discussion. 

The divergence of the velocity field
$\u = (u,v)$ is approximated at cell centers by
\begin{align}
\label{eq_div_fd}
& \vD \cdot \u = D^x u + D^y v, \\
&(D^x u)_{i,j} = \frac{u_{i+\half, j} - u_{i-\half, j}}{\dx}, \\
&(D^y v)_{i,j} = \frac{v_{i, j+\half} - v_{i, j-\half}}{\dy}. 
\end{align}
The gradient of cell-centered quantities (i.e., $p$) is approximated at cell faces by
\begin{align}
\label{eq_grad_fd}
& \G p = (G^x p, G^y p), \\
&(G^x p)_{i-\half,j} = \frac{p_{i,j} - p_{i-1,j}}{\dx}, \\
&(G^y v)_{i,j - \half} =\frac{p_{i,j} - p_{i,j-1}}{\dy}. 
\end{align}
The continuous strain rate tensor form of the viscous term is
\begin{equation}
\label{eq_visc_cont}
\div \left[\mu \left(\grad \u + \grad \u^\intercal\right) \right] = 
\left[
\begin{array}{c}
 2 \D{}{x}\left(\mu \D{u}{x}\right) + \D{}{y}\left(\mu\D{u}{y}+\mu\D{v}{x}\right) \\
 2 \D{}{y}\left(\mu \D{v}{y}\right) + \D{}{x}\left(\mu\D{v}{x}+\mu\D{u}{y}\right) \\
\end{array}
\right],
\end{equation}
which couples the velocity components through spatially variable viscosity 
\begin{equation}
\label{eq_visc_discrete}
\Lmu \u= 
\left[
\begin{array}{c}
 (\Lmu \u)^x_{i-\half,j} \\
 (\Lmu \u)^y_{i,j-\half}  \\
\end{array}
\right].
\end{equation}
The viscous operator is discretized using standard second-order, centered finite differences
\begin{align}
 (\Lmu \u)^x_{i-\half,j} &= \frac{2}{\dx}\left[\mu_{i,j}\frac{u_{i+\half,j} - u_{i-\half,j}}{\dx} -
					        \mu_{i-1,j}\frac{u_{i-\half,j} - u_{i-\3half,j}}{\dx}\right] \nonumber \\ 
                    &+ \frac{1}{\dy}\left[\mu_{i-\half, j+\half}\frac{u_{i-\half,j+1} - u_{i-\half,j}}{\dy} - 
					         \mu_{i-\half, j-\half}\frac{u_{i-\half,j} - u_{i-\half,j-1}}{\dy}\right] \nonumber \\
	            &+ \frac{1}{\dy}\left[\mu_{i-\half, j+\half}\frac{v_{i,j+\half} - v_{i-1,j+\half}}{\dx} - 
					         \mu_{i-\half, j-\half}\frac{v_{i,j-\half} - v_{i-1,j-\half}}{\dx}\right] \label{eq_viscx_fd} \\				         
 (\Lmu \u)^y_{i,j-\half} &= \frac{2}{\dy}\left[\mu_{i,j}\frac{v_{i,j+\half} - v_{i,j-\half}}{\dy} -
					        \mu_{i,j-1}\frac{v_{i,j-\half} - v_{i,j-\3half}}{\dy}\right] \nonumber \\ 
                    &+ \frac{1}{\dx}\left[\mu_{i+\half, j-\half}\frac{v_{i+1,j-\half} - v_{i,j-\half}}{\dx} - 
					         \mu_{i-\half, j-\half}\frac{v_{i,j-\half} - v_{i-1,j-\half}}{\dx}\right] \nonumber \\
	            &+ \frac{1}{\dx}\left[\mu_{i+\half, j-\half}\frac{u_{i+\half,j} - u_{i+\half,j-1}}{\dy} - 
					         \mu_{i-\half, j-\half}\frac{u_{i-\half,j} - u_{i-\half,j-1}}{\dy}\right] \label{eq_viscy_fd},
\end{align}
in which viscosity is required on both cell centers and \textit{nodes} of the staggered grid (i.e., $ \mu_{i\pm\half, j\pm\half}$).
Node centered quantities are obtained via interpolation by either arithmetically or harmonically averaging
the neighboring cell centered quantities. In three dimensions the viscosity is required on both cell
centers and \textit{edges} of the staggered grid. Arithmetic averaging is more accurate (second-order accurate interpolation), 
whereas harmonic averaging provides a better convergence rate at the expense of solution accuracy for solvers dealing with large 
contrasting material properties, such as density, viscosity, and thermal conductivity. A discussion of this topic can be found in the 
classic CFD textbook of Patankar~\cite{patankar2018numerical}. We have also observed this in our previous work related to two phase flows~\cite{Nangia2019MF}. In this work we use arithmetic averaging.

The linear operators described above are needed to fully discretize the continuous equations of motion.
An additional approximation to a variable-coefficient Laplacian is required for the
projection preconditioner described in Sec.~\ref{sec_proj_pc}
\begin{align}
\label{eq_lap_fd}
(\Lrhochi \ p)_{i,j} =  \frac{1}{\dx}\left[\frac{1}{\rho_{\chi \ i+\half,j}} \frac{p_{i+1,j} - p_{i,j}}{\dx} - \frac{1}{\rho_{\chi \ i-\half,j}} \frac{p_{i,j} - p_{i-1,j}}{\dx}\right]
		     + \frac{1}{\dy}\left[\frac{1}{\rho_{\chi \ i,j+\half}} \frac{p_{i,j+1} - p_{i,j}}{\dy} - \frac{1}{\rho_{\chi \ i,j-\half}} \frac{p_{i,j} - p_{i,j-1}}{\dy}\right],
\end{align}
which requires sum of density and Brinkman penalty coefficient, denoted $\rho_\chi$, on faces of the staggered grid ($\rho_{\chi \ i \pm \half,j}$ and $\rho_{\chi \ i,j \pm \half}$  in Eq.~\eqref{eq_lap_fd}). These can also be computed using either the arithmetic or harmonic averages of density from the two adjacent cell centers. It is important to note that our formulation utilizes the interpolated face-centered density within the preconditioner, which does not affect solution accuracy or stability as long as the linear system of equations converges. This means that any suitable interpolation scheme for density works equally well for the preconditioner.

Evaluation of finite difference operators near boundaries of the computational domain requires specification of abutting ``ghost" values. We also use an adaptive mesh refinement (AMR) framework to reduce computational costs of 3D simulations. For further details on the spatial discretization and boundary conditions on uniform and spatially adaptive grids, see our prior work~\cite{Nangia2019MF}.


\subsection{Density and viscosity specification} \label{sec_heaviside_density_viscosity}

Smoothed Heaviside functions are used to transition between liquid-gas and fluid-solid interfaces $\Gamma(t)$ and $S_b(t)$, respectively\footnote{Due to the scale of problems we are interested in, such as simulating wave energy converters and objects slamming into air-water interfaces, we do not impose any contact angle condition at the three material points, since this does not affect the rigid dynamics and the hydrodynamic forces acting on them. Furthermore, this detail also does not affect the spectrum of linear operators, which is the main area of concern for the preconditioner.} The material properties in the transition area are smoothly varied by using $\ncells$ grid cells on either side of the interface. As an example, to calculate a given material property $\Im$, such as density or viscosity, the \textit{flowing} phase property (i.e., gas and liquid) is calculated first
\begin{equation}
	\Im^{\text{flow}}_{i,j} = \Im^{\textrm{L}}+ (\Im^{\textrm{G}} - \Im^{\textrm{L}}) \widetilde{H}^{\text{flow}}_{i,j},
\label{eqn_ls_flow}
\end{equation}
and later correcting $\Im^{\text{flow}}$ to account for the solid body by 
\begin{equation}
	\Im_{i,j}^{\text{full}} = \Im^{\textrm{S}} + (\Im^{\text{flow}}_{i,j} - \Im^\textrm{S}) \widetilde{H}^{\text{body}}_{i,j}.
\label{eqn_ls_solid}
\end{equation}
Here, $\Im^{\text{full}}$ is the final scalar material property field throughout $\Omega$. For the transition specified by  Eqs.~\ref{eqn_ls_flow} and~\ref{eqn_ls_solid}, the usual numerical Heaviside functions are used:
\begin{align}
\widetilde{H}^{\text{flow}}_{i,j} &= 
\begin{cases} 
       0,  & \sigma_{i,j} < -\ncells \, h,\\
        \frac{1}{2}\left(1 + \frac{1}{\ncells \, h} \sigma_{i,j} + \frac{1}{\pi} \sin\left(\frac{\pi}{ \ncells \, h} \sigma_{i,j}\right)\right) ,  & |\sigma_{i,j}| \le \ncells \, h,\\
        1,  & \textrm{otherwise}.
\end{cases}       \label{eqn_Hflow} \\
\widetilde{H}^{\text{body}}_{i,j} &= 
\begin{cases} 
       0,  & \psi_{i,j} < -\ncells \, h,\\
        \frac{1}{2}\left(1 + \frac{1}{\ncells \, h} \psi_{i,j} + \frac{1}{\pi} \sin\left(\frac{\pi}{ \ncells \, h} \psi_{i,j}\right)\right) ,  & |\psi_{i,j}| \le \ncells \, h,\\
        1,  & \textrm{otherwise}.  \label{eqn_Hbody}
\end{cases}
\end{align}
in which $h$ is a suitable measure of the cell size (e.g., $\Delta x$). By convention, we define $\sigma$ and $\psi$ to be negative (positive) in the liquid (gas) and solid (fluid) regions. In all simulations performed in this study, the number of transition cells $\ncells = 1$ for both air-water and fluid-solid interfaces, unless mentioned otherwise. 

The VP method is a diffuse interface method in which all quantities (pressure, velocity, etc.) vary smoothly over the transition region (which is $2\ncells$ wide in our model), and therefore, do not jump across the interface. In addition, the transition region remains incompressible, since the same kinematic constraint  is imposed  on the velocity (Eq.~\eqref{eqn_continuity}) in this region as well. Diffuse interface methods differ from their sharp interface counterparts, such as the immersed interface method~\cite{li2001immersed,kolahdouz2021sharp} and the ghost fluid method~\cite{gibou2002second}, which assume the interface thickness is zero and explicitly impose jump conditions for the quantities of interest.

\subsection{Temporal discretization}

A fixed-point iteration scheme with $\ncycles$ cycles per time step is used to evolve quantities from time level $t^n$ to time level $t^{n+1} = t^n + \Delta t$. The cycle number of the fixed-point iteration scheme is denoted with a $k$ superscript.  At the beginning of each time step, the solutions from the previous time step are used to initialize cycle $k = 0$: $\u^{n+1,0} = \u^{n}$, $p^{n+\half,0} = p^{n-\half}$, $\sigma^{n+1,0} = \sigma^{n}$, and $\psi^{n+1,0} = \psi^{n}$. The physical quantities at the initial time $n = 0$ are prescribed via initial conditions.  Unless mentioned otherwise, we use $\ncycles = 2$ in all of the test cases.

\subsubsection{Level set advection}

The two level set/signed distance functions $\sigma$ and $\psi$ are advected using an explicit advection scheme as follows
\begin{align}
\frac{\sigma^{n+1,k+1} - \sigma^{n}}{\dt} + Q\left(\u^{n+\half,k}, \sigma^{n+\half,k}\right) &= 0, \label{eq_ls_gas_discrete} \\
\frac{\psi^{n+1,k+1} - \psi^{n}}{\dt} + Q\left(\u^{n+\half,k}, \psi^{n+\half,k}\right) &= 0, \label{eq_ls_solid_discrete}
\end{align}
in which $Q(\cdot,\cdot)$ represents an explicit piecewise parabolic method (xsPPM7-limited) approximation to the linear advection terms on cell centers~\cite{Griffith2009, Rider2007}.


\subsubsection{Mitigating mass/volume loss with the level set method} \label{sec_volume_loss}

With the geometries considered in this study, we are able to reset the solid level set function $\psi(\x,t)$ analytically. The analytical resonstruction preserves the mass/volume of the body while not distorting $\psi$'s signed distance property following the linear advection Eq.~\eqref{eq_ls_solid_discrete}. In contrast, $\sigma$ cannot be reinitialized analytically after its signed distance property is disrupted by the linear advection Eq.~\eqref{eq_ls_gas_discrete}. To restore its signed distance property, a reinitialization strategy is required.   Let $\widetilde{\sigma}^{n+1}$ denote the level set function following an advection procedure after time stepping through the interval $\left[t^{n}, t^{n+1}\right]$. We aim to reinitialize it to obtain a signed distance function ${\sigma}^{n+1}$.  As proposed by Sussman et al.~\cite{Sussman1994}, this can be achieved by computing a steady-state
solution to the Hamilton-Jacobi equation
\begin{align}
&\D{\sigma}{\tau} + \sgn\left(\widetilde{\sigma}\right)\left(\|\grad \phi \| - 1\right) = 0, \label{eq_eikonal} \\
& \sigma(\x, \tau = 0) = \widetilde{\sigma}(\x), \label{eq_eikonal_init}
\end{align}
in which we have dropped the $n+1$ superscript because this process is agnostic to the particular time step under consideration.
At the end of a physical time step, Eq.~\eqref{eq_eikonal} is evolved in pseudo-time $\tau$, which, at steady state, produces
a signed distance function satisfying the Eikonal equation $\|\grad \phi \|  = 1$. Here, $\sgn$ denotes the sign of
$\widetilde{\sigma}$, which is either $1$, $-1$, or $0$. The discretization of Eq.~\eqref{eq_eikonal} from the pseudo-time 
interval  $\left[\tau^{m}, \tau^{m+1} \right]$ yields

\begin{equation}
\label{eq_discretized_eikonal}
\frac{\sigma^{m+1} - \sigma^{m}}{\Delta \tau} +
\sgn\left(\widetilde{\sigma}_{i,j}\right) \left[\HG\left(D^{+}_x \sigma_{i,j}, D^{-}_x \sigma_{i,j}, D^{+}_y \sigma_{i,j}, D^{-}_y \sigma_{i,j}\right) - 1 \right] = 0,
\end{equation}
in which $\HG$ denotes a discretization of $\|\grad \sigma \|$ using the Godunov-Hamiltonian, and $D^{\pm}_x$ and $D^{\pm}_y$
denote one-sided discretizations of $\D{\sigma}{x}$ and $\D{\sigma}{y}$, respectively. These are typically discretized using high-order
essentially non-oscillatory (ENO) or weighted ENO (WENO) schemes~\cite{Shu1998}.

It is well known that continually applying Eq.~\eqref{eq_discretized_eikonal} will cause the interface to shift as a function of
$\tau$~\cite{Russo2000}, which will eventually shrink closed interfaces and lead to substantial spurious changes in the volume of each phase. To mitigate the spurious volume loss associated with Eq.~\eqref{eq_eikonal}, we employ second-order ENO finite differences combined with a subcell-fix method described by Min~\cite{Min2010}.
Briefly, the subcell-fix method uses $\widetilde{\sigma}$ to estimate the interface location (i.e., where $\widetilde{\sigma} = 0$) by fitting 
a high-order polynomial and computing an improved estimate of the one-sided derivatives $D^{\pm}_x$ and $D^{\pm}_y$ from the 
polynomial fit. A dimension-by-dimension approach is followed to fit the high-order polynomial. 
 After iterating Eq.~\ref{eq_discretized_eikonal} (using an appropriate time-stepping scheme, e.g., TVD RK2) to some desired convergence criteria, the level set function $\sigma^{n+1}$ is updated, and the next physical time step is carried out. In the present work, we always reinitialize the level set every time step and declare convergence when the $L^2$ norm between subsequent pseudo-time iterations is smaller than some tolerance (taken to be $10^{-6}$ in the present work) or when a maximum number of iterations (taken to be 50) have been carried out --- whichever happens first.
 

All simulations that involve liquid-gas interfaces report percentage mass/volume changes. Additionally, a two-phase dam break problem and two- and three-phase Rayleigh-Taylor instability problems are simulated in Appendix Secs.~\ref{sec_dam_break} and~\ref{sec_RTI}, respectively . We compare the percentage volume change of conserved phases with prior numerical studies that also employ the standard level set methodology. Our level set method implementation achieves acceptable mass/volume loss, as shown in the results. We note that mitigating mass/volume loss issues with the level set method is an active area of research; see for example, Howard and Tartakovsky~\cite{howard2021conservative} who recently proposed a conservative level set method for $N$-phase flows that preserves the volume of every phase simultaneously. Our current work focuses primarily on solving the stiff system of equations that result from discretizing volume penalized multiphase flow equations. The proposed preconditioner is agnostic to level set implementation details and applies equally to conservative level set and volume of fluid methods.


\subsubsection{The discrete multiphase equations}

The discretized form of the multiphase incompressible Navier-Stokes Eqs.~\eqref{eqn_momentum} and \eqref{eqn_continuity} in conservative form reads as
\begin{align}
	\frac{\breve{\V \rho}^{n+1,k+1} \u^{n+1,k+1} - { \V \rho}^{n} \u^n}{\dt} + \C^{n+1,k} &= -\grad_h \, p^{n+\half, k+1}
	+ \left(\L_{\mu} \u\right)^{n+\half, k+1}
	+  \f^{n+\half,k+1}  \nonumber \\
	& \qquad +  \frac{\widetilde{\V{\chi}}^{n+1,k+1}}{\kappa}\left(\u_b^{n+1,k+1} - \u^{n+1,k+1}\right), \label{eqn_c_discrete_momentum}\\
	 \grad \cdot \u^{n+1,k+1} &= \V{0}, \label{eqn_c_discrete_continuity}
\end{align}
in which $\C^{n+1,k} = \C(\rho^{n+1,k}, \u^{n+1,k})$ is the discretized version of the convective term $\div (\rho \, \u \otimes \u)$ and the density approximation $\breve{\V \rho}^{n+1,k+1}$ is computed by integrating the (auxiliary) mass balance equation to achieve mass and momentum transport consistency at the discrete level. The consistent mass/momentum transport scheme ensures the numerical stability of cases involving high density contrast between solid, liquid and gas phases. This is discussed in greater  detail in our previous work~\cite{Nangia2019MF}. 

The viscous strain rate  in Eq.~\eqref{eqn_c_discrete_momentum} is handled using the Crank-Nicolson (trapezoidal rule) approximation: $\left(\L_{\mu} \u\right)^{n+\half, k+1} =  \half\left[\left(\L_{\mu} \u\right)^{n+1,k+1} + \left(\L_{\mu} \u\right)^n\right]$, in which $\left(\L_{\mu} \u \right)^{n+1} = \grad_h \cdot \left[\mu^{n+1} \left(\grad \u + \grad \u^\intercal\right)^{n+1}\right]$.
The newest approximation to the viscosity $\mu^{n+1,k+1}$ is obtained using the two-stage process as described in Sec.~\ref{sec_heaviside_density_viscosity}.


\subsubsection{Fluid-structure coupling} \label{sec_fsi_coupling}

Next, we describe the Brinkman penalization term that imposes the rigidity constraint in the solid region,
and the overall fluid-structure coupling scheme. The Brinkman penalization term is treated implicitly and computed as

\begin{align}
\V {\lambda}^{n+1,k+1} = \frac{\widetilde{\V{\chi}} }{\kappa}\left(\u_b^{n+1,k+1} - \u^{n+1,k+1}\right),  \label{eqn_bp_discrete}
\end{align}
in which the discretized indicator function $\widetilde{\V{\chi}} = \V{1} - \widetilde{\V{H}}^{\text{body}}$ is $1$ only inside the body 
domain and defined using the structure Heaviside function $ \widetilde{H}^{\text{body}}$ from Eq.~\eqref{eqn_Hbody}.
With $\Xcom$ denoting the position of the center of mass of the body, the rigid body velocity $\u_b = \U_r + \W_r \times \left(\x-\Xcom\right)$ can be expressed as a sum of translational $\U_r$ and rotational $\W_r$ velocities. The rigid body velocities can be obtained by integrating Newton's second law of motion
	\begin{align}
		M_b \frac{\U_r^{n+1,k+1} - \U_r^n}{\dt} &=  \cF^{n+1,k} + M_b \g,  \label{eq_newton_u} \\
		 \frac{(\I_b\W_r)^{n+1,k+1} - (\I_b\W_r)^n}{\dt} &=  \cM^{n+1,k},  \label{eq_newton_w}
	\end{align}
in which $M_b$ is the mass, $\I_b$ is the moment of inertia, $\cF$ is the net hydrodynamic force, $\cM$ is the net 
hydrodynamic torque and $M_b \g$ is the net gravitational force acting on the body. Eqs.~\eqref{eq_newton_u} and~\eqref{eq_newton_w} 
are integrated using an explicit forward Euler scheme to compute $\U_r^{n+1,k+1}$, $\W_r^{n+1,k+1}$ and $\Xcom^{n+1,k+1}$. 
In practice we employ quaternions to integrate Eq.~\eqref{eq_newton_w} in the initial reference frame, which avoids 
recomputing $\I_b$ as the body rotates in a complex manner in three spatial dimensions.

The multiphase FSI simulations presented in this work consider immersed bodies with only one unlocked translational degree of freedom. In a previous work~\cite{BhallaBP2019}, simultaneous free translation and rotational motions of the body have been considered with the VP approach. 

\section{Solution methodology}\label{sec_wave_eqs}
\subsection{Fully-coupled Brinkman penalized Stokes system}
Solving for $\u^{n+1,k+1}$ and $p^{n+\half, k+1}$ in Eqs.~\eqref{eqn_c_discrete_momentum} and~\eqref{eqn_c_discrete_continuity} requires the solution of the following block linear system

\begin{equation}
\label{eq_full_bp_stokes}
\left[
\begin{array}{cc}
 \frac{1}{\dt}\breve{\vrho}^{n+1,k+1}  + \frac{1}{\kappa} \widetilde{\V{\chi}}^{n+1,k+1}  - \half \Lmu^{n+1,k+1} & \G\\
 -\vD\cdot & \mathbf{0} \\
\end{array}
\right]
\left[
\begin{array}{c}
 \u^{n+1,k+1}\\
 p^{n+1,k+1} \\
\end{array}
\right] =
\left[
\begin{array}{c}
 \fu\\
\V0\\
\end{array}
\right],
\end{equation}
in which $\breve{\vrho}^{n+1,k+1}$ and $\widetilde{\V{\chi}}^{n+1,k+1} = \V{1} -\widetilde{\V{H}}^{\text{body}}$  are diagonal matrices of face-centered densities and body characteristic function corresponding to each velocity
degree of freedom, respectively. The right-hand side of the momentum equation is lumped into $\fu$, which reads as
\begin{equation}
\fu = \left(\frac{1}{\dt}  \vrho^{n} + \half \Lmu^{n}\right)\u^n + \left(\frac{\widetilde{\V{\chi}}}{\kappa} \right)^{n+1,k+1} \u_b^{n+1,k+1} - 
	\C^{n+1,k}+ \f^{n+\half}.
\end{equation}
The operator on the left-hand side of Eq.~\eqref{eq_full_bp_stokes} is the time-dependent,
incompressible staggered Stokes operator with an additional Brinkman penalty term in the (1,1) block.
We call this the Brinkman penalized Stokes operator or \textit{Stokes-BP} operator for short.
In the next section, we describe the solution of Eq.~\eqref{eq_full_bp_stokes} via
the GMRES or flexible GMRES (FGMRES) Krylov solver~\cite{Saad93} that is preconditioned with an inexact projection solver.

\subsection{Projection solver for the Brinkman penalized Stokes system} \label{sec_projection_solver}

The most popular approach to solving the incompressible Stokes system is the fractional-step projection method. 
Bergmann and Iolla~\cite{Bergmann2011} solved the Stokes-BP system (Eq.~\eqref{eq_full_bp_stokes}) using the projection solver by considering spatially uniform  
density and viscosity in the domain. Recently, Sharaborin et al.~\cite{Sharaborin2021} solved the Stokes-BP system for variable density and viscosity flows. 
However, in both works~\cite{Bergmann2011, Sharaborin2021} the pressure Poisson equation (PPE) did not include the Brinkman penalty term $ \widetilde{\V{\chi}}/\kappa$. Our tests suggest that including the penalty  term in the projection algorithm ensures robust convergence of the monolithic fluid solver, particularly  when $\kappa$ values are small. To see how the penalty term appears in PPE, the algorithmic derivation of projection method for variable density and viscosity Stokes-BP system is presented next. The special case of spatially uniform density (and viscosity) is also discussed.

The Stokes-BP system of Eq.~\eqref{eq_full_bp_stokes}  can be succinctly written in the following form
\begin{align}
\M \; \x &= \b \nonumber \\
\label{eq_stokes_system} 
\left[
\begin{array}{cc}
 \A & \G\\
 -\vD\cdot & \mathbf{0} \\
\end{array}
\right]
\left[
\begin{array}{c}
  \xu\\
  \xp \\
\end{array}
\right] & =
\left[
\begin{array}{c}
 \bu\\
 \bp \\
\end{array}
\right]
\end{align}
in which $\M$ denotes the Stokes-BP operator, $\A =  \frac{1}{\dt} \breve{\V {\rho}}^{n+1,k+1}  + \frac{1}{\kappa} \widetilde{\V{\chi}}^{n+1,k+1}  - \half \Lmu^{n+1,k+1}$ denotes
the discretization of the temporal, Brinkman penalty and viscous operator, $\xu$ and $\xp$ denote the velocity and pressure
degrees of freedom, and $\bu$ and $\bp = \V 0$ denote the velocity and pressure right-hand sides. 

Formally, the projection method can be derived by approximating the inverse of the Schur complement of the saddle-point system Eq.~\eqref{eq_stokes_system}. This is shown in the Appendix Sec.~\ref{sec_projection_derivation}.  Algorithmically, in the first step of the projection method, an intermediate approximation to
$\u$ is computed by solving
\begin{equation}
\label{eq_frac_vel}
\A \widehat{\x}_{\u} = \bu.
\end{equation}
Note that this approximation does not in general satisfy the discrete continuity equation
i.e., $-\vD \cdot \widehat{\x}_{\u} \ne \bp$. This condition can be satisfied by introducing an auxiliary
scalar field $\vvarphi$ and writing out a fractional timestep
\begin{align}
& \left( \frac { \breve{\V {\rho}} } {\dt} +  \frac{ \widetilde{\V{\chi}}}{\kappa}   \right) (\x_{\u}-\widehat{\x}_{\u}) = -\G \vvarphi, \label{eq_frac_timestep} \\
& -\vD \cdot \xu = \bp \label{eq_frac_continuity}.
\end{align}
Multiplying Eq.~\eqref{eq_frac_timestep} by 
\begin{equation} 
\vrhochi^{-1} = \left( \breve{\V \rho} +  \frac{\widetilde{\V \chi} \dt}  {\kappa} \right)^{-1}, \label{eq_rhochi}
\end{equation}
taking the discrete divergence $\vD \cdot$, and substituting
in Eq.~\ref{eq_frac_timestep} yields the \textit{density and Brinkman penalty}-weighted Poisson problem
\begin{equation}
\label{eq_frac_div}
-(\vD \cdot \vrhochi^{-1} \G) \vvarphi  = -\Lrhochi \ \vvarphi = -\frac{1}{\dt}\left(\bp + \vD \cdot \widehat{\x}_{\u}  \right).
\end{equation}
The updated velocity solution can be computed as
\begin{equation}
\label{eq_frac_up_vel}
\xu = \widehat{\x}_{\u} - \dt \vrhochi^{-1} \G \vvarphi,
\end{equation}
and that of pressure can be computed as
\begin{equation}
\label{eq_frac_up_pressure}
\xp = \vvarphi.
\end{equation}

The main difference between our projection method and that of Bergmann and Iolla and Sharaborin et al. is that we include the stiff Brinkman penalty term  $ \widetilde{\V{\chi}} / \kappa$ in the pressure Poisson Eq.~\eqref{eq_frac_div} and the velocity update Eq.~\eqref{eq_frac_up_vel}. This is a small but a crucial detail that leads to robust convergence of the monolithic velocity-pressure solver, particularly when $\kappa$ is small. In Sec.~\ref{sec_results_and_discussion}, we careful study the effect of the Brinkman penalty term on the solver convergence rate. Note that in Bergmann and Iolla~\cite{Bergmann2011}  a variable coefficient Poisson solver was not used to solve the PPE. This is because for constant density and without the Brinkman penalty $\vrhochi^{-1} = \frac{1}{\rho} \V{I}$ trivializes to a scalar multiple of the identity matrix $\V{I}$.

\subsection{Projection preconditioner for the Brinkman penalized Stokes system} \label{sec_proj_pc}

Although the Stokes-BP system of Eq.~\eqref{eq_stokes_system} can be solved using the projection method, we do not follow this approach here. Instead we use the projection method as a preconditioner to solve the coupled velocity-pressure system.   There are several advantages to using the projection method as a preconditioner rather than as a solver. These are discussed in more detail in Griffith~\cite{Griffith2009} and Cai et al.~\cite{Cai2014}. Below is a summary of the main ones.
\begin{itemize}
\item The projection method solves the Stokes system by splitting the velocity and pressure degrees of freedom. The operator-splitting approach requires specifying artificial boundary conditions for the velocity (Eq.~\eqref{eq_frac_vel}) and pressure (Eq.~\eqref{eq_frac_div}) fields. This split affects the global  order of accuracy of the solution~\cite{Brown2001}. For example, it is not possible to impose normal traction boundary condition in the projection solver because this requires a linear combination of discretized 
pressure and velocity variables. This combination can be accounted for in the (Brinkman penalized) Stokes operator $\M$ directly, and the projection preconditioner can still use the artificial boundary conditions in the split velocity and pressure solves. A preconditioner based on artificial boundary conditions does not affect the final outcome of the discretized system it only affects the solver's convergence rate. In Sec.~\ref{sec_mms} we consider a test problem with spatially-varying normal traction boundary conditions. The projection preconditioner uses homogeneous Dirichlet and Neumann boundary conditions for the pressure and normal component of velocity, respectively when the normal traction boundary condition is imposed.   

\item The projection solver is derived under the assumption that certain operators commute; see Appendix Sec.~\ref{sec_projection_derivation} for operator commutations associated with the projection method. These assumptions are typically only satisfied by constant-coefficient operators defined on periodic domains. There is an unavoidable commutator error associated with using the projection method as a solver for variable-coefficient operators.


\item Furthermore, using the projection method as a preconditioner is  no less efficient than using it as a solver, as Griffith~\cite{Griffith2009}, Cai et al.~\cite{Cai2014}, and Nangia et al.~\cite{Nangia2019MF} demonstrate.
\end{itemize}

It is relatively straightforward to use the projection solver discussed in Sec.~\ref{sec_projection_solver} as a projection preconditioner. In this approach, an outer Krylov solver (e.g., GMRES or FGMRES) is employed that generates a Krylov subspace by applying the action of matrix $\M$ on vectors. When the Krylov solver is preconditioned with the projection solver, it also requires the action of the preconditioner on residual vectors to get estimates on velocity and pressure errors. For the projection preconditioner, the unknowns $\xu$ and $\xp$ defined in Eq.~\eqref{eq_stokes_system} should be interpreted as errors in velocity and pressure degrees of freedom, respectively, and the right-hand side vectors $\bu$ and $\bp \ne \V0$ as residuals of momentum and continuity constraint equations, respectively. The projection preconditioner computes the error in velocity ($\xu$) and pressure ($\xp$) only approximately. This is achieved by solving Eqs.~\eqref{eq_frac_vel} and~\eqref{eq_frac_div} in an inexact manner. Specifically, we solve the velocity and pressure subdomain problems using a Richardson solver that is preconditioned with a single V-cycle of a geometric multigrid solver~\cite{Mccormick1986}. For both velocity and pressure problems, $3$ iterations of Gauss-Seidel smoothing are performed on each multigrid level.

For the first-order accurate projection method, the pressure solution can
be approximated as $\xp = \vvarphi$ as written in Eq.~\eqref{eq_frac_up_pressure}; see Brown et al.~\cite{Brown2001}. In the presence of a spatially-varying viscosity, a more accurate approximation to the pressure solution can be obtained as
\begin{equation}
\label{eq_frac_up_pressure2}
\xp = \left(\I - \dt \  \vmu \ \Lrhochi  \right) \vvarphi,
\end{equation}
in which $\vmu$ is a diagonal matrix of cell-centered viscosities corresponding to each pressure
degree of freedom. The above form of $\xp$ is derived by approximating the inverse of the  
Schur complement of the saddle-point system Eq.~\eqref{eq_stokes_system} as done in Cai et al.~\cite{Cai2014}.  
Appendix Sec.~\ref{sec_projection_derivation} provides the derivation. In the projection preconditioner we update pressure using Eq.~\eqref{eq_frac_up_pressure2} instead of Eq.~\eqref{eq_frac_up_pressure}.  Note that in Cai et al. $\Lrho$ is the density-weighted Laplace operator, whereas in this work $\Lrhochi$ is the density and Brinkman penalty-weighted Laplace operator (Eq.~\eqref{eq_frac_div}). 


\section{Software implementation}     
     
The volume penalization algorithm and multiphase fluid solver described here are implemented
within the IBAMR library~\cite{IBAMR-web-page}, an open-source C++
software enabling simulation of immersed boundary-like methods with
adaptive mesh refinement (AMR). The code is hosted on GitHub at \url{https://github.com/IBAMR/IBAMR}.
IBAMR relies on SAMRAI \cite{HornungKohn02, samrai-web-page} for Cartesian grid 
management and the AMR framework. Solver support in IBAMR is provided by the 
PETSc library~\cite{petsc-efficient, petsc-user-ref, petsc-web-page}. 

All solvers and preconditioners have been implemented matrix-free to improve computational efficiency, especially on dynamically adaptive grids. The only exception is the ``bottom" solver of the geometric multigrid preconditioners, which explicitly forms the $\A$ and $\Lrhochi$ matrices for velocity and pressure subdomain problems, respectively. 
    
\section{Validation of the multiphase VP method: Free falling wedge slamming into an air-water interface} \label{sec_wedge}

\begin{figure}[]
\centering
\includegraphics[scale = 0.55]{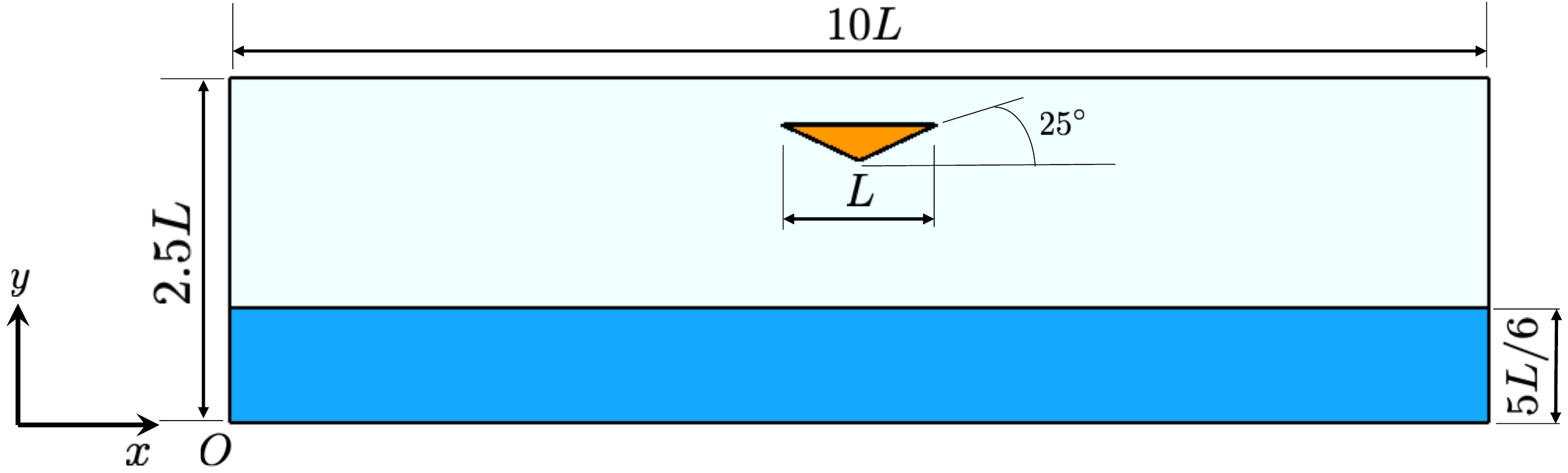}
  \caption{Schematic of the 2D wedge slamming on an air-water interface.}
  \label{fig_wedge_schematic}
\end{figure}

As a validation case, we simulate the free fall of a two-dimensional and a three-dimensional wedge slamming against an air-water interface.
The top length of the 2D wedge is $L$ = 1.2 m, which is positioned within a computational domain of extents $\Omega = [0, 10L] \times [0, 5L/2]$. The 3D wedge simulation is performed in a computational domain of extents $\Omega = [0, 10L] \times [0, 5L/3] \times [0, 5L/2]$. The top surface of the 3D wedge is a square of extents $L \times L$. 
The origin $O$ of the computational domains are at the bottom left corner; see Fig.~\ref{fig_wedge_schematic}.
The initial coordinates of the lower-most vertex of the 2D wedge is $(x_0,y_0)$ = $(5L, 23L/12)$ and that of the 3D wedge is $(x_0, y_0, z_0)$ = $(5L, 5L/6, 23L/12)$.  
The wedge makes an angle of 25$^\circ$ with the horizontal and its free fall height is $\Delta s = 13L/12$. There is a distance of $5L/12$ between the wedge vertical sides and the lateral wall of the 3D computational domain.
The depth of the water column is $5L/6$ and the remainder of the domain is occupied by air. We assume the density of water is $\rhol$ = 1000 kg/m$^3$ and the viscosity is $\mul$ = 10$^{-3}$ Pa$\cdot$s. For air, the density is assumed to be $\rhog$ = 1.2 kg/m$^3$ and viscosity to be $\mug$ = 1.8$\times 10^{-5}$ Pa$\cdot$s. Wedge density is assumed to be $\rhos$ = 466.6 kg/m$^3$ and its fictitious viscosity $\mus$ is the same as that of water. 

The computational domain for the 2D case is discretized into a uniform grid of size $1200 \times 300$, which corresponds to 120 grid cells per wedge length. The uniform mesh spacing is $\dx = \dy = 0.01$ m. A similar grid size in 3D would be computationally very expensive; therefore, we employ an adaptive mesh refinement framework to keep the mesh resolution high in only a few select regions. Among these are regions containing immersed bodies, air-water interfaces, and vorticity of large magnitude. The 3D domain is discretized using  $\ell = 3$ grid levels with a refinement ratio $n_\text{ref} = 2$. The mesh spacing on the coarsest level is $\dx_0 = \dy_0 = \dz_0 = 0.04$ m and on the finest level is $\dx_\text{min} = \dy_\text{min} = \dz_\text{min} = 0.01$ m. A constant time step size of $\Delta t = 6.25 \times 10^{-5}$ s is used in both cases.
As determined by a grid resolution study (see Fig.~\ref{fig_grid_convergence_2dwedge}), the mesh and time step size used are adequate to resolve the FSI dynamics of the freely falling wedge.

\begin{figure}
  \centering
  \subfigure[Vertical postion]{
    \includegraphics[scale = 0.32 ]{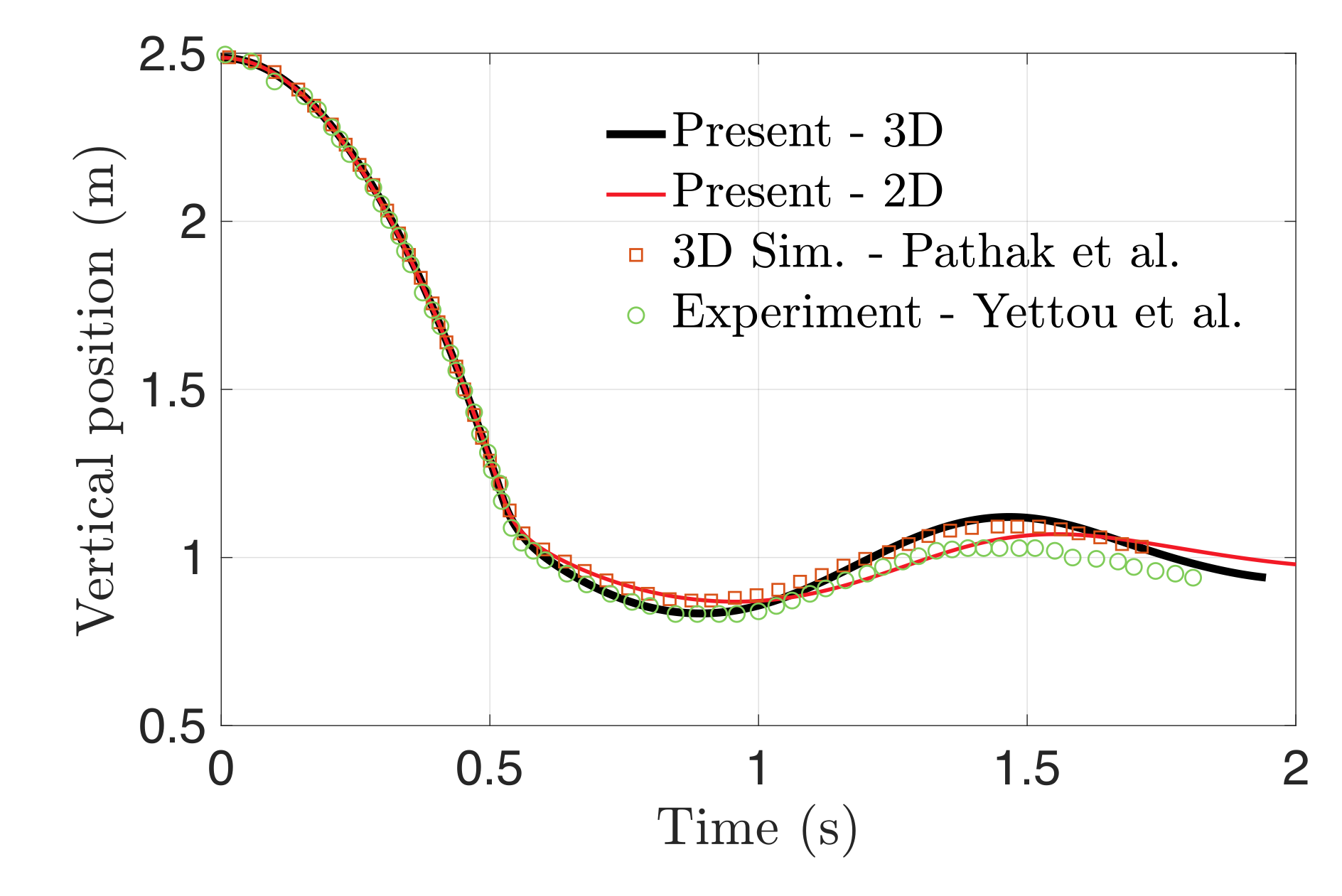} 
    \label{fig_wedge_position}
  }
   \subfigure[Vertical velocity]{
    \includegraphics[scale = 0.32]{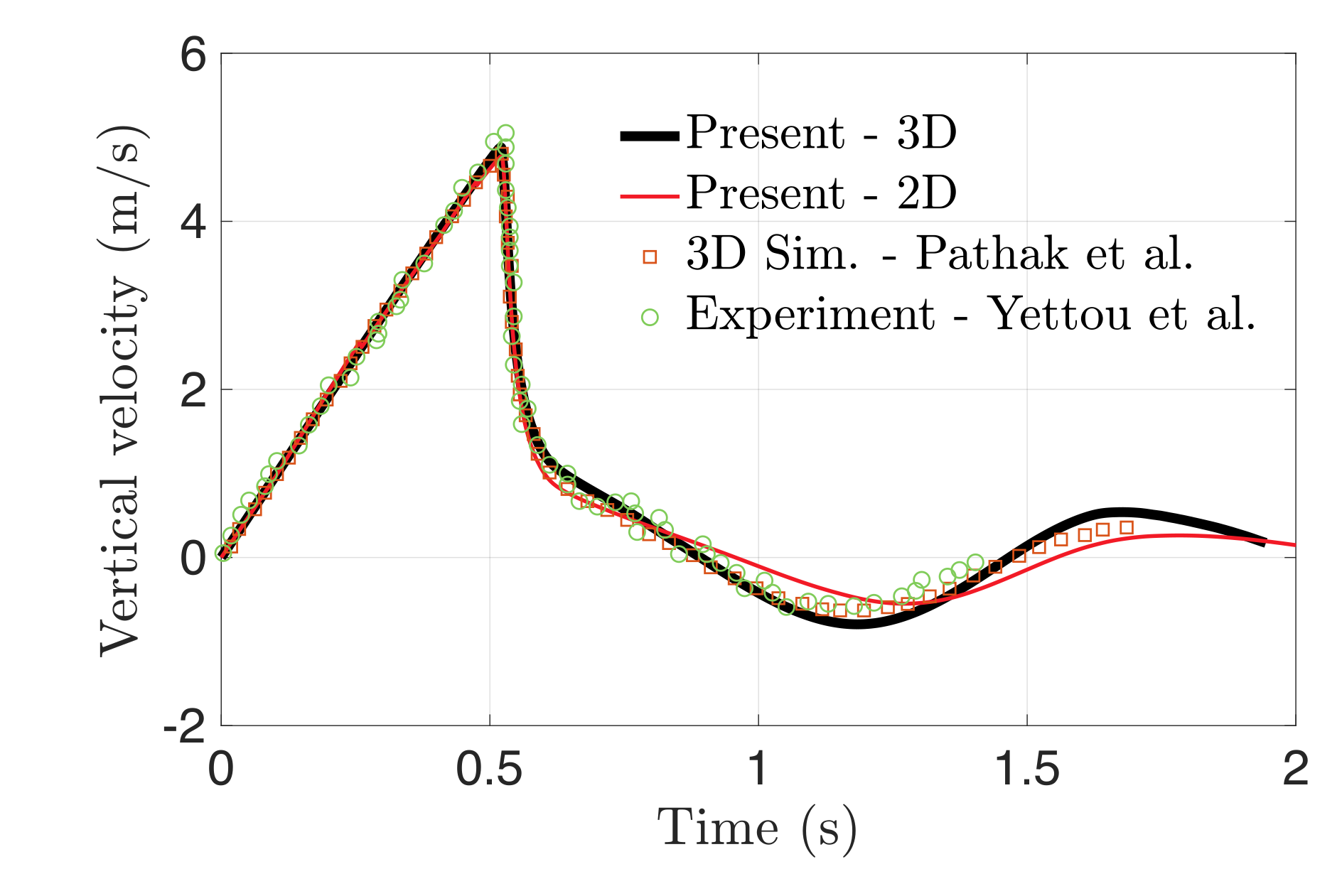}
    \label{fig_wedge_velocity}
  }
    \subfigure[Vertical force]{
    \includegraphics[scale = 0.32]{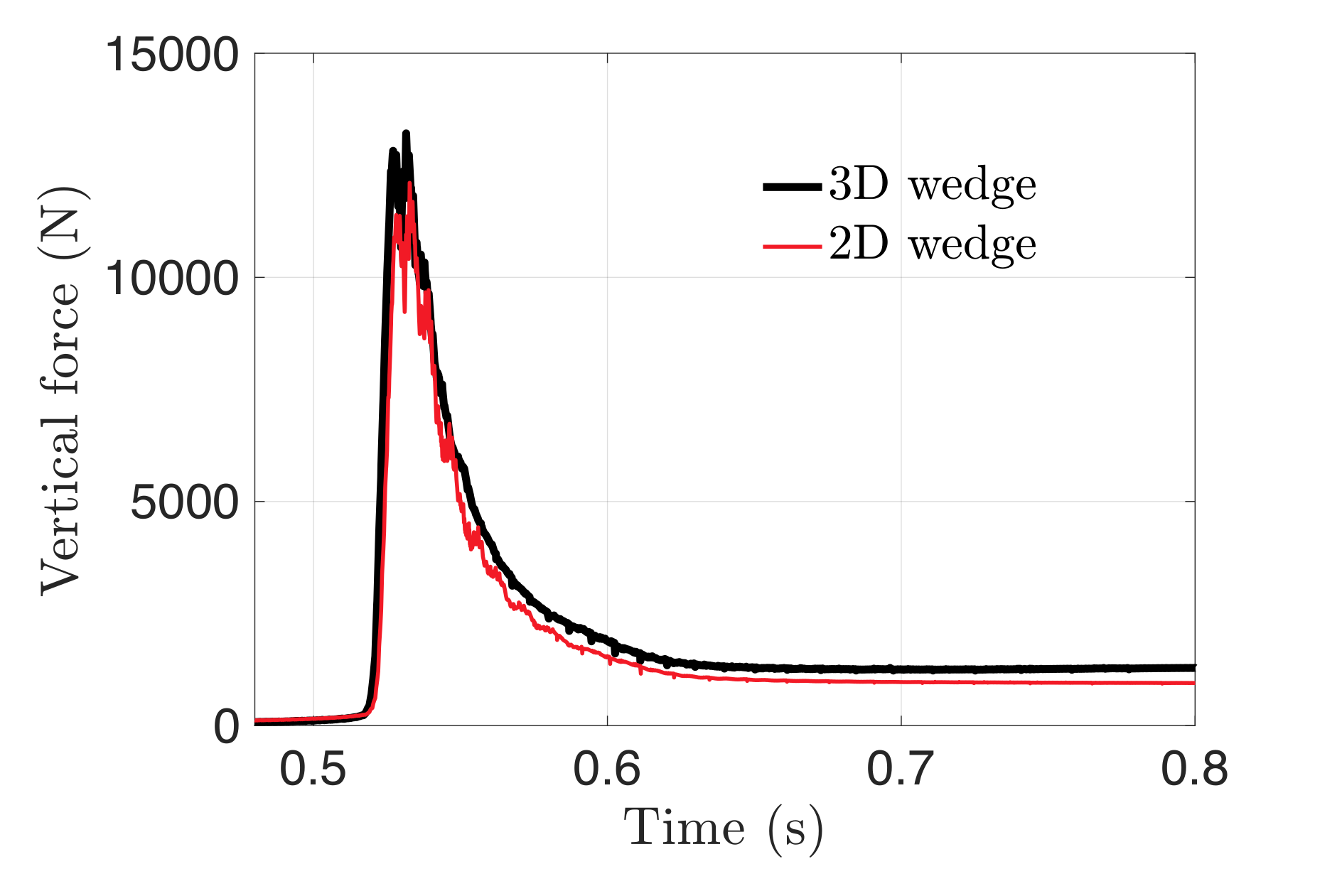} 
    \label{fig_wedge_hydro_force}
  }   
  \caption{Temporal evolution of a free falling wedge slamming against an air-water interface: \subref{fig_wedge_position} vertical position; \subref{fig_wedge_velocity} vertical velocity; and \subref{fig_wedge_hydro_force} vertical component of the hydrodynamic force for 2D and 3D wedge case. The simulation results are compared against previous 3D volume of fluid simulations of Pathak et al.~\cite{Pathak16} and experimental study of Yettou et al.~\cite{Yettou2006}.
}
\label{fig_temporal_evol_cyl_decay}
\end{figure}

\begin{figure}
  \centering
    \subfigure[$t$ = 0.4375 s]{
    \includegraphics[scale = 0.25 ]{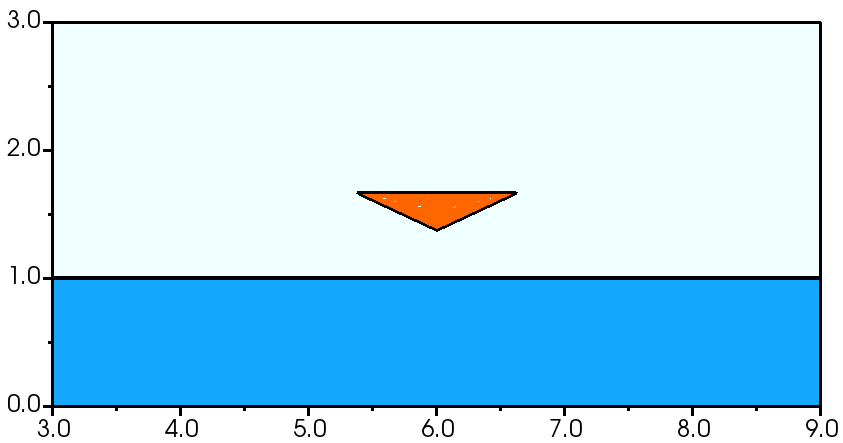} 
    \label{fig_wedge_rho_t1}
  }
    \subfigure[$t$ = 0.4375 s]{
    \includegraphics[scale = 0.25 ]{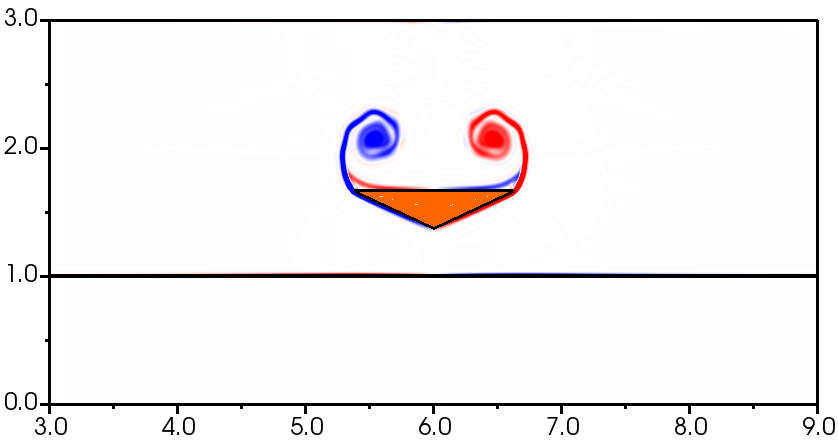} 
    \label{fig_wedge_omega_t1}
  }
    \subfigure[$t$ = 0.5625 s]{
    \includegraphics[scale = 0.25]{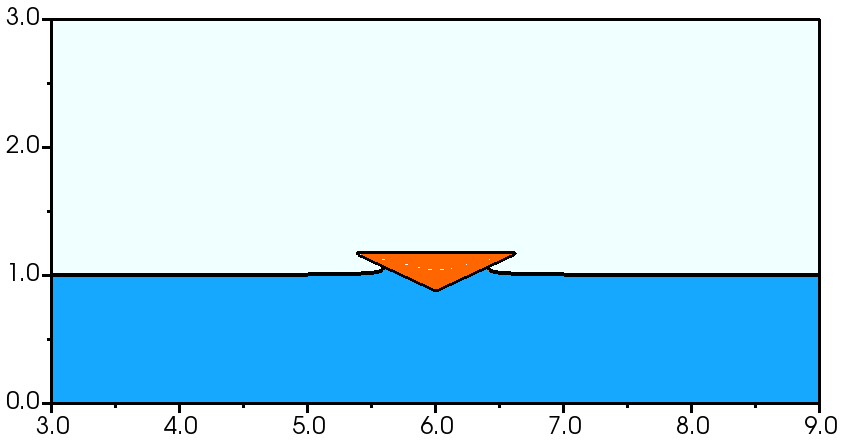}
    \label{fig_wedge_rho_t2}
  }
   \subfigure[$t$ = 0.5625 s]{
    \includegraphics[scale = 0.25]{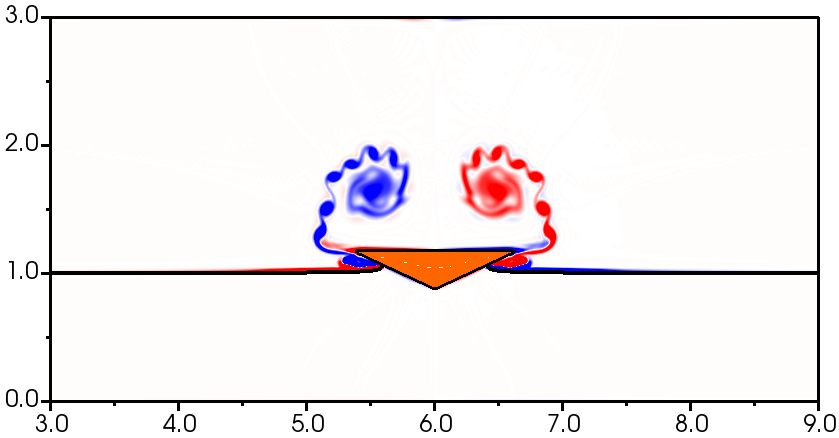}
    \label{fig_wedge_omega_t2}
  }
    \subfigure[$t$ = 0.875 s]{
    \includegraphics[scale = 0.25]{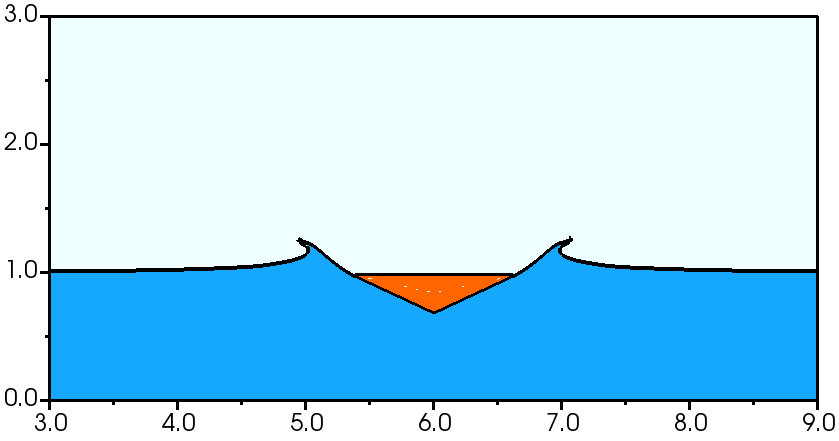} 
    \label{fig_wedge_rho_t3}
  }
    \subfigure[$t$ = 0.875 s]{
    \includegraphics[scale = 0.25]{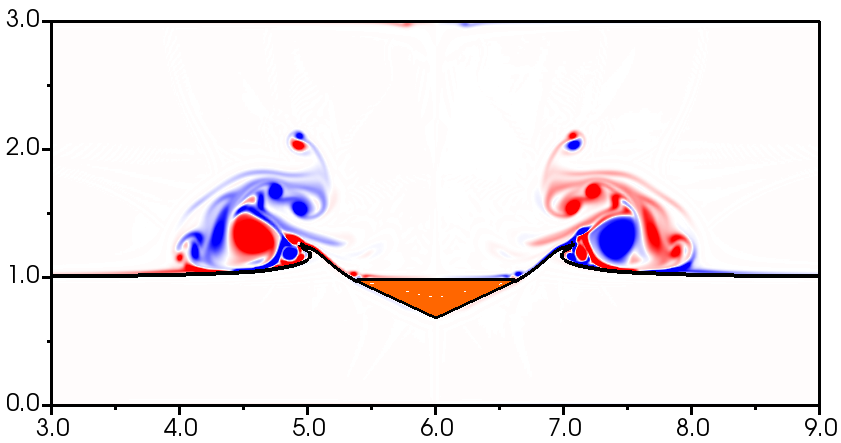} 
    \label{fig_wedge_omega_t3}
  }   
   \subfigure[$t$ = 1.25625 s]{
    \includegraphics[scale = 0.25]{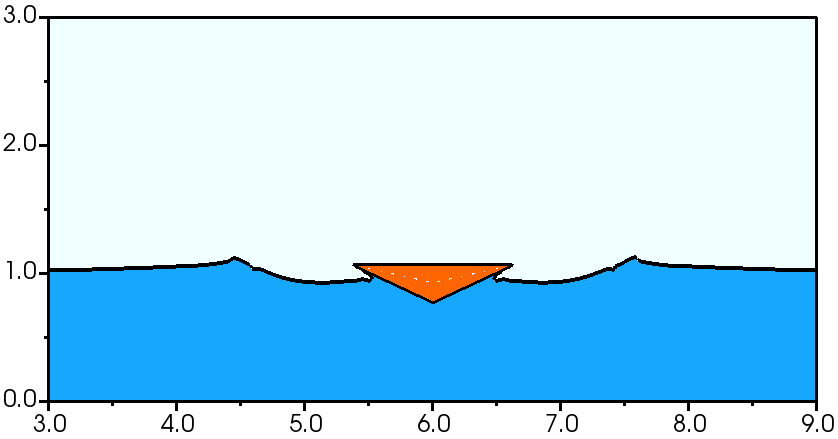} 
    \label{fig_wedge_omega_t3}
  }
    \subfigure[$t$ = 1.25625 s]{
    \includegraphics[scale = 0.25]{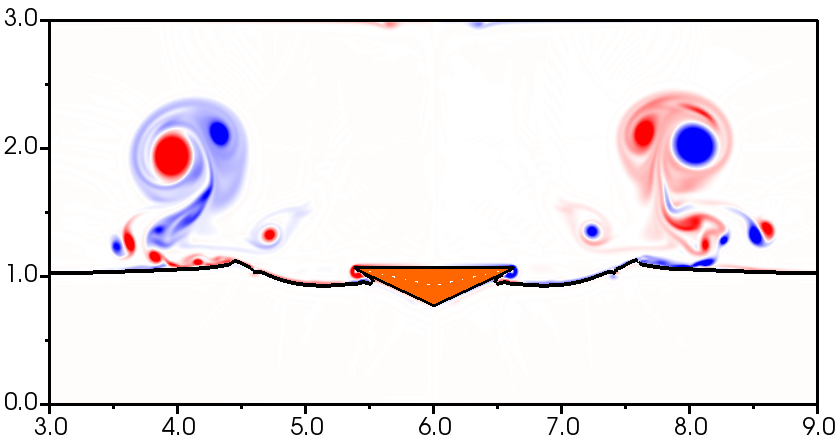} 
    \label{fig_wedge_omega_t3}
  }
  \caption{Temporal evolution of free falling 2D wedge slamming into an air-water interface: (left) density and (right) vorticity generated in the range -100 to 100 s$^{-1}$.}
\label{fig_temporal_evol_wedge_decay}
\end{figure}

\begin{figure}
  \centering
    \subfigure[$t = 0.525$ s]{
    \includegraphics[scale = 0.25]{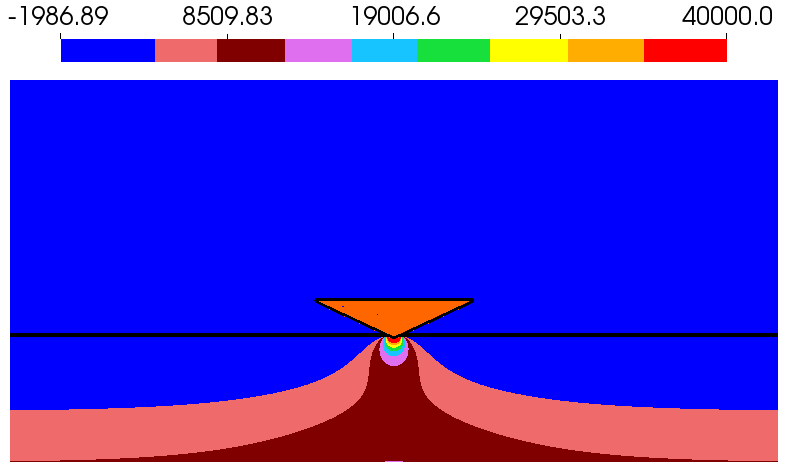} 
    \label{fig_cone_t1}
  }
    \subfigure[$t = 0.55$ s]{
    \includegraphics[scale = 0.25]{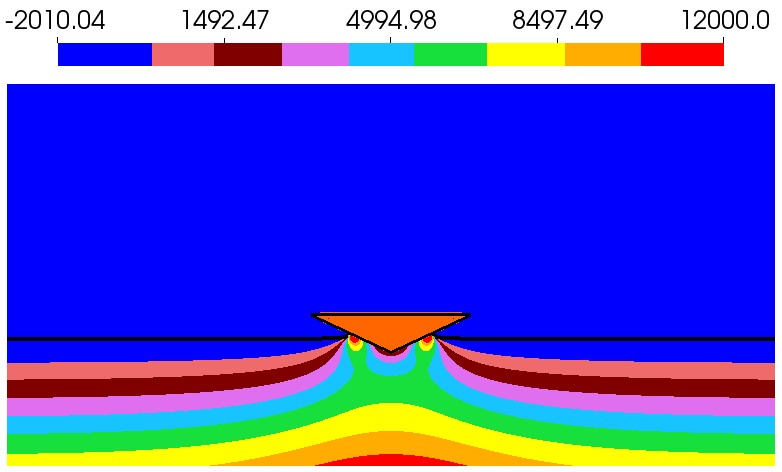} 
    \label{fig_cone_t2}
  }
    \subfigure[$t = 0.65$ s]{
    \includegraphics[scale = 0.25]{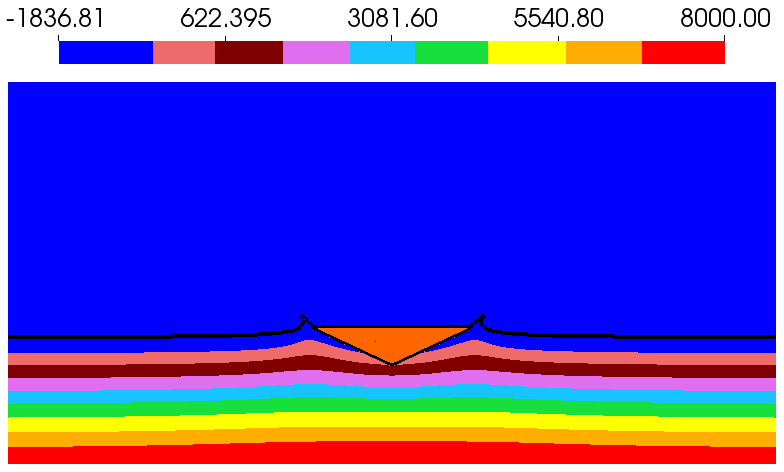}
    \label{fig_cone_t3}
  }    \subfigure[$t = 1.25$ s]{
    \includegraphics[scale = 0.25]{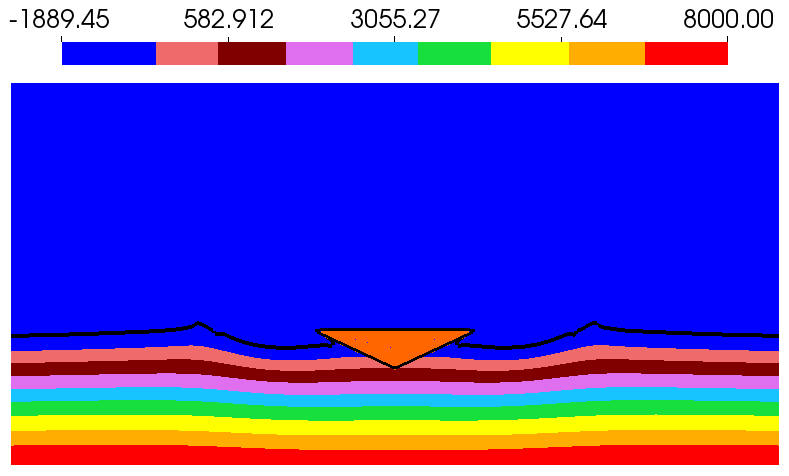}
    \label{fig_cone_t3}
  }
  \caption{Temporal evolution of pressure in the domain for a free falling 2D wedge slamming on an air-water interface.
}
\label{fig_temporal_evol_wedge_pressure}
\end{figure}

Figs.~\ref{fig_wedge_position} and~\ref{fig_wedge_velocity} illustrate the temporal evolution of the wedge's vertical position and velocity, respectively. The results are in good agreement with previous 3D volume of fluid simulations of Pathak et al.~\cite{Pathak16}~\footnote{Pathak et al. also impose a contact angle condition at the three material points in their volume of fluid simulation.} and experimental study of Yettou et al.~\cite{Yettou2006}. There is some mismatch between the 2D simulations and experimentally measured dynamics during later times. In contrast, the present 3D simulation agrees better with the 3D simulations of Pathak et al.~\cite{Pathak16} and the experimental results of Yettou et al.~\cite{Yettou2006}. Fig.~\ref{fig_wedge_velocity} illustrates how the vertical velocity of the wedge decreases as the wedge penetrates the water and how it reverses directions due to buoyancy forces. Fig.~\ref{fig_wedge_hydro_force} compares the vertical hydrodynamic force (viscous and pressure forces) on the wedge surface over time for both 2D and 3D cases. Some differences are observed. A peak load around the time of slamming ($t \approx 0.52$ s) can be observed in the figure.  
Fig.~\ref{fig_temporal_evol_wedge_decay} shows the evolution of wake behind the falling 2D wedge and the vortex shedding upon slamming. This type of wake has also been observed in experiments with falling cones~\cite{Hamed2015}. 
In Fig.~\ref{fig_temporal_evol_wedge_pressure} we show pressure in the domain at various times.  As the 2D wedge impacts, a high pressure region forms at its bottom tip. When the wedge penetrates further into the water, the high pressure region shifts to its inclined surface. After a period of time, the pressure in the domain becomes hydrostatic (increases linearly with depth).

\begin{figure}
  \centering
    \includegraphics[scale = 0.5]{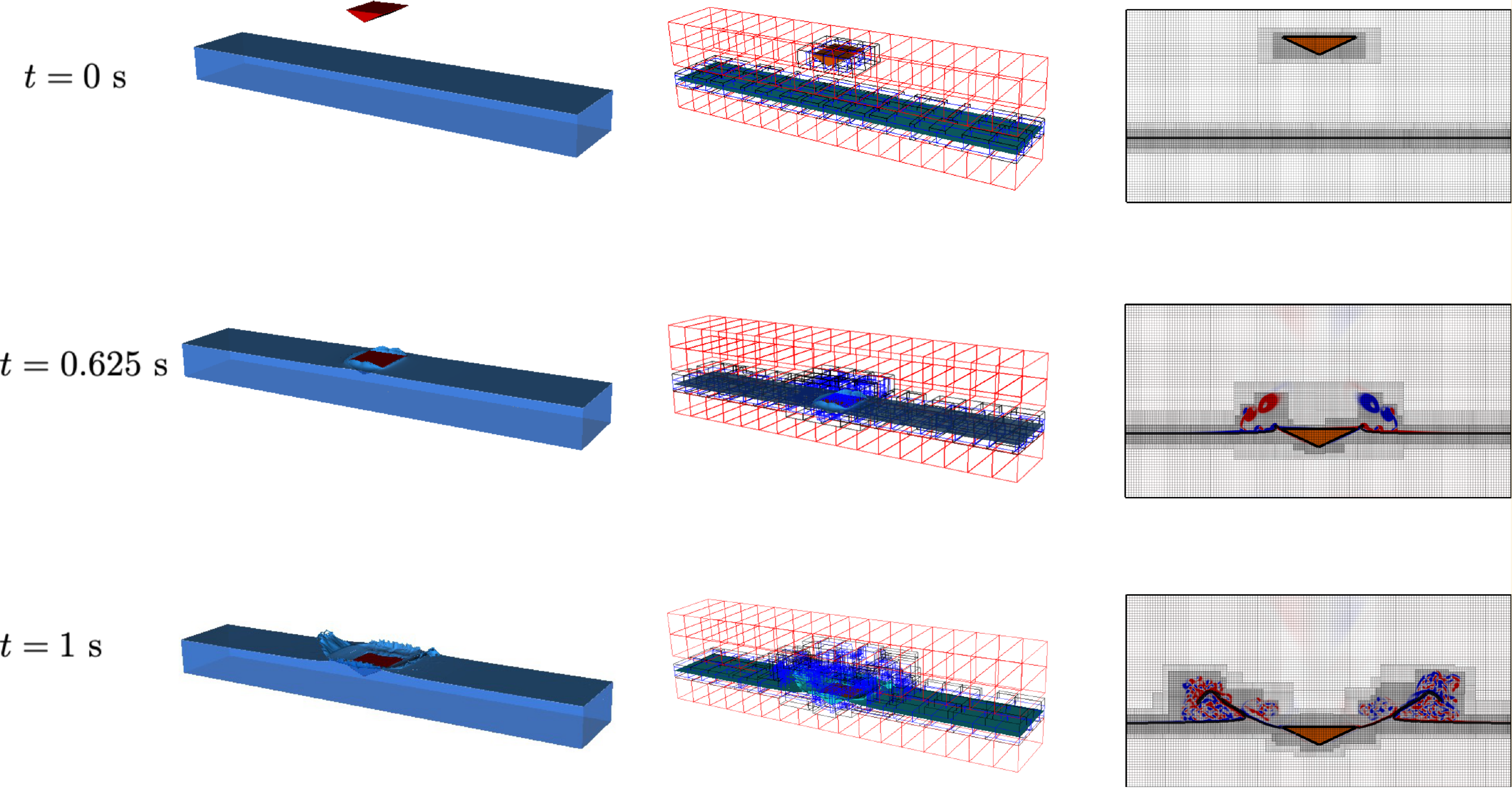} 
  \caption{Temporal evolution of a free falling 3D wedge slamming on an air-water interface using AMR: (left) density; (center) different mesh levels with the coarsest level shown in red boxes and the finest level shown in blue boxes; and (right) vorticity generated in the air region shown on a 2D slice of the domain at $y=5L/6$. Vorticity is in the range -50 to 50 $\text{s}^{-1}$.
}
\label{fig_temporal_evol_3dwedge}
\end{figure}

Fig.~\ref{fig_temporal_evol_3dwedge} shows the fluid-structure interaction of the 3D wedge case simulated using AMR at three distinct time instances: (left column) density plot; (center column) mesh levels; (right column) shed vortex structures shown on a 2D slice of the domain taken at $y=5L/6$ . Initially, the finest mesh level captures only the air-water interface and the region around the 3D wedge. More fine mesh regions are dynamically generated when the wedge slams the air-water interface, so that larger vortex structures shed in the air phase can be resolved accurately.

\begin{figure}
  \centering
    \subfigure[Percentage change in liquid volume]{
    \includegraphics[scale = 0.08]{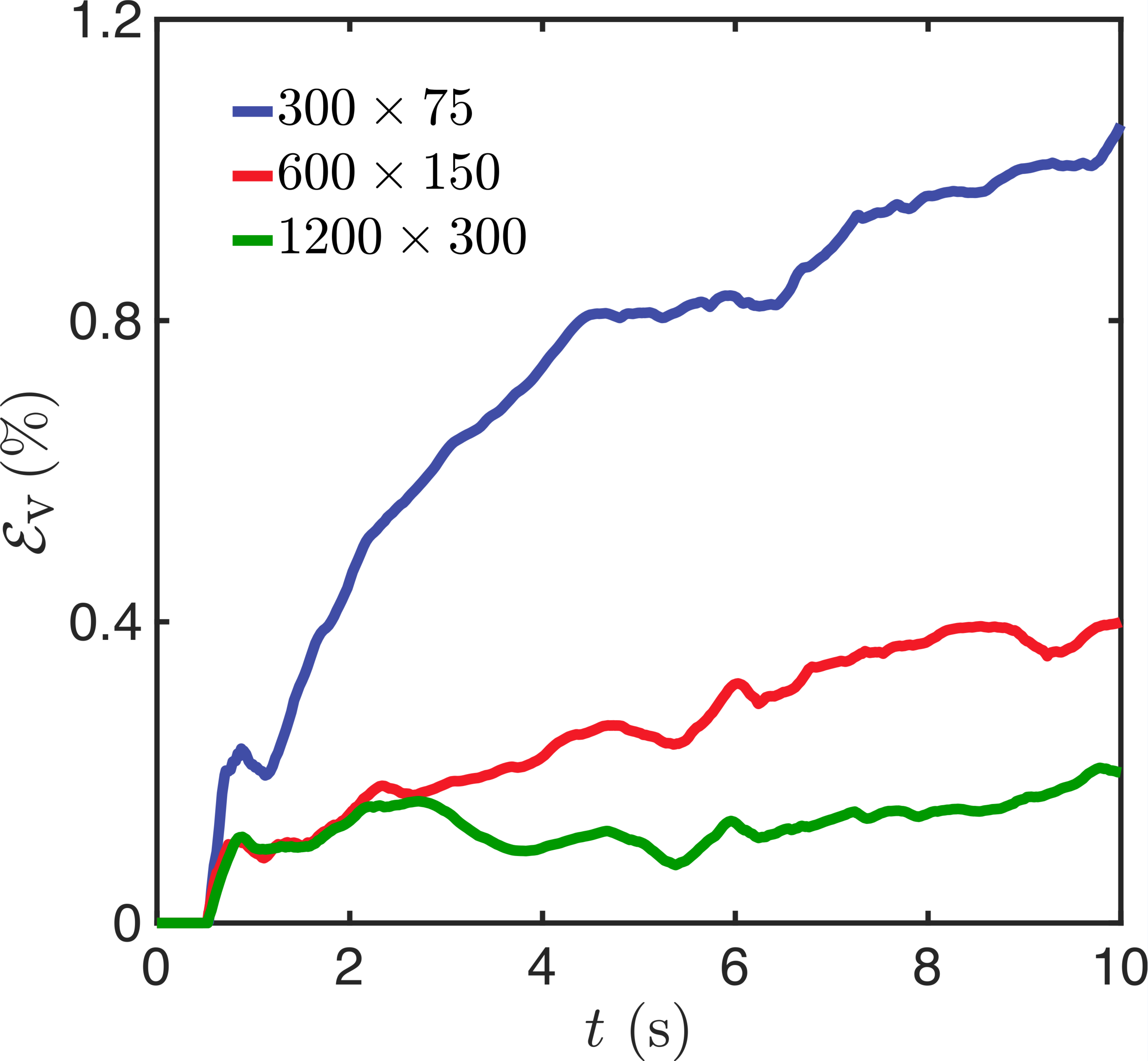} 
    \label{fig_wedge_liquid_volume}
  }
  \subfigure[Percentage change in gas volume]{
    \includegraphics[scale = 0.08]{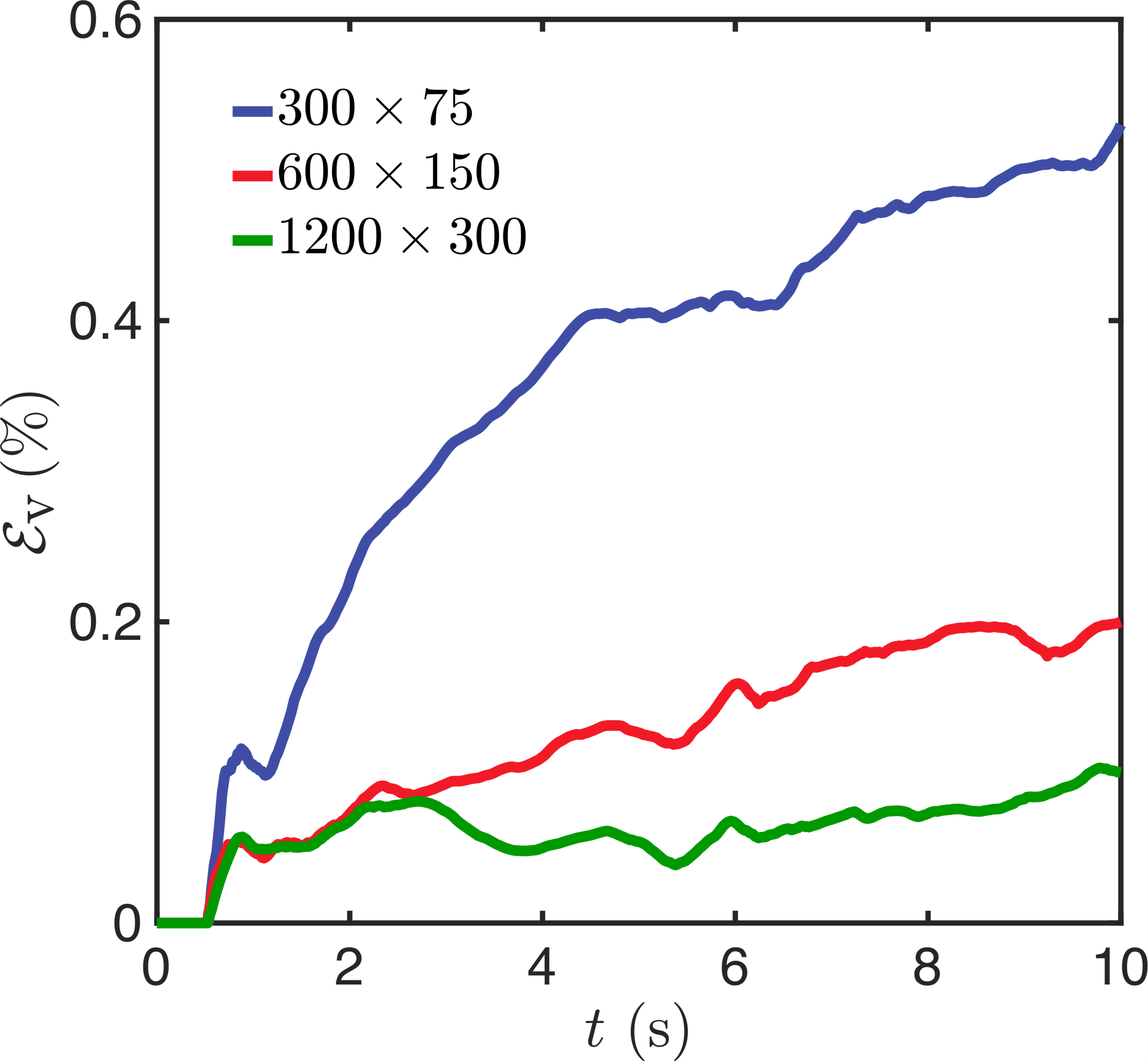} 
    \label{fig_wedge_gas_volume}
  }
 \caption{Percentage volume (or mass) change $\mathcal{E}_\textrm{v}$ of liquid and gas as a function of time for the two-dimensional wedge case  under grid refinement. 
}
\label{fig_wedge_volume_error}
\end{figure}

Fig.~\ref{fig_wedge_volume_error} shows the percentage volume (or mass) change for the liquid and gas phases in the 2D wedge case. The percentage error $\mathcal{E}_\textrm{v}$ is defined as $\mathcal{E}_\textrm{v} = \frac{|v(t) -v_0|}{v_0} \times 100$, in which $v(t) = \int_\Omega H^{\rm flow} \, \d \Omega$ is the volume occupied by the gas phase at time $t$,  and $v_0 = v(t = 0)$ is its volume at the beginning of the simulation. The liquid volume/mass can be defined similarly. The errors are below 1\% and they decrease as the grid refines.

\subsection{Choosing the right $\kappa$ value for the multiphase FSI model and its effect on the contact line}  \label{sec_kappa_effect_wedge}

\begin{figure}[]
\centering
\includegraphics[scale = 0.4]{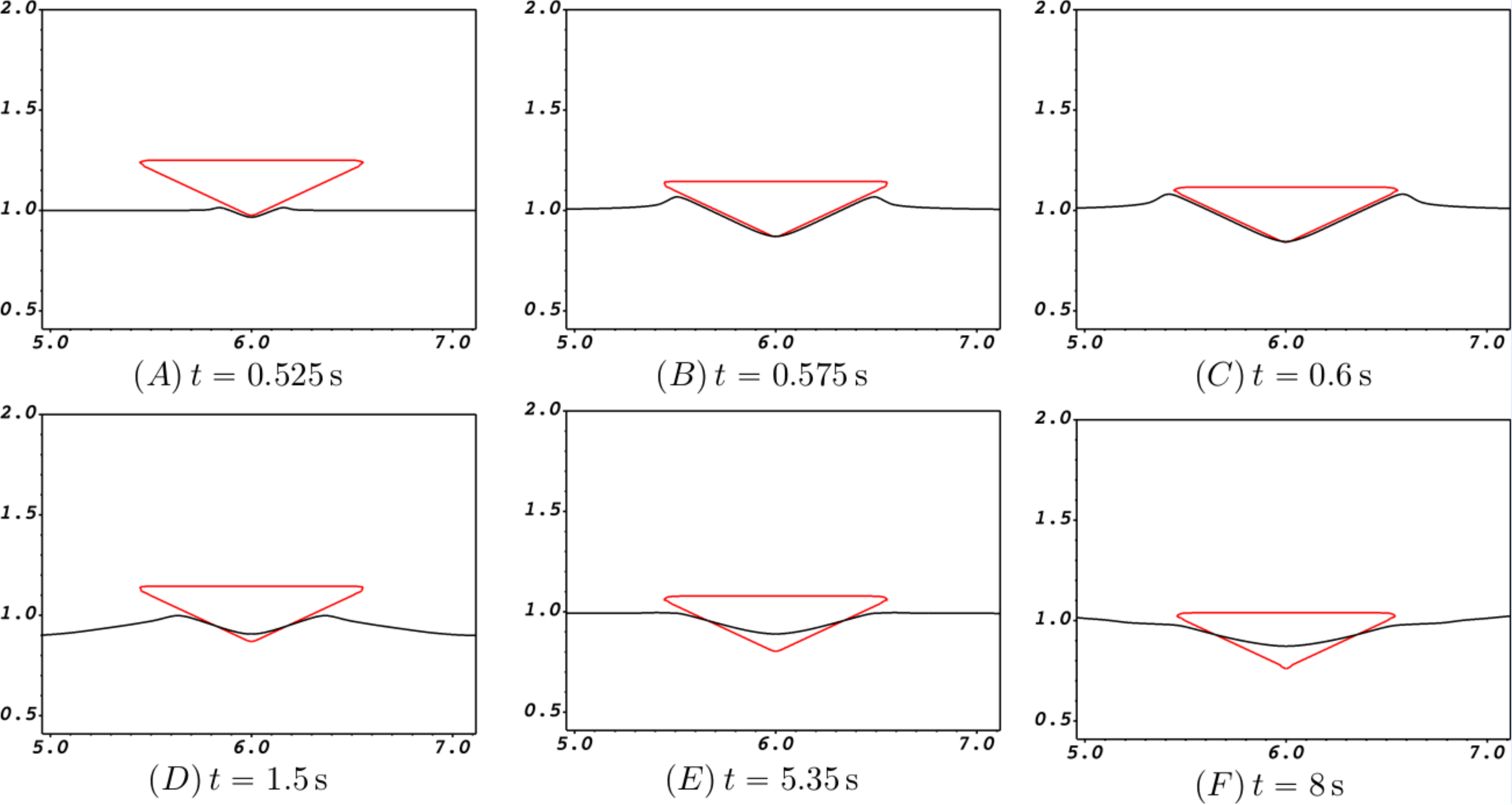}
  \caption{Evolution of the zero contours of fluid $\sigma$ (black line) and solid $\psi$ (red line) level set functions. }
  \label{fig_wedge_ls}
\end{figure}

In both 2D and 3D wedge problems considered in this section, the permeability coefficient is taken to be $\kappa = \Delta t/\rhos \sim \mathcal{O}(10^{-8})$. While this value is small, it does not approach machine precision. In the case of a small but non-zero value of $\kappa$, the Brinkman penalty term only weakly imposes the no-slip condition on the solid surface. Weak no-slip conditions on solid surfaces can also allow fluid to penetrate into porous solids and wet them. It is also possible to observe this wetting phenomenon in Fig.~\ref{fig_wedge_ls}, which illustrates the temporal evolution of the zero contours of the fluid and solid level set functions. When the wedge impacts the water surface, the liquid-gas interface deforms and conforms to the wedge shape. Over time, the liquid-gas interface gradually wets the wedge to conserve mass and to achieve force balance (weight of the wedge = buoyancy force). Once the system reaches mechanical equilibrium (when velocity is zero everywhere after a long time), the air-gas interface flattens out and returns to its original shape. 

Fluid leakage into an immersed structure is a common trait of immersed boundary (IB) methods that use penalty forces to enforce no-slip conditions. Most IB methods in this category use delta functions proposed by Peskin; for a description of the leakage problem for single phase IB methods see~\cite{Peskin02,Griffith2012vol,Kallemov16,bale2021one,cheron2023hybrid}. By lowering $\kappa$, it is possible to control flow leakage within the Brinkman/volume penalization approach. However, this will make the system of equations stiff. For the proposed preconditioner, the stiff system of equations is not an issue. The solver remains robust even at very low values of $\kappa \sim \mathcal{O}(10^{-12})$, as demonstrated in  the next Sec.~\ref{sec_results_and_discussion}. Nevertheless, very small $\kappa$ values break a key assumption behind the multiphase FSI model employed here: to conserve mass and to achieve force balance, we need the fluid to penetrate the structure. Moreover, we need to relax the no-slip boundary condition for the tangential velocity components to avoid singularities in tangential stress and pressure at the contact line. Considering the falling wedge problem, we can understand the need for normal penetration of the fluid into the impacting wedge as follows:
 
The enclosed domain consists of liquid and gas phases at $t = 0$. As far as material mass is concerned, the wedge is not present (recall that the structure domain is fictitious and the immersed body interacts only through the penalty term in the momentum equation). Imagine that the wedge comes to rest after a very long time, and we have imposed the no-penetration boundary condition perfectly throughout the simulation. For the tangential components of velocity we can assume either a slip or no-slip condition for the purpose of this discussion.  Eventually, the air-water interface will conform to the wedge geometry (also called the ``dry" contact line or the 180 degree contact angle condition), and both the stationary water level and wedge will be in equilibrium. However, this implies that the wedge has displaced the liquid and placed air in the region where it is submerged. At the stationary air-water interface comprising both horizontal and wedge-shaped regions, the hydrostatic pressure level is the same. Physically, this is not possible as the horizontal and curved-down regions of the interface are at different elevations. Instead, the method allows liquid and gas to penetrate the wedge, conserving both phases' mass and achieving the same (hydrostatic) pressure at the air-water interface. In the model, if $\kappa$ is not too small, this mechanism is permitted. Fluid leakage also means contact angle conditions cannot be controlled directly, and they emerge numerically from mass and force balance conditions.  Our simulations demonstrate this in Fig.~\ref{fig_wedge_ls}. Initially, the contact angle appears to be 180 degrees. However, over time, the fluid penetrates into the porous solid at some numerical or apparent angle less than 180 degrees.  A further discussion on the issue of fluid entering the immersed structure is provided at the end of this section.

\begin{figure}
  \centering
  \subfigure[Vertical postion 2D wedge]{
    \includegraphics[scale = 0.35 ]{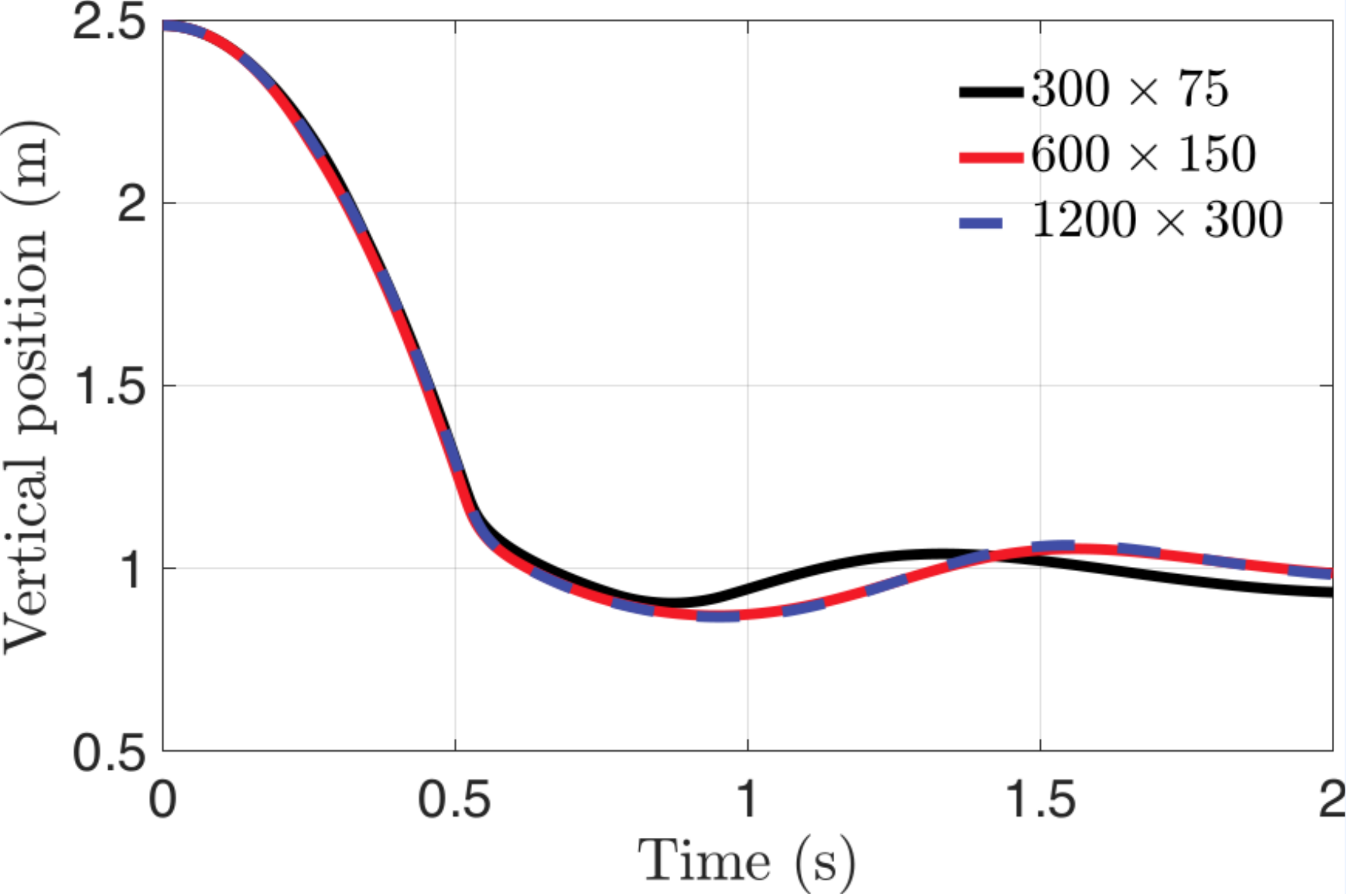} 
    \label{fig_wedge_position_convg}
  }
   \subfigure[Vertical velocity 2D wedge]{
    \includegraphics[scale = 0.35]{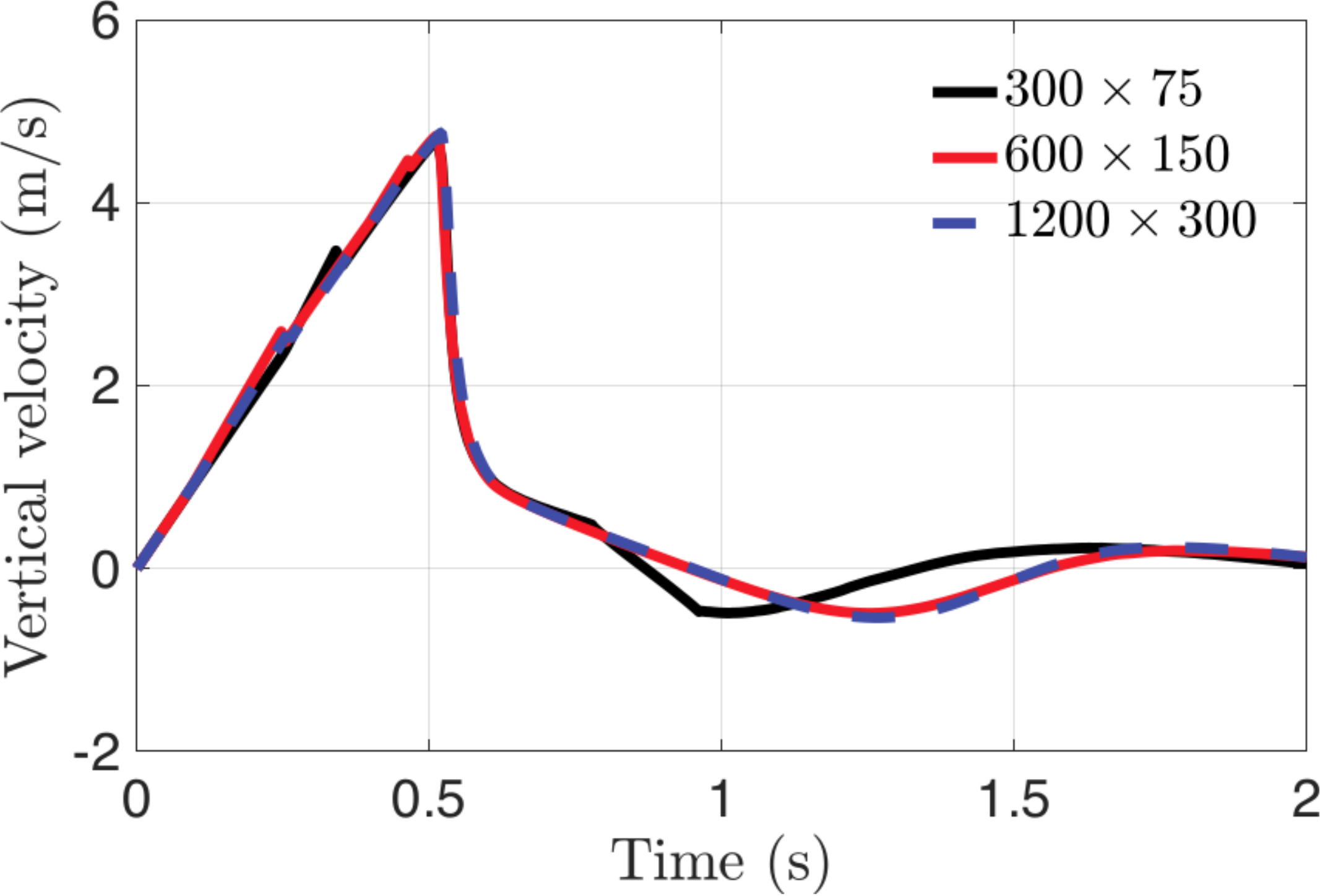}
    \label{fig_wedge_velocity_convg}
  }  
  \subfigure[Liquid-gas interface]{
    \includegraphics[scale = 0.5]{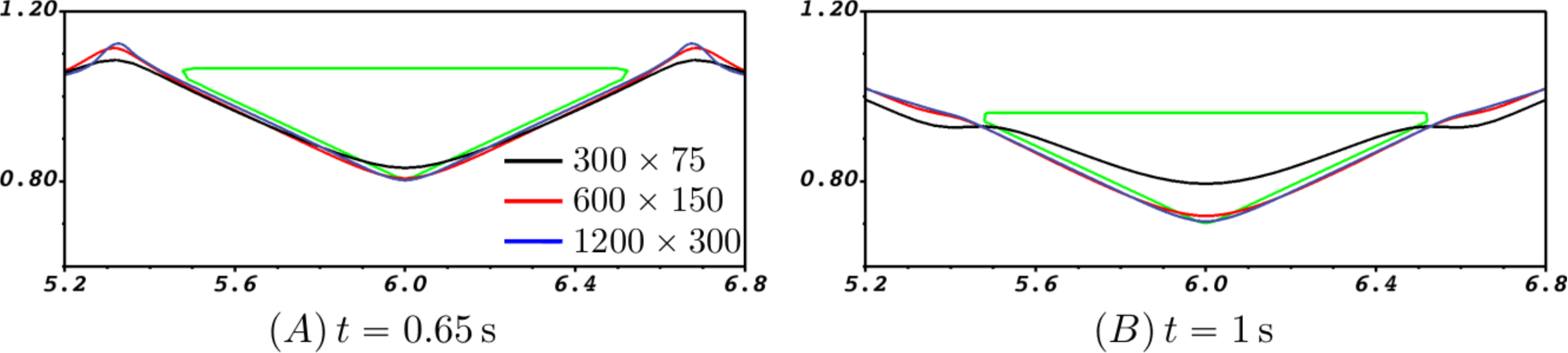}
    \label{fig_liquid_gas_interface_convg}
  }  
  \caption{Grid convergence study for the free falling 2D wedge case: \subref{fig_wedge_position_convg} vertical position; \subref{fig_wedge_velocity_convg} vertical velocity;  and \subref{fig_liquid_gas_interface_convg} liquid-gas interface at two different time instants. $\kappa = \Delta t/\rhos$ is used as the penalty factor. A uniform time step size of $\Delta t = 2.5 \times 10^{-4}$ s  is employed for the coarse grid, which is halved for each successively refined grid.  
}
\label{fig_grid_convergence_2dwedge}
\end{figure}

Based on empirical tests, $\kappa = \Delta t/\rhos$ represents a robust choice for multiphase FSI models. This results in a continuous air-water interface near the triple points and within the solid, as well as converged FSI dynamics. In Fig.~\ref{fig_grid_convergence_2dwedge}, we show the convergence of 2D wedge velocity and position, and liquid-gas interface at three grid resolutions: coarse, medium, and fine. Numerical instability occurs when $\kappa$ is lowered by a factor of 2 or more. In the simulations we observe a sudden large rise in pressure and velocity magnitude at triple points when the wedge meets the air-water interface; very low $\kappa$ values do not affect the stability of the simulations when the wedge is completely in the air phase. This is because the no-slip condition for the tangential velocity components at the contact line leads to singularities in shear stress and pressure. Huh and Scriven analyzed this situation analytically in their 1971 paper~\cite{huh1971hydrodynamic}, where they considered a solid driven into phase B from phase A (similar to the wedge problem considered in this section). The authors showed that the classical no-slip condition (for the tangential velocity) breaks down as it leads to infinite stress and pressure at the contact line and remarked that ``\textit{not even Herkales [Hercules] could sink a solid if the physical model [i.e., the no-slip model] were entirely valid, which it is not.}" As a possible means to remedy this situation, Huh and Scriven suggest using a slip model (see Eq. 37 of \cite{huh1971hydrodynamic}) in the tangential direction, which is what the Brinkman penalty term is also doing. Furthermore, the authors in~\cite{huh1971hydrodynamic} also suggest relaxing the no-slip condition in the normal direction, where they remark ``\textit{Relaxing the normal component of the adherence condition, Eqs. [2] and [5], in the immediate vicinity of the contact line---to allow for fracture, for example--is another possibility, but seems to demand slip anyway, along the fluid interface.}" Thus, low but non-zero $\kappa$ values in the volume penalization approach allow tangential slip and mass flux into the solid. The simulations presented in this section confirm Huh and Scriven's theoretical analysis. Multiple viewpoints (e.g., force and mass balance at the final equilibrium position and singularities in shear stress and pressure) lead to the same conclusion that $\kappa$ should not be taken too low.  Further physical insights on the effect of $\kappa$ on the contact line/angle are provided in Sec.~\ref{sec_heaving_cyl}, where we consider a different multiphase FSI problem (a rigid cylinder heaving on an air-water interface).    

Last but not least, we discuss the usefulness of our multiphase FSI model for certain classes of problems (e.g., wave energy converters, water entry/exit of marine structures), despite the fact that a specified contact angle condition cannot be directly imposed and that fluid penetrates the structure.  For the former issue, at the scale of these ocean engineering problems, the contact angle boundary condition, or for that matter, the surface tension force as a whole does not affect the device/structure dynamics. As a concrete example, let's consider the water entry of the wedge simulated in this section. The structure length scale is $L = 1.2$ m, whereas the capillary length scale is $l_c \propto \sqrt{\frac{\gamma}{(\rho^{\rm L} - \rho^{\rm G}) g}} \approx 2.7$ mm. The structure is $\approx$ 500x larger than the capillary length. The grid resolution that we used (same as that of Pathak et al.~\cite{Pathak16}, who impose static contact angle conditions [the validity of this assumption is questionable during the dynamic phase of impact] over the surface of the impacting wedge) to capture the rigid body dynamics of the wedge is $\Delta x = \Delta y= \Delta z = 0.01 $ m or 10 mm, which is 4x larger than the capillary length scale. Now imagine that the wedge comes to rest after a long period of time and the entrained air near the contact line has either dissolved or escaped. In this scenario the static contact angle condition becomes valid. Assuming the wedge is metallic (hydrophilic surface with $\theta_{\rm static} = 60^\circ < \pi/2$) the air-water interface/meniscus will rise by a height of $l_c \cot(\theta_{\rm static}) \approx 1.55$ mm, which is $\approx 777$x smaller than the wedge dimensions, and falls in the subgrid scale ($\approx$ 10x smaller than the cell size). In more practical ocean engineering problems, marine structures typically span a length of 25 - 500 m. Therefore, it is computationally unfeasible/impractical to resolve capillary length scales/triple points along with structure and wave dynamics. Furthermore, the dynamic contact angle conditions under these highly turbulent and unsteady conditions are not well known/studied. 

Regarding the issue of fluid penetrating the structure, technically speaking, this should not happen. If this happens it means that some fluid is taken from the surrounding reservoir. For ocean engineering problems, the reservoir volume is very large in comparison to how much fluid penetrates the structure. Fluid-structure interaction is typically not affected by this. There are specialized numerical techniques (e.g., cut-cell methods) that prevent this from happening~\cite{xie2020three,van2023two}. These methods also employ discontinuous interface capturing methods like the volume of fluid method to truncate the air-water interface on the surface of the body. The VP method employing the level set method assumes the air-water interface is continuous which does not truncate on the structure surface. It remains a future endeavor to develop a VP/fictitious domain method that (i) does not permit fluid from entering the structure; (ii) satisfies mass balance and momentum equations; (iii) imposes a specified contact angle condition; and (iv) keeps the air-water interface continuous/smooth.  To summarize, the current multiphase FSI model can only be used for problems that do not depend critically upon imposing a specific contact angle condition (i.e., problems with length scale much larger than the capillary length scale), and where the reservoir volume is much larger than the structure volume.  As a final note on the multiphase FSI model, there have been several studies in the literature which have ignored contact angle conditions at material triple points, and have allowed the liquid-gas interface to penetrate/exist within the solid region; see for example~\cite{Zhang2010,Patel2018,Calderer2014,Sharaborin2021,bergmann2022numerical,sanders2011new}.  In~\cite{Patel2018} the authors mention that ``\textit{In the present work, we assume that the solids are filled with a ``virtual" fluid with density and viscosity equal to the largest among all fluids in the domain.}" In~\cite{Sharaborin2021} the authors show the penetration of the liquid-gas interface into the initially dry solid for their coupled volume of fluid and Brinkman penalization approach.  And in~\cite{sanders2011new} the authors show the air-water interface passing through a floating buoy in a numerical wave tank. A coupled level set and immersed boundary method (similar to the VP method) was used in \cite{sanders2011new}.


\section{Results and discussion}
\label{sec_results_and_discussion}

In this section, we consider two nontrivial test problems to demonstrate the efficacy of the projection preconditioner to solve the coupled velocity-pressure system written in Eq.~\eqref{eq_stokes_system}. In the first problem, a uniform density and viscosity flow is considered in a complex domain. Using the method of manufactured solutions, we compute the spatial order of accuracy of the solution ($\u$ and $p$), and monitor the number of iterations the outer FGMRES solver takes to converge with decreasing values of $\kappa$.   In the second case, we study the free-decay of a rigid cylinder heaving on an air-water interface. The heave displacement of the cylinder is compared against literature to assess the accuracy of the FSI solution. The number of iterations taken by the Krylov solver to converge are monitored for this case as well. In contrast to the first problem, the density and viscosity of the three phases (solid, liquid, gas) differ by orders of magnitude in the second problem.  

In the tests, the outer FGMRES solver is deemed to be converged if a value of $10^{-9}$ or below is reached for the norm of the relative residual 
\begin{equation}  
\mathcal{R} = \frac{|| \r || } { || \b || }   =  \frac {|| \b - \M  \x|| } { || \b || }.  \label{eq_residual_def}
\end{equation} 
For velocity and pressure subdomain problems, the inner Richardson solver (that is preconditioned with a single multigrid V-cycle) is set to use only a single iteration. We remark that if the preconditioner employs a fixed number of Richardson iterations (as considered here, which is equal to one), then it is also possible to use the more memory-efficient GMRES solver as the outer Krylov solver. However,  here we report the convergence rate of the FGMRES solver (which requires roughly twice the amount of memory compared to GMRES), as it exhibits more uniform convergence behavior across a wide range of thermo-physical parameters, i.e., $\mu$ and $\rho$.

\subsection{Uniform density and viscosity flow in a complex domain}
\label{sec_mms}

\begin{figure}[]
\centering
\includegraphics[scale = 0.08]{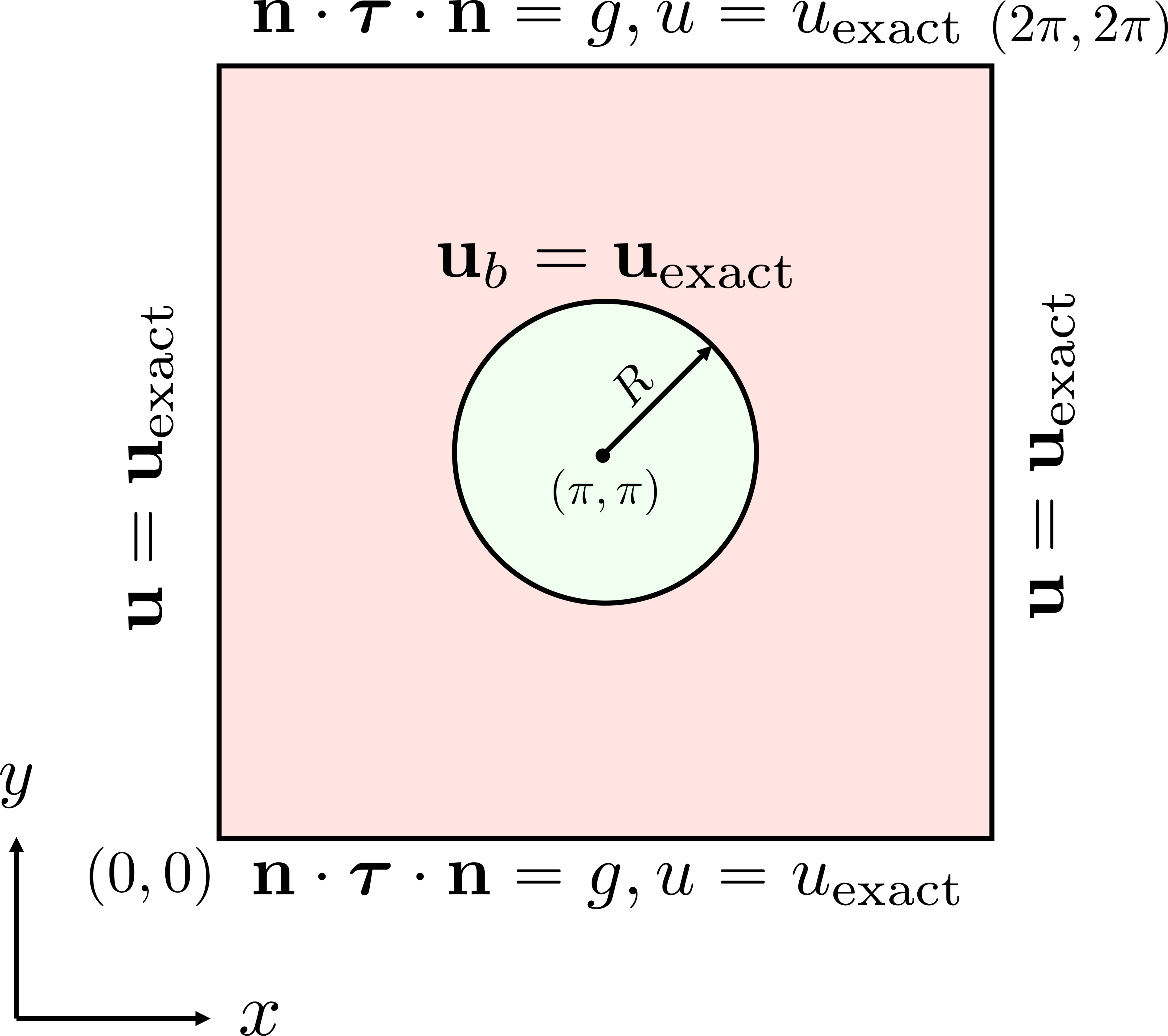}
  \caption{Schematic of the computational domain. Here, $g = -p + 2\mu \frac{\partial v}{\partial y}  = \sin(x)\sin(y) - 2\cos(x)\cos(y)$ is the imposed normal traction on the top and bottom boundaries. For the top boundary $\n = (0, 1)$ and for the bottom boundary $\n = (0, -1)$. }
  \label{fig_mms_schematic}
\end{figure}

Consider a computational domain $\Omega \in [0, 2\pi]^2$ which embeds a circular cylinder of radius $R = 1.5$ at its center $(\pi,\pi)$ as shown in Fig.~\ref{fig_mms_schematic}. A steady state manufactured solution for velocity $\u$ and pressure $p$
\begin{align}
&\uexact (\x, t \rightarrow \infty) = \sin(x) \cos (y), \label{eq_mms_u} \\
& \vexact (\x, t \rightarrow \infty) = - \cos (x) \sin (y),  \label{eq_mms_v} \\
& \pexact(\x, t \rightarrow \infty) = \sin(x)\sin(y) \label{eq_mms_p},
\end{align}
is used to drive a constant density $\rho(\x,t) \equiv 1$ and viscosity $\mu(\x,t) \equiv 1$ flow in the domain. Specifically,  Eqs.~\eqref{eq_mms_u}-\eqref{eq_mms_p} are plugged into the momentum Eq.~\eqref{eqn_momentum} to determine the body force $\f$ that drives the flow. The manufactured solution is also used to impose boundary conditions on $\partial \Omega$ and inside the fictitious cylinder $\Omega_b$. On the left and right ends of the domain, we impose velocity boundary conditions $u= \uexact$ and $v = \vexact$. On the top and bottom boundaries a combination of normal traction $\n\cdot \V{\tau} \cdot\n = g =  -p + 2\mu \frac{\partial v}{\partial y} = \sin(x)\sin(y) - 2\cos(x)\cos(y)$ and tangential velocity $u = \uexact$ condition is imposed. Here, $\V{\tau} = -p \V{I} + \mu \left(\grad \u + \grad \u^\intercal \right)$ denotes the hydrodynamic stress tensor. The velocity inside the cylinder is prescribed to be $\u_b = \u_\text{exact}$.  The initial conditions for velocity and pressure are taken to be zero. 


Five grid sizes $N \times N = \{64^2, 128^2, 256^2, 512^2, 1024^2 \}$ are used to run the simulations starting from $t = 0$ till steady state is reached. $\ncells = 1$ is used in Eq.~\eqref{eqn_Hbody} to smear the fluid-solid interface.  A constant time step size of $\Delta t = 1\times10^{-3}$ is used in all simulations, which maintains convective CFL number below 0.35 for all grid sizes $N$. The permeability coefficient $\kappa$ should be kept small, but not too small to avoid the plateauing of spatial discretization errors~\cite{Angot1999, Sharaborin2021}. To study the effect of $\kappa$ on the spatial order of accuracy of $\u$ and $p$ solutions, as well as on the solver convergence rate, we consider three different values for $\kappa =   \{ \Delta t/\rho, \, \Delta t/100 \rho, \, \Delta t/10000 \rho \}$ in the numerical experiments. Note that  $\Delta t/\rho$ is the maximum value which $\kappa$ can or should take as per the inertial scale $\chi/\kappa \sim \rho/\Delta t$.  Another possibility is to select $\kappa$ based on the viscous scale $\chi/\kappa \sim \mu/h^2$, in which $h$ is the uniform cell size. Here, our strategy is to start with the maximum value of $\kappa$ based on the inertial scale, and then reduce it progressively till no further improvement in the solution is observed.  

\begin{figure}[]
  \centering
  \subfigure[$\kappa= \Delta t/\rho$]{
    \includegraphics[scale = 0.08]{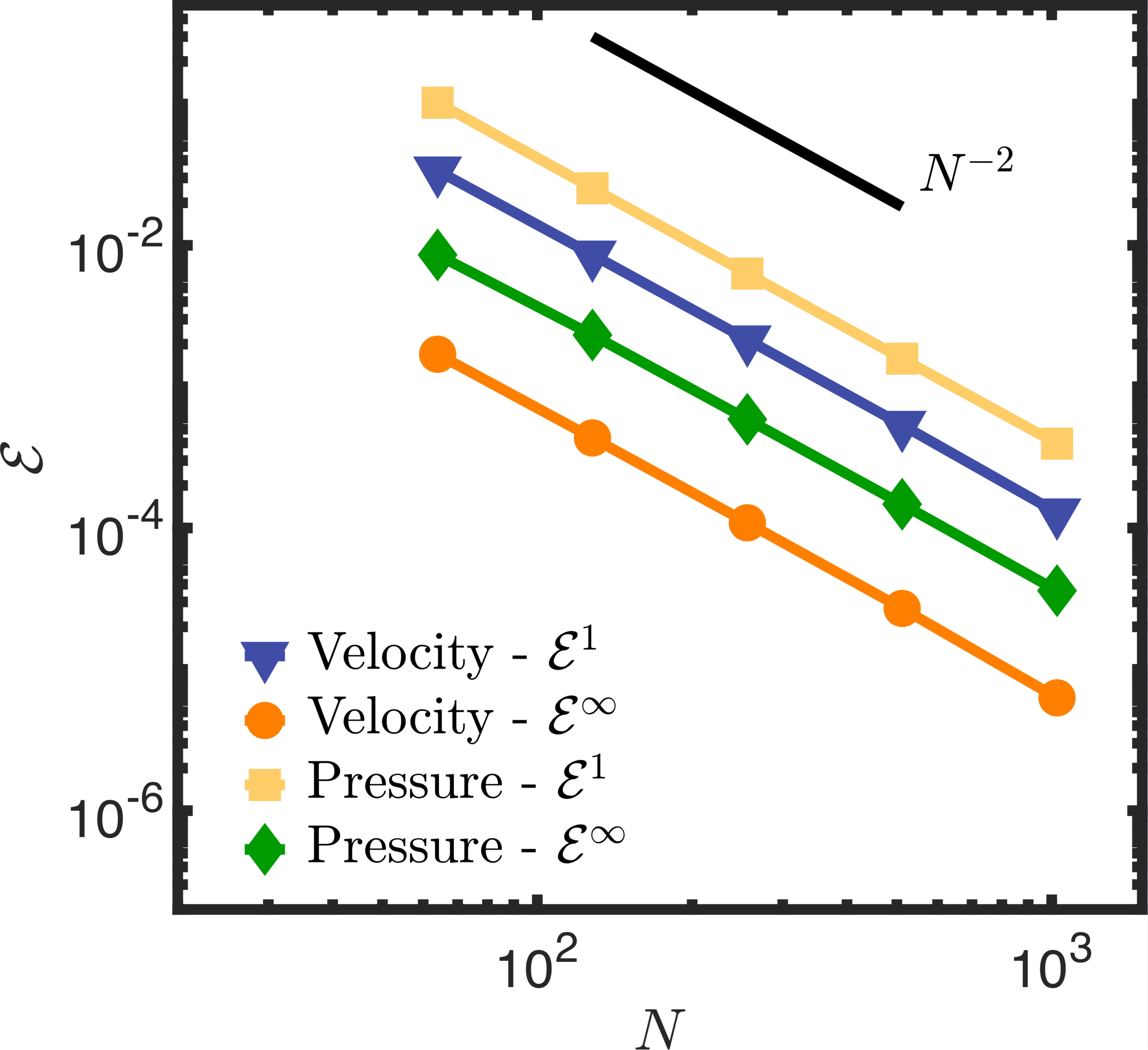} 
    \label{fig_ooa_k1e0}
  }
   \subfigure[$\kappa=\Delta t/100\rho$]{
    \includegraphics[scale = 0.08]{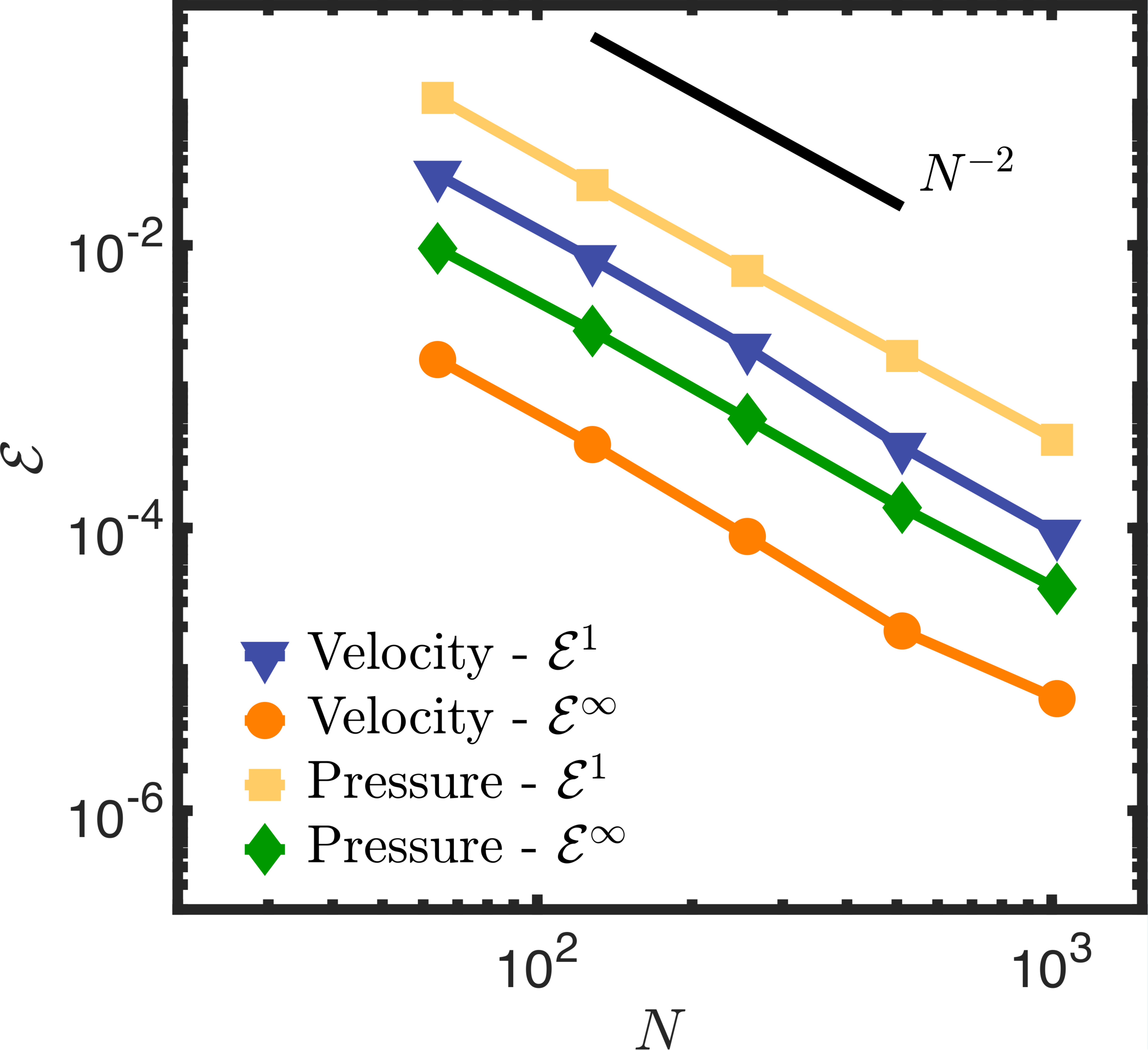}
    \label{fig_ooa_k1e2}
  }
  \subfigure[$\kappa=\Delta t/10000\rho$]{
    \includegraphics[scale = 0.08]{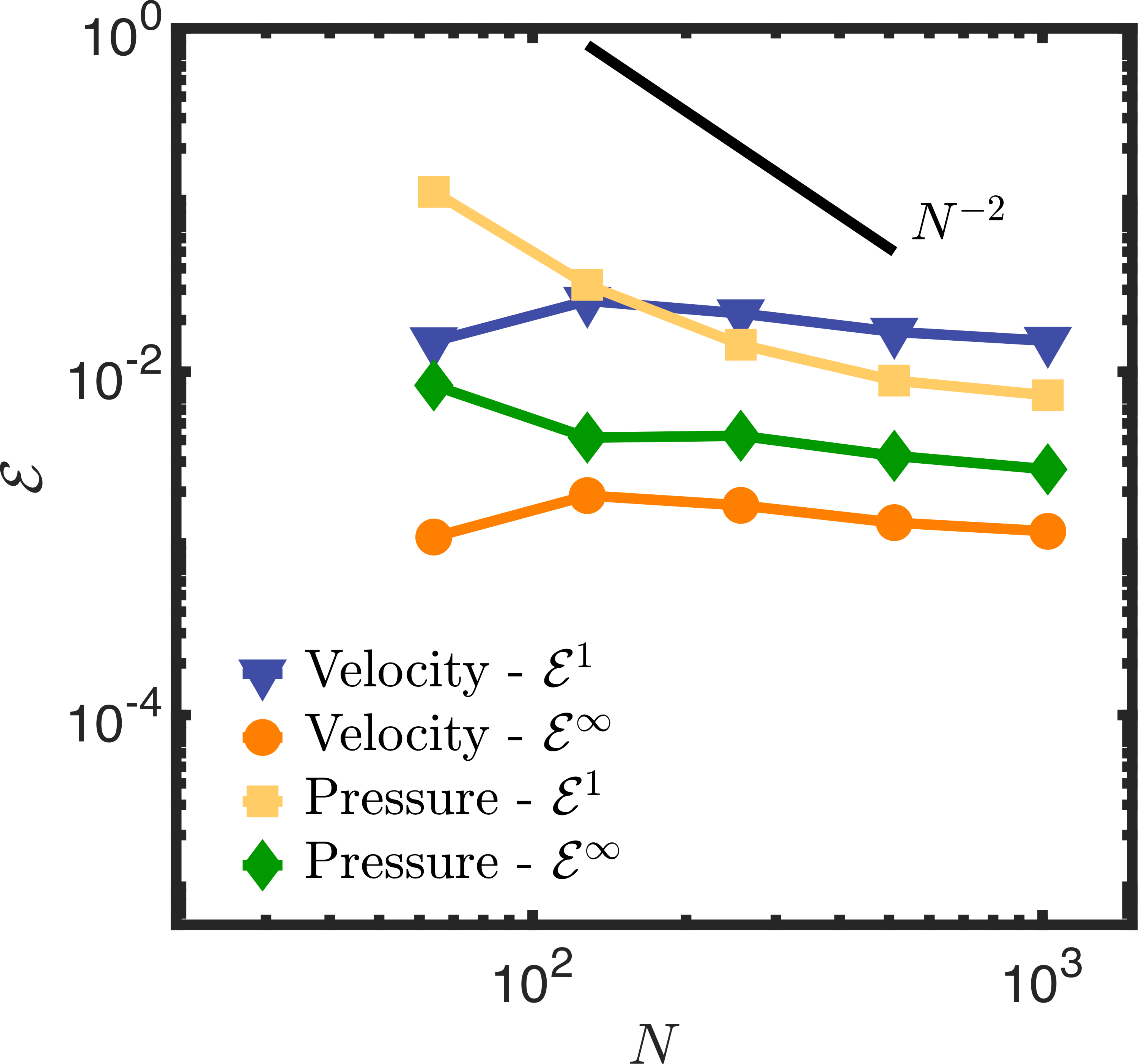}
    \label{fig_ooa_k1e4}
  }
  \caption{Spatial order of convergence of the volume penalized Navier-Stokes system using manufactured solutions and different values of $\kappa$.}
\label{fig_ooa}
\end{figure}

\begin{figure}[]
\centering
\includegraphics[scale = 0.08]{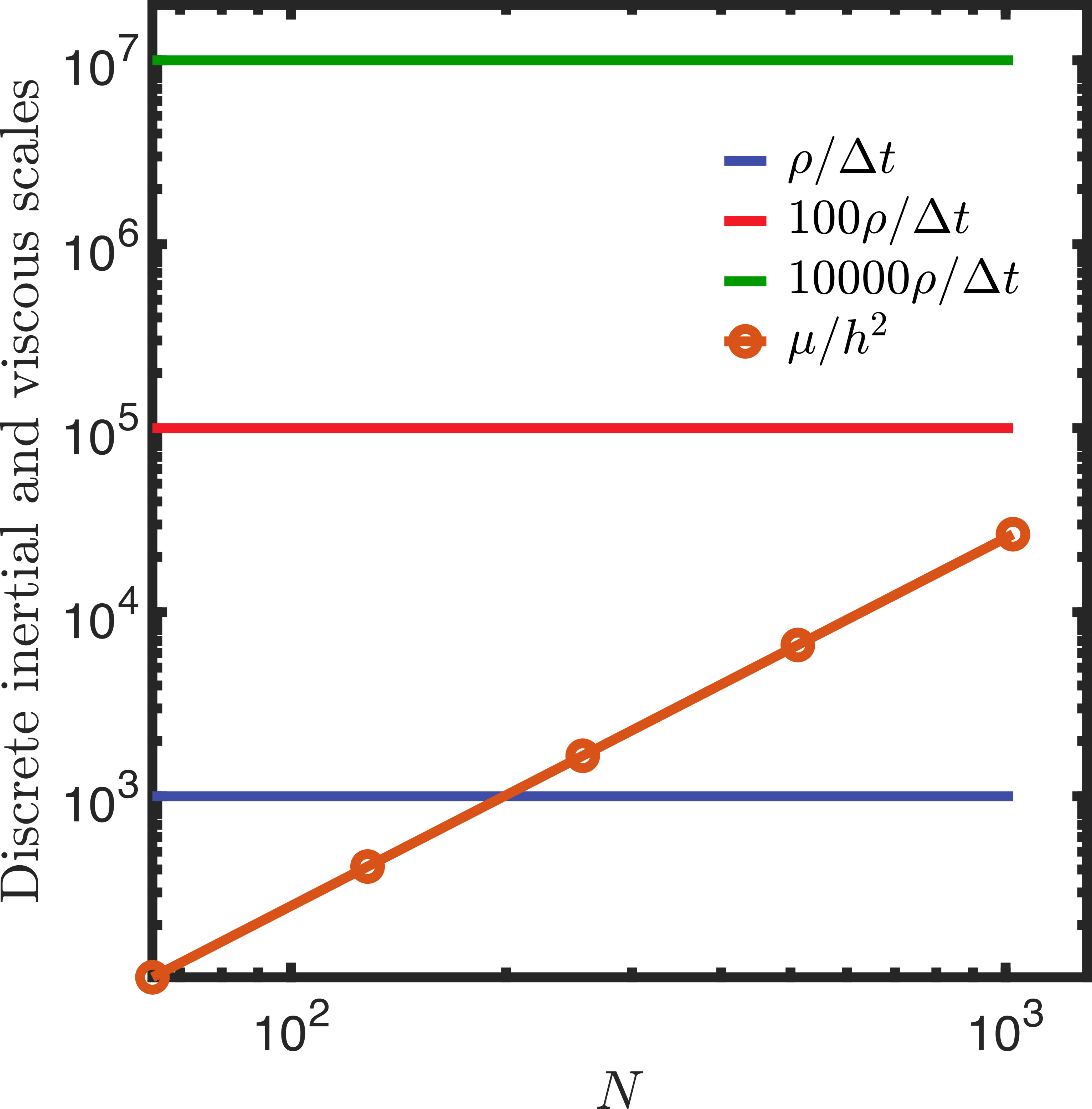}
  \caption{Discrete inertial and viscous scales as a function of grid size $N$. }
  \label{fig_N_vs_kappa}
\end{figure}

We first present the spatial order of accuracy of the Brinkman penalized $\u$ and $p$ solutions. The error between steady state numerical and analytical solutions is denoted $\mathcal{E}$. As can be observed in Fig.~\ref{fig_ooa}, second-order pointwise ($L^\infty$-norm) convergence rate is obtained for both velocity and pressure errors when $\kappa = \Delta t/\rho, \; \Delta t/100 \rho$.  For a very small value of $\kappa = \Delta t/10000\rho = 1 \times 10^{-7}$, the error in velocity and pressure saturates after a certain grid size ($N = 128$ in Fig.~\ref{fig_ooa_k1e4}). To understand this trend, we compare the magnitudes of the Brinkman penalty coefficient $\chi/\kappa$ and the discrete inertial and viscous scales, by plotting the latter two as a function of grid size $N$ in Fig.~\ref{fig_N_vs_kappa}. The discrete inertial scale $\rho/\Delta t$ remains constant (because of the constant time step size $\Delta t$), whereas the discrete viscous scale varies quadratically (linearly on a log scale) with $N$. The magnitude of the inertial scale is larger than the viscous scale for the first two grids, and vice versa for the remaining grids.  For the largest value of $\kappa = \Delta t/\rho$, although the  penalty  coefficient $\chi/\kappa = \rho/\Delta t$ is not always larger than the viscous scale, second-order convergence is still observed.  When $\kappa = \Delta t/100\rho$, the Brinkman penalty coefficient is significantly larger than the inertial scale and comparable with the viscous scale. Second-order pointwise convergence is obtained for this $\kappa$ value as well. However, when $\kappa$ is decreased further to $\Delta t /10000\rho$, the Brinkman penalty becomes four orders larger than the inertial scale and two orders larger than the viscous scale. This large penalty value causes the errors to saturate.  Note that in this problem we imposed spatially-varying traction boundary condition on the top and bottom boundaries of $\partial \Omega$, and obtained second-order convergence rates for velocity and pressure solutions. This is not possible to achieve if pressure and velocity are solved in a split manner using projection method as a solver.

Next, we study the impact of $\kappa$ on the convergence rate of the preconditioned FGMRES solver. For this problem, the first time step poses the most difficultly for the (iterative) Krylov solver as the initial zero guess for velocity and pressure is far-off from the true solution at $t = \Delta t$.  Therefore, it suffices to monitor the solver performance at the first time step only to evaluate the efficacy of the preconditioner. We run this test case with and without the Brinkman penalty term $ \widetilde{\V{\chi}} / \kappa$ in the projection preconditioner.  In practice, we re-define $\vrhochi^{-1}$ with the help of a boolean parameter $\theta$  
\begin{equation} 
\vrhochi^{-1} = \left( \breve{\V \rho} +  \theta \frac{\widetilde{\V \chi} \dt}  {\kappa} \right)^{-1}, \label{eq_rhochi_theta}
\end{equation}
so that by setting $\theta = 1$ in $\vrhochi^{-1}$ we obtain the new projection method and by setting $\theta = 0$ we revert to the projection methods of Bergmann and Iolla~\cite{Bergmann2011} and Sharaborin et al.~\cite{Sharaborin2021}.

\begin{figure}
  \centering
  \subfigure[Relative residual when $\theta = 0$]{
    \includegraphics[scale = 0.08]{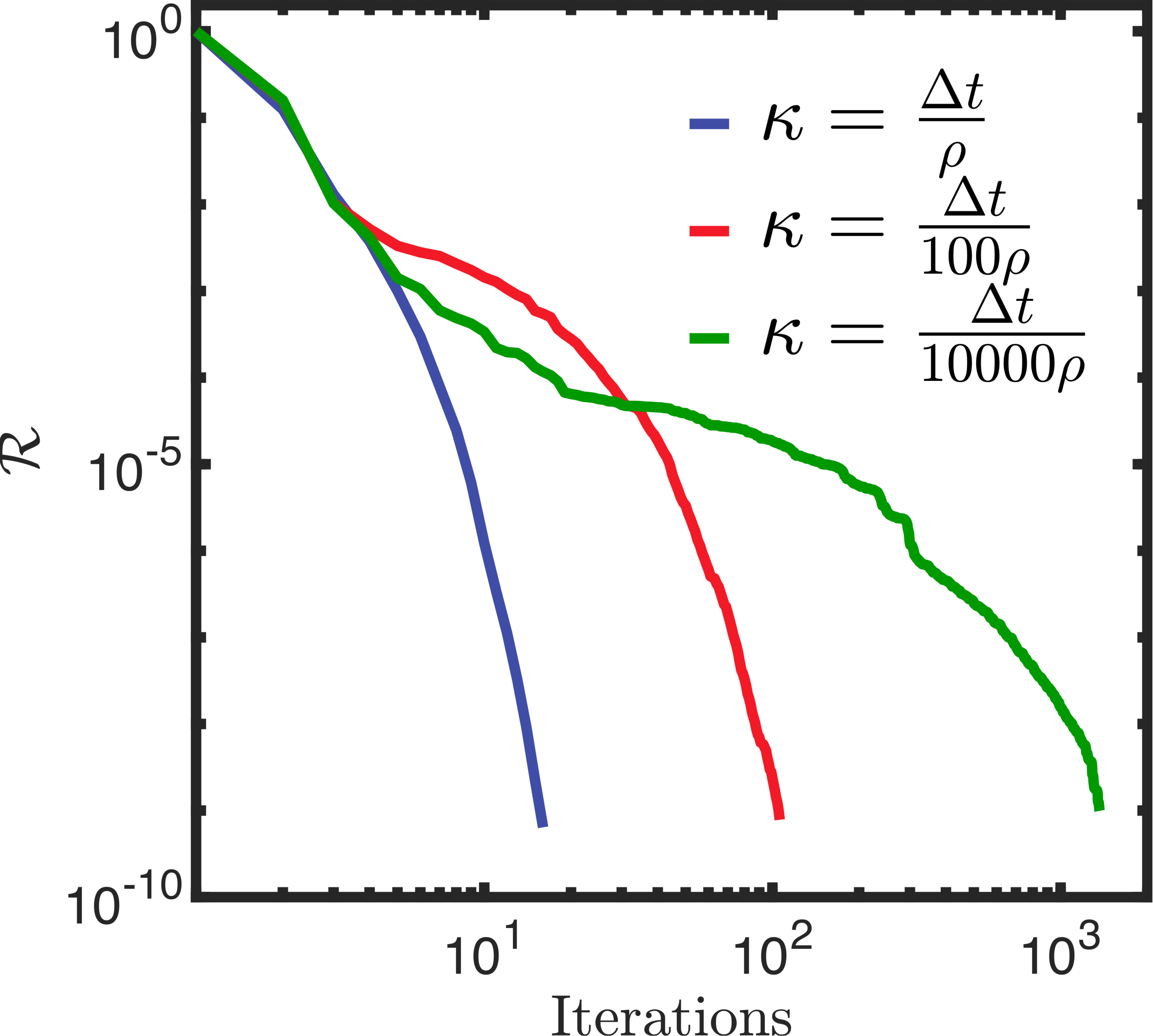} 
    \label{fig_residual_without_brinkman_in_PPE}
  }
   \subfigure[Relative residual when $\theta = 1$]{
    \includegraphics[scale = 0.08]{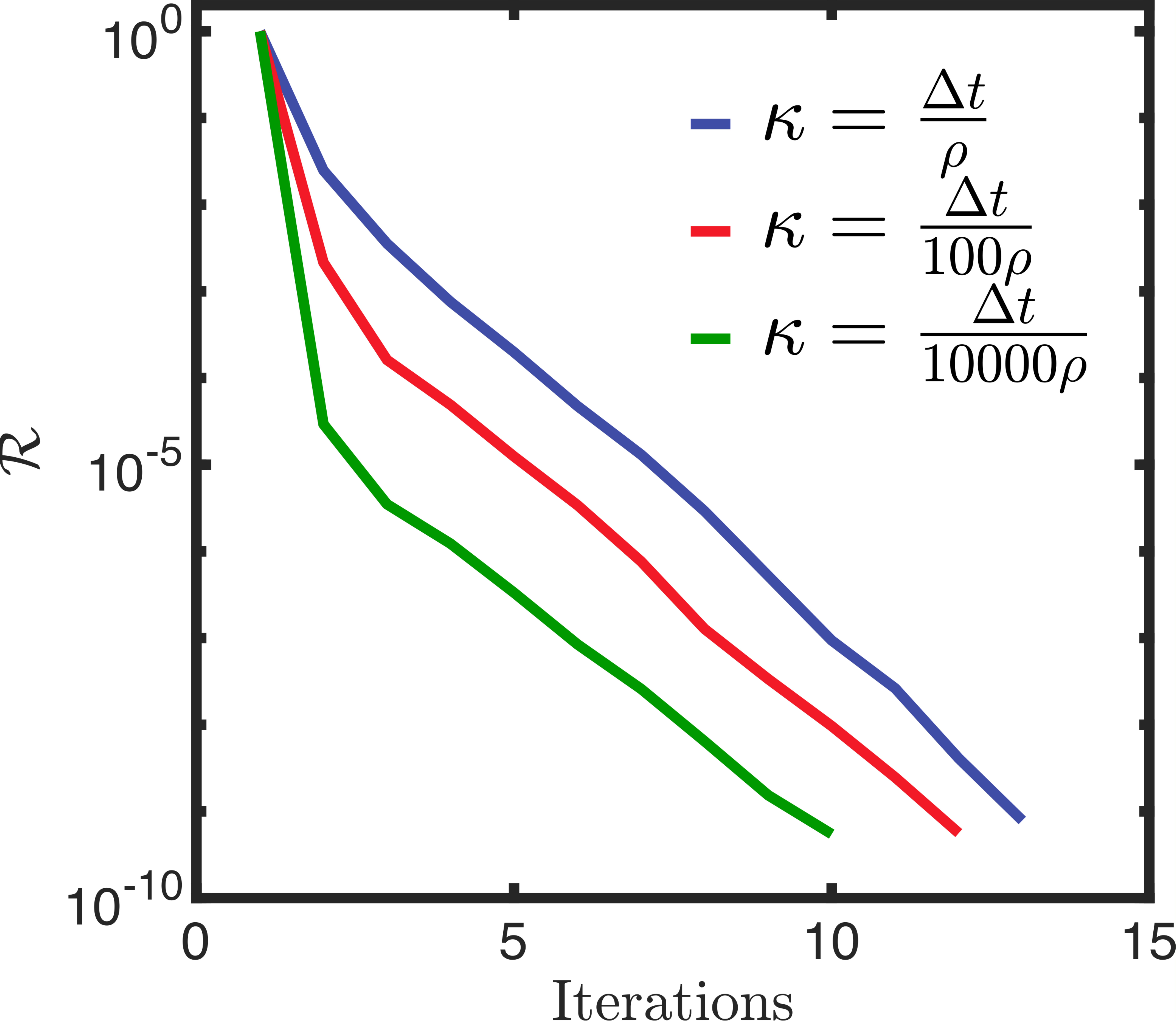}
    \label{fig_residual_with_brinkman_in_PPE}
  }
  \caption{Convergence rate of the preconditioned FGMRES solver during the first time step of the simulation \subref{fig_residual_without_brinkman_in_PPE} without and \subref{fig_residual_with_brinkman_in_PPE} with the Brinkman penalty term in the projection preconditioner. The grid size is $N = 256$.}
\label{fig_residuals}
\end{figure}

Fig.~\ref{fig_residuals} compares the convergence rate of the projection method preconditioned FGMRES solver for different $\kappa$ values. The   grid size is taken to be $N = 256$. When the Brinkman penalty is excluded from the projection step (i.e., $\theta = 0$ in Eq.~\eqref{eq_rhochi_theta}),  we note from Fig.~\ref{fig_residual_without_brinkman_in_PPE} that the number of iterations required to convergence to a relative residual of $\mathcal{R} = 10^{-9}$ increases approximately by a factor of 10 with decreasing values of  $\kappa$. However, with the proposed projection method (i.e., $\theta = 1$ in Eq.~\eqref{eq_rhochi_theta}) the convergence of the FGMRES solver remains robust. This can be observed in  Fig.~\ref{fig_residual_with_brinkman_in_PPE} where the solver converges with approximately 10 iterations for all three $\kappa$ values. This clearly demonstrates the importance of including the Brinkman term in the projection method.  In fact, with decreasing $\kappa$ (or increasing penalty) values, the convergence rate of the solver actually improves. This can be attributed to $\A$ matrix of Eq.~\eqref{eq_frac_vel} which becomes diagonally dominant when the penalty coefficient is larger than the viscous scale. This in turn makes the velocity subdomain problem ``easier" to solve. 

\begin{figure}[]
\centering
\includegraphics[scale = 0.08]{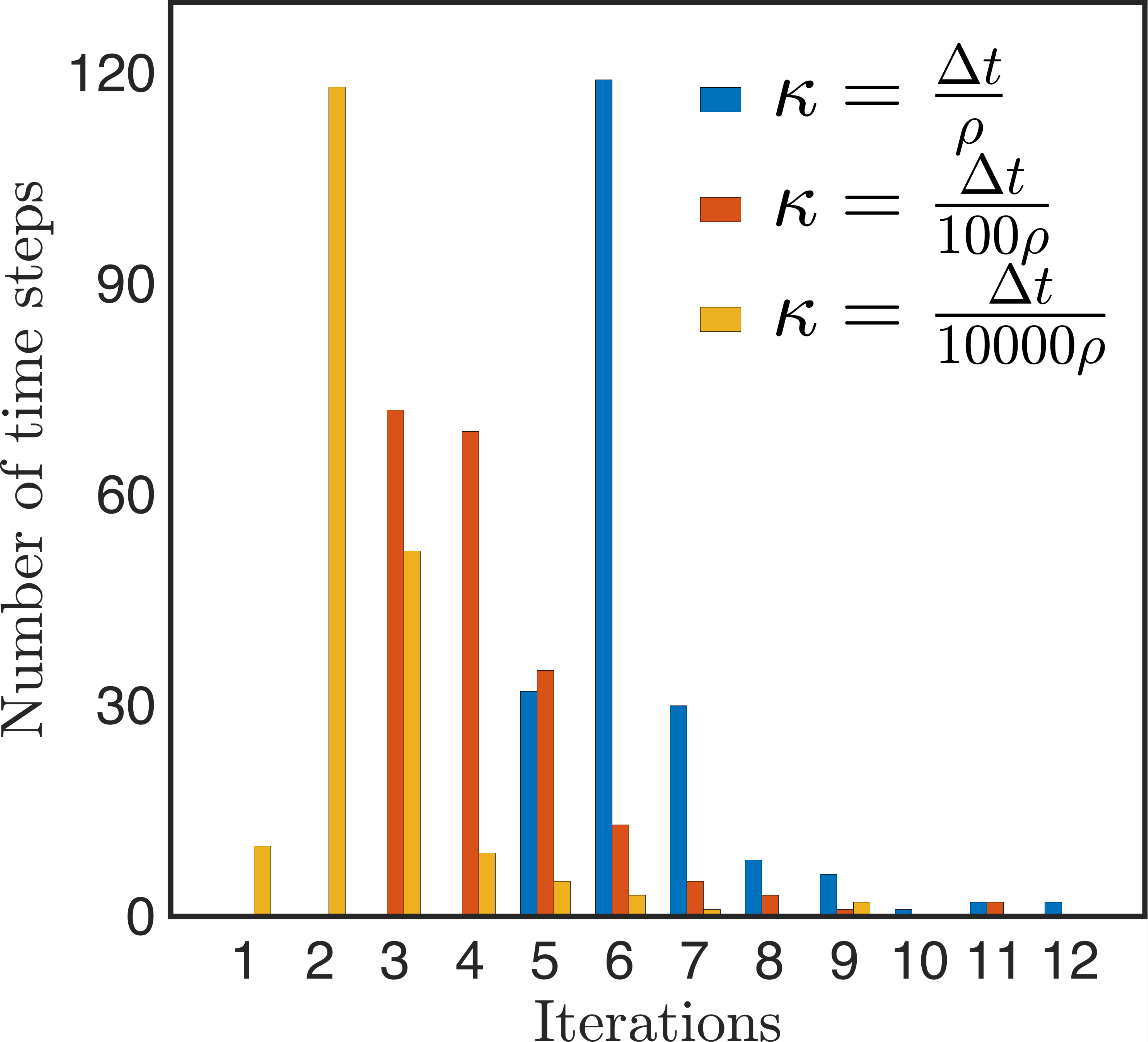}
 \caption{Performance of the preconditioned ($\theta = 1$) FGMRES solver during the first 200 time steps of the simulations. The average number of iterations required to converge for $\kappa = \Delta t / \rho$, $\kappa = \Delta t /100 \rho$, and $\kappa = \Delta t /10000 \rho$ are 6, 4 and 2, respectively. The grid size is $N = 256$. Here, only a single-cycle of fixed-point iterations is considered. }
 \label{fig_mms_bar_chart}
\end{figure}

To present a more complete picture of the solver performance as the simulation progresses,  Fig.~\ref{fig_mms_bar_chart} presents the number of iterations to converge for the first 200 time steps. Here, only a single cycle of fixed-point iterations is employed as the number of FGMRES iterations reduces (substantially) at iteration 2 and beyond. Thus, the ``worst-case-scenario" is considered. The grid size is taken to be $N = 256$ and the same three $\kappa$ values are considered. It is clearly seen that with decreasing $\kappa$ values or conversely with increasing penalty values,  the average number of iterations to converge decreases.

\begin{figure}[]
\centering
\includegraphics[scale = 0.08]{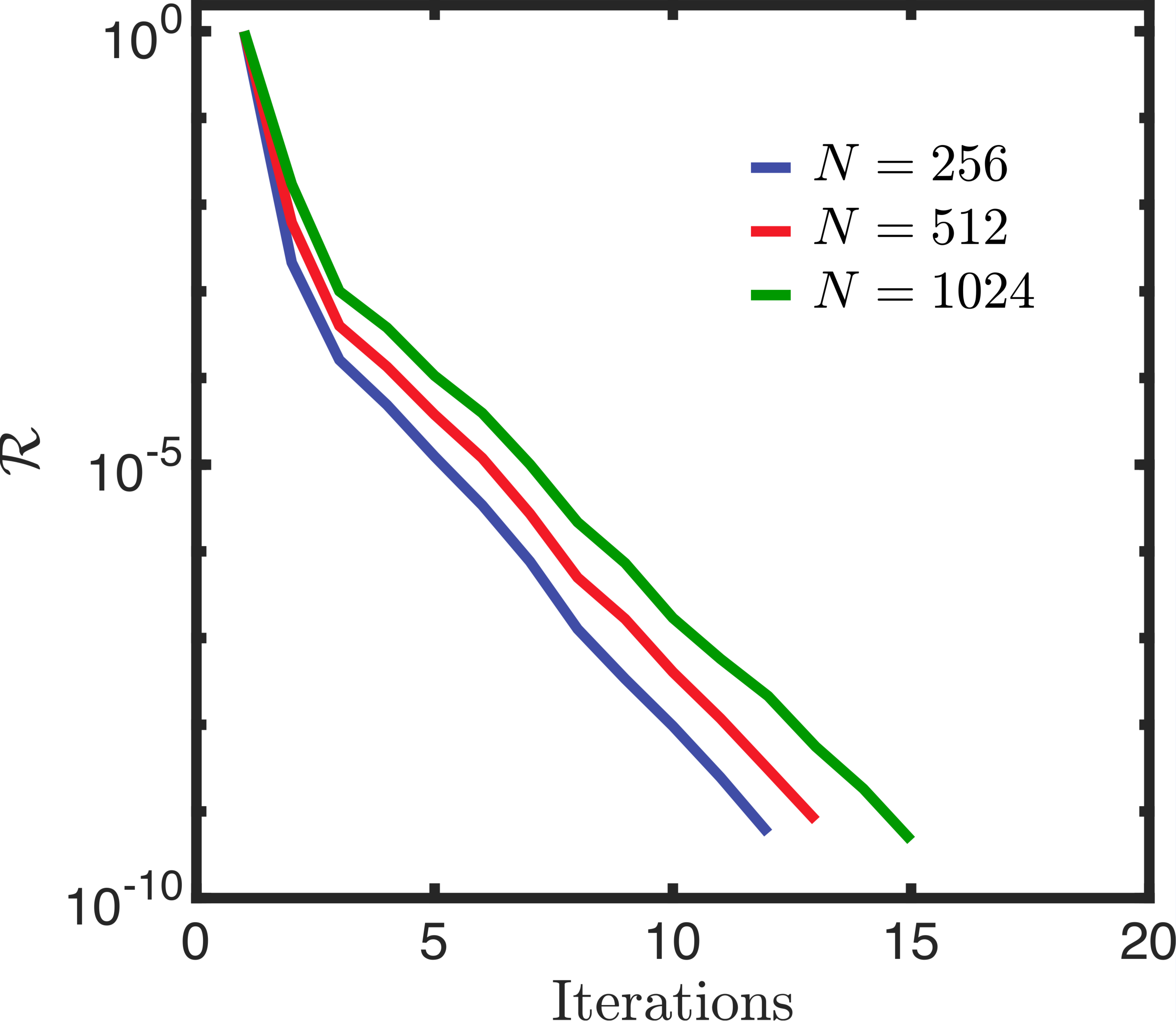}
  \caption{Convergence rate of the preconditioned ($\theta = 1$) FGMRES solver during the first time step of the simulation on three grids:  $N = 256, 512, 1024$. The permeability parameter is taken to be $\kappa = \Delta t / 100 \rho$.}
  \label{fig_residual_N}
\end{figure}

Thus far in this section, we demonstrated the efficacy of the new projection preconditioner on a single grid $N = 256$. To demonstrate that the proposed preconditioner is scalable, Fig.~\ref{fig_residual_N} reports the convergence rate of the preconditioned FGMRES solver on three grids: $N = 256, 512, 1024$. The permeability parameter is taken to be $\kappa = \Delta t / 100 \rho$, which is larger than both inertial and viscous scales as discussed earlier. In Fig.~\ref{fig_residual_N} it can be observed that the solver's convergence rate remains approximately the same, even when the degrees of freedom, $\xu$ and $\xp$, increase substantially with increasing grid size. 

Based on the results presented in this section we conclude that: (\textbf{1}) the proposed projection method is a scalable preconditioner for the volume penalized incompressible Navier-Stokes system; (\textbf{2}) it is possible to achieve pointwise second-order accuracy in velocity and pressure solutions with nontrivial traction boundary conditions without sacrificing computational efficiency; and (\textbf{3}) a reasonable starting value for the permeability parameter is $\kappa = \Delta t/\rho$. In the next section we consider a three phase FSI problem to demonstrate that the proposed projection preconditioner remains effective even with spatially varying $\rho$ and $\mu$.  

\subsection{Free-decay of a rigid cylinder heaving on an air-water interface} \label{sec_heaving_cyl}

\begin{figure}
\centering
\includegraphics[scale = 0.45]{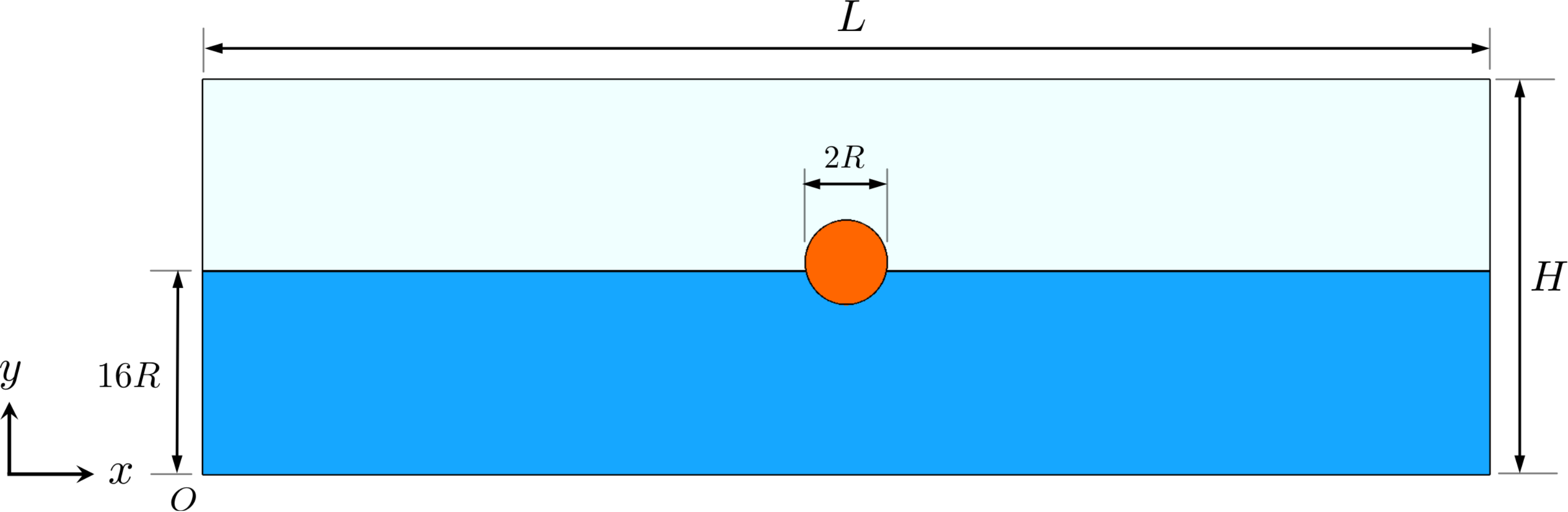}
  \caption{Schematic of a 2D cylinder heaving on an air-water interface.}
  \label{fig_schematic_cyl_decay}
\end{figure}

We consider a two-dimensional computational domain of extents $\Omega \in [0, L] \times [0, H]$ to simulate the heaving motion of a 2D cylinder on an air-water interface; see Fig.~\ref{fig_schematic_cyl_decay}. The length of the domain is $L$ = 10 m, height is $H$ = 0.2$L$, and the origin $O$ is located at the bottom left corner. The radius of the cylinder is $R$ = 0.0762 m. Water of density $\rhol$ = 1000 kg/$\text{m}^3$ and viscosity $\mul = 10^{-3}$ Pa$\cdot$s occupies the computational domain from $y$ = 0 to $y$ = $16R$ and air of density $\rhog$ = 1 kg/$\text{m}^3$ and viscosity $\mug = 1.8 \times 10^{-5}$ Pa$\cdot$s occupies the domain from $y = 16R$ to $y = H$. The solid cylinder is half-buoyant and has a density of $\rho^\textrm{S}$ = 500 kg/$\text{m}^3$. Its fictitious viscosity $\mus$ is taken to be same as that of water.  The initial center of the cylinder is located slightly above the air-water interface $(x_c,y_c) = (L/2, 16R + R/3)$ from where it is released. The cylinder decays freely under the action of gravity and hydrodynamic forces. The cylinder's heave (vertical) degree of freedom is free,  whereas its surge (horizontal) and pitch (rotational) motions are locked. No-slip (zero-velocity) boundary conditions are imposed along $\partial \Omega$. To smoothly transition between different material properties while keeping the liquid-gas and fluid-solid interfaces sharp, one grid cell on either side of the interface  is considered ($\ncells = 1$ in Eqs.~\eqref{eqn_Hflow} and~\eqref{eqn_Hbody}).  The surface tension of water does not affect the dynamics of the cylinder at this scale and is neglected in the simulation.

\begin{figure}[]
\centering
\includegraphics[scale = 0.45]{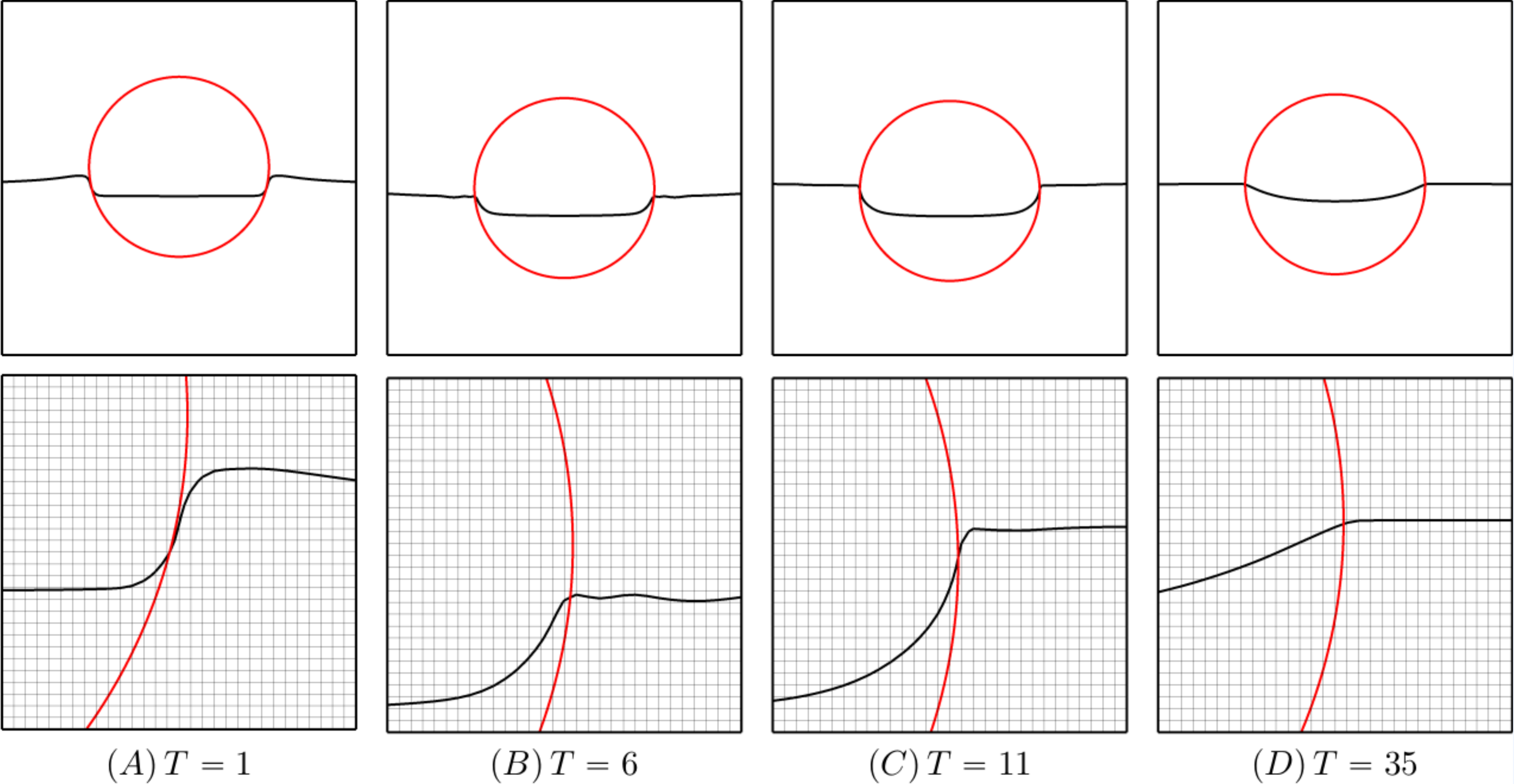}
  \caption{Evolution of the zero contours of fluid $\sigma$ (black line) and solid $\psi$ (red line) level set functions. An enlarged view of the liquid-gas interface near the cylinder's right side is shown in the bottom row.}
  \label{fig_heaving_cyl_ls}
\end{figure}

The domain is discretized with a uniform grid of size 6000 $\times$ 1200 such that there are 46 grid cells per radius  of the cylinder. A constant time step of $\Delta t = 10^{-3}$ s is used which maintains the convective CFL number below 0.5. The permeability coefficient is taken to be $\kappa = \Delta t/\rho^\textrm{S}$. Newton's second law is used to compute the rigid body velocity of the cylinder $\u_b$, which requires integrating hydrodynamic traction over the irregular surface of the immersed body. This procedure is explained in our prior works~\cite{BhallaBP2019,Dafnakis2020}.  Fig.~\ref{fig_temporal_evol_cyl_decay} shows the temporal evolution of  the heaving cylinder at various time instances. The vertical displacement of the cylinder is shown in Fig.~\ref{fig_heave_disp_cyl_decay} and compared with prior immersed boundary (IB) simulations of  Nangia et al.~\cite{Nangia2019WSI}. Here,  $T = t\sqrt{g/R}$ and $\bar{y}= 3y/R$ denote the non-dimensional time and vertical displacement of the cylinder, respectively.  An excellent agreement is obtained with the prior study.  After a while, the heaving motion of the cylinder ceases and its center coincides with the free water surface. An illustration of the temporal evolution of the zero contours of the fluid $\sigma$ and solid $\psi$ level set functions can be found in Fig.~\ref{fig_heaving_cyl_ls}. Within the cylinder and near the material triple points, the air-water interface remains continuous.

 \begin{figure}
\centering
\includegraphics[scale = 0.6]{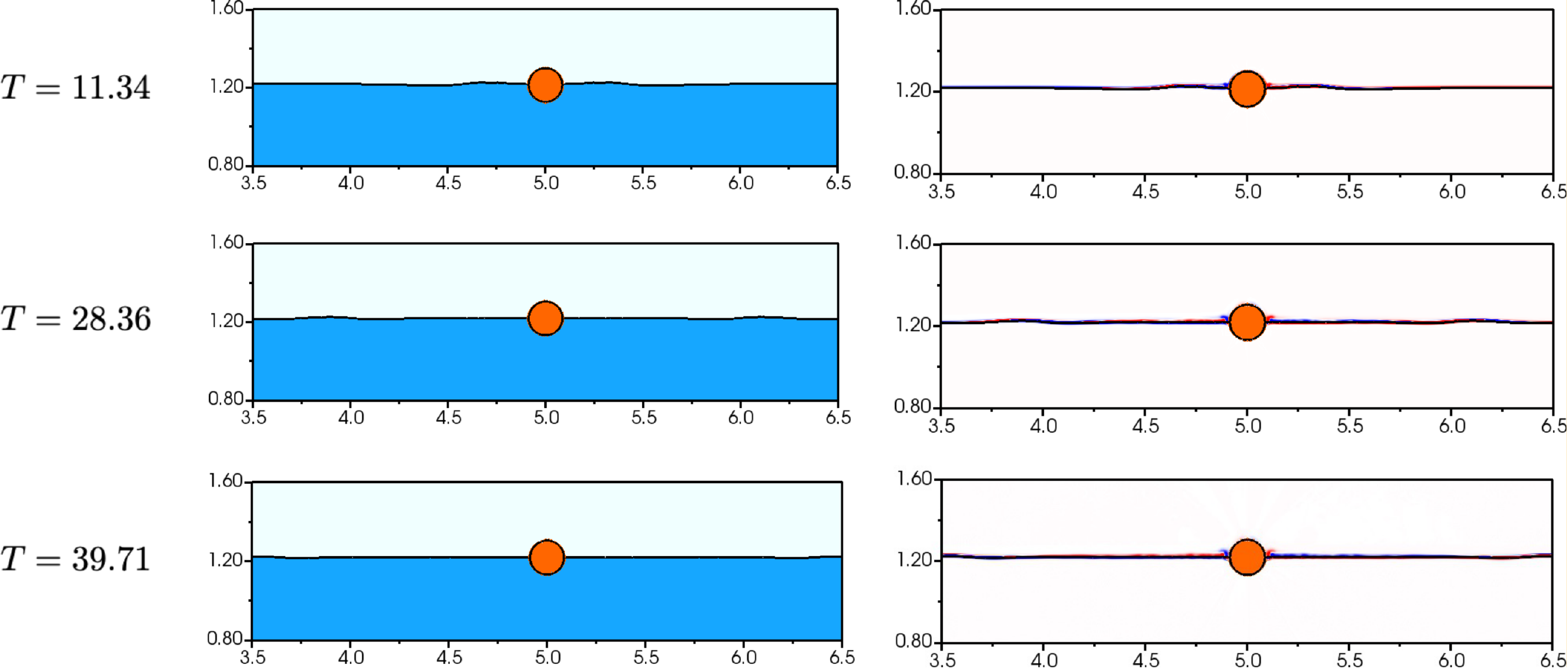}
  \caption{Temporal evolution a cylinder heaving on an air-water interface: (left) density  and (right) vorticity generated in the range -1  to 1 $\text{s}^{-1}$.}
  \label{fig_temporal_evol_cyl_decay}
\end{figure}

\begin{figure}
\centering
\includegraphics[scale = 0.4]{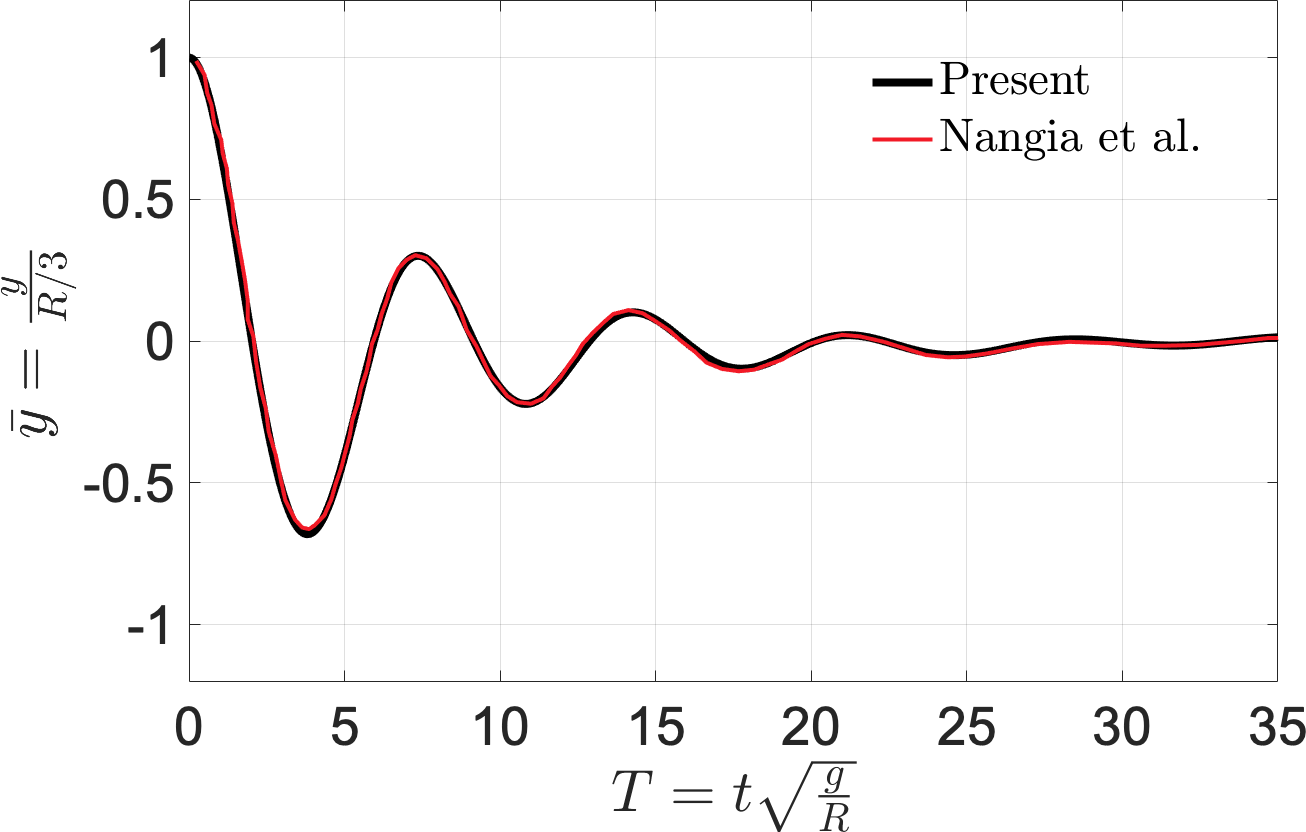}
  \caption{Temporal evolution of the non-dimensional vertical displacement $\bar{y}$ of the cylinder. The heave displacement of the cylinder is compared against the prior simulations of  Nangia et al.~\cite{Nangia2019WSI}}
  \label{fig_heave_disp_cyl_decay}
\end{figure} 

\begin{figure}
  \centering
    \subfigure[Percentage change in liquid volume]{
    \includegraphics[scale = 0.08]{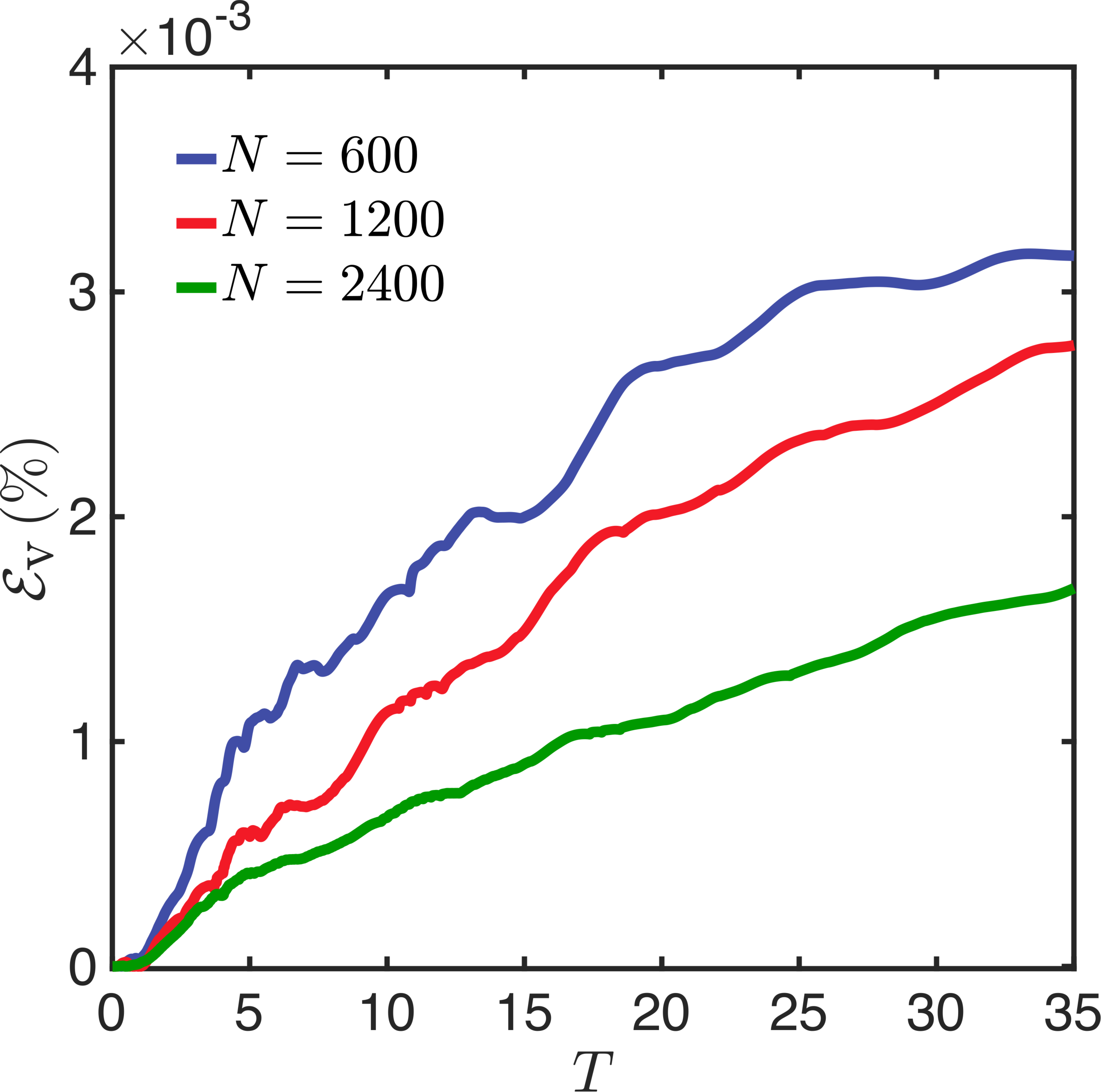} 
    \label{fig_wedge_liquid_volume}
  }
  \subfigure[Percentage change in gas volume]{
    \includegraphics[scale = 0.08]{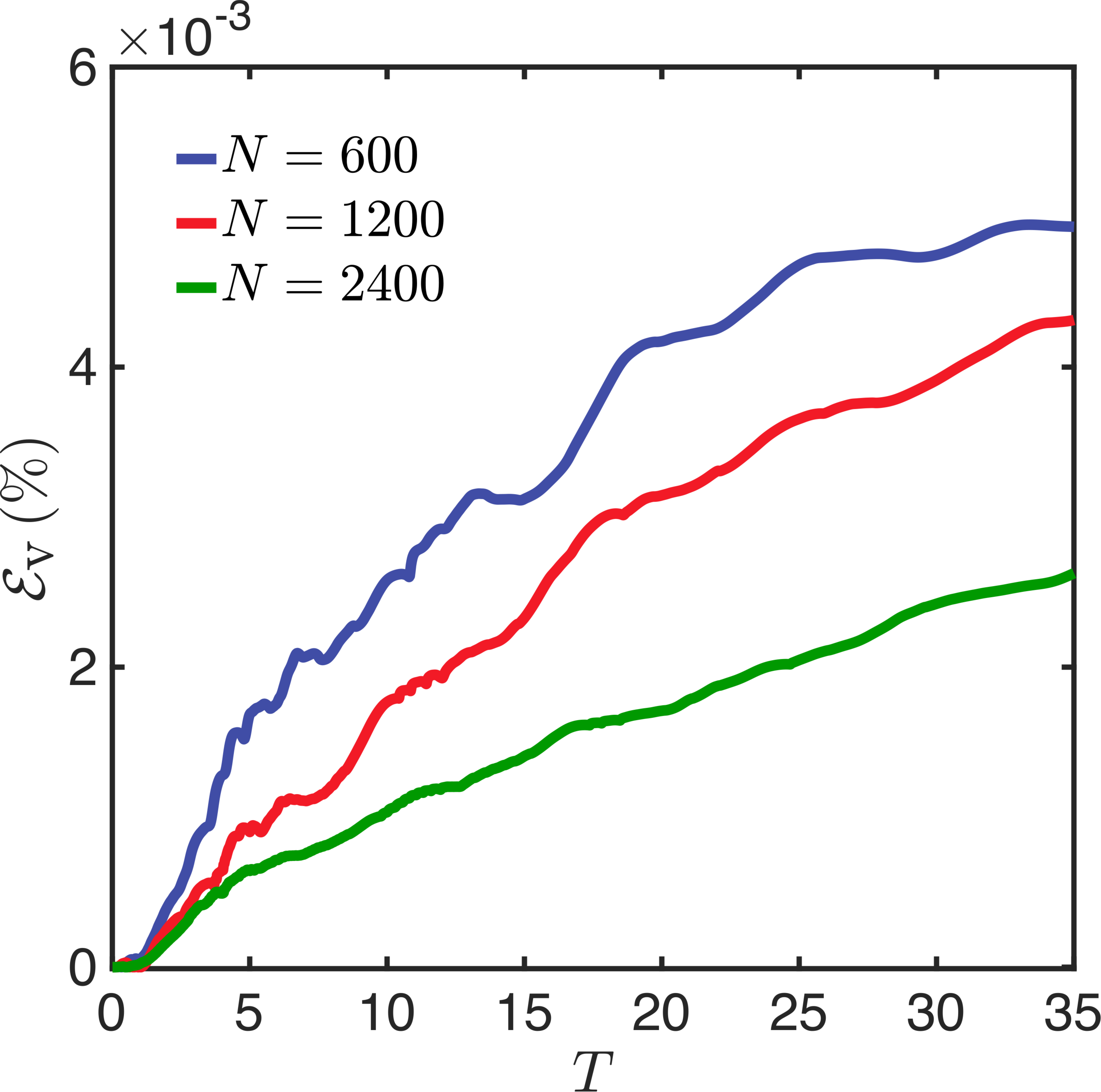} 
    \label{fig_wedge_gas_volume}
  }
 \caption{Percentage volume change $\mathcal{E}_\textrm{v}$ of liquid and gas over time for the heaving cylinder case. Three different grid resolutions are considered. A uniform time step size of $\Delta t = 2\times10^{-3}$ is used for the coarsest grid $N = 600$ and it is halved for each successive finer grid. The permeability coefficient is taken to be $\kappa=\Delta t/\rho^\textrm{S}$. 
}
\label{fig_heaving_cyl_volume_error}
\end{figure}

Fig.~\ref{fig_heaving_cyl_volume_error} shows the percentage volume/mass change of liquid and gas over time.  $\kappa=\Delta t/\rho^\textrm{S}$ is taken and $\mathcal{E}_\textrm{v}$ is computed numerically as described in Sec.~\ref{sec_wedge}.  There is a very small percentage change in volume, less than 0.006\%. The $\mathcal{E}_\textrm{v}$ value decreases further under grid refinement.

\subsubsection{Effect of $\kappa$ on the contact line motion and solver performance} 

\begin{figure}[]
\centering
\includegraphics[scale = 1]{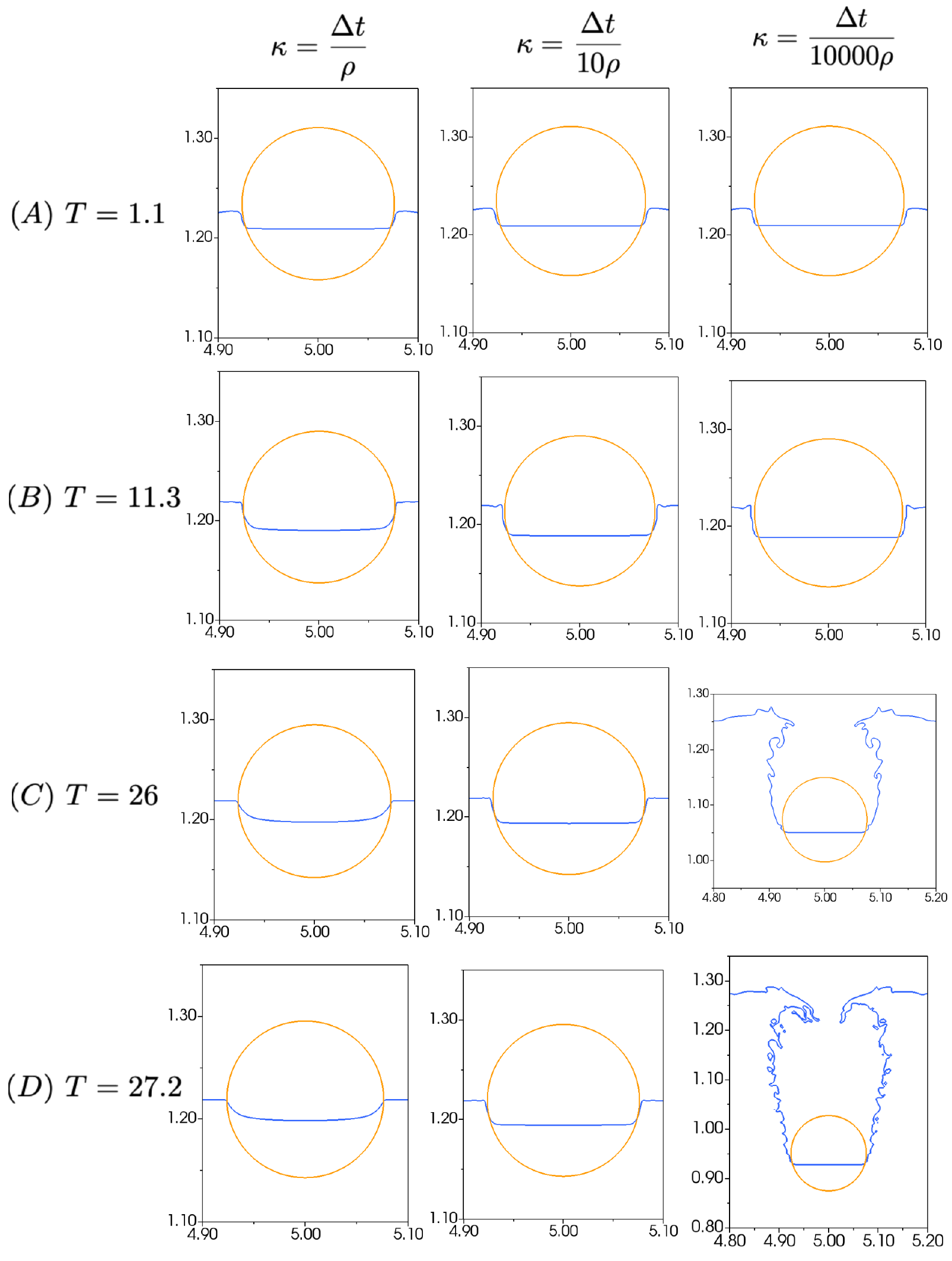}
  \caption{Evolution of the zero contours of fluid $\sigma$ (blue line) and solid $\psi$ (mustard line) level set functions using three different values of $\kappa = \{\Delta t/\rhos, \Delta t/10\rhos, \Delta t/10000\rhos\}$.}
  \label{fig_cyl_kappa_variation_interface}
\end{figure}

\begin{figure}[]
\centering
\includegraphics[scale = 0.45]{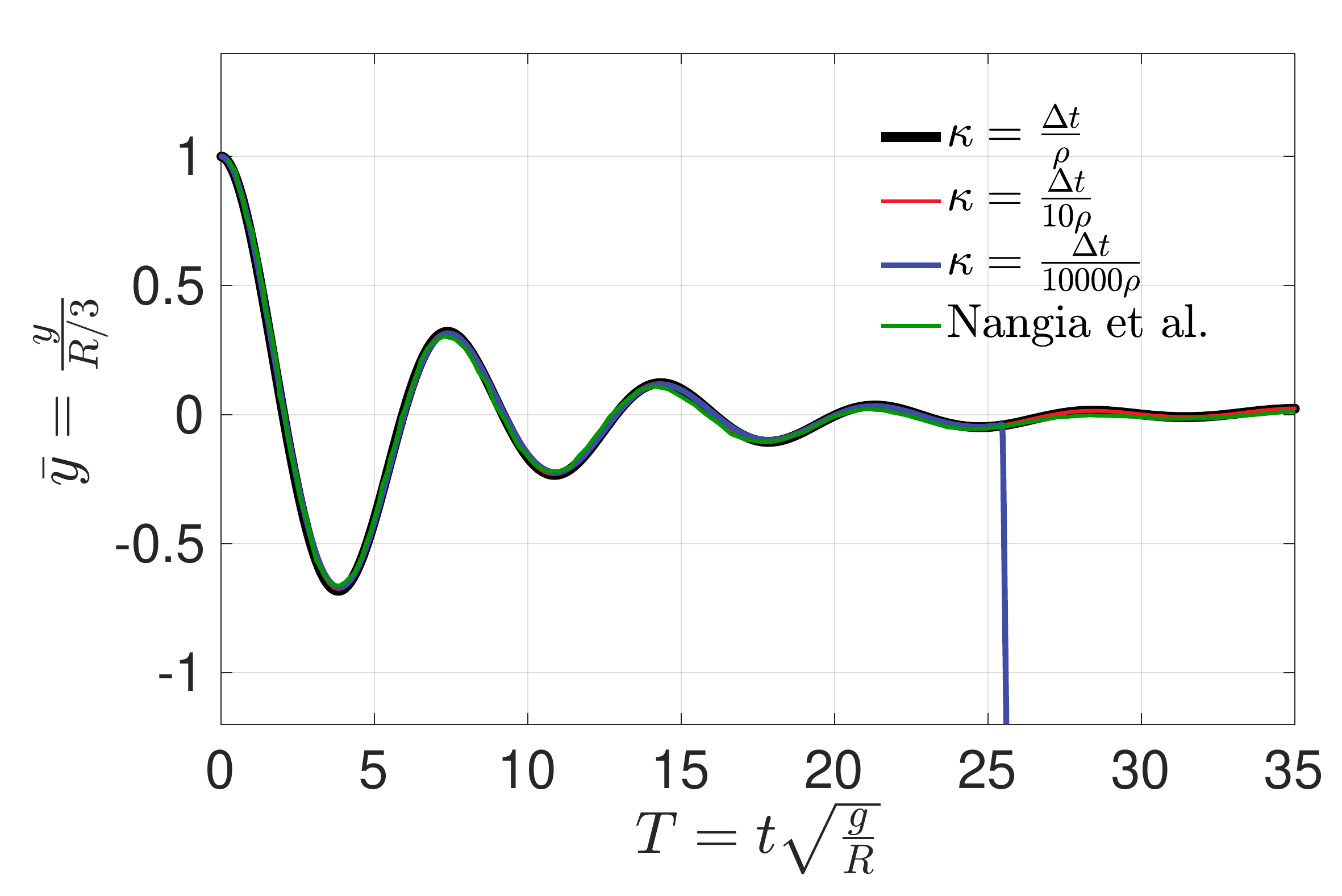}
  \caption{Temporal evolution of the non-dimensional vertical displacement $\bar{y}$ of the cylinder using three different values of $\kappa = \{\Delta t/\rhos, \Delta t/10\rhos, \Delta t/10000\rhos\}$. The heave displacement of the cylinder is compared against the prior study of  Nangia et al.~\cite{Nangia2019WSI}. The sudden drop in the cylinder displacement is due numerical instabilities that arise for the lowest $\kappa$ case. }
  \label{fig_heave_kappa_compare}
\end{figure}

Unlike the falling wedge problem considered in Sec.~\ref{sec_wedge}, where the solid evolved from a non-wet state to a wet state, here the rigid cylinder starts out in the wet state at $T = 0$. Recall that the liquid-gas interface could not penetrate the wedge when $\kappa $ was lowered below $\Delta t/ \rhos$ (by a factor of 2 or more) and numerical instability occurred due to the breakdown of the classical no-slip condition for the contact line. The heaving cylinder case also presents a similar scenario. The wet cylinder attempts to reach a dry state by conforming the contact line around its outer periphery. This is observed when $\kappa$ is further lowered to $\kappa = \{\Delta t/10\rhos, \Delta t/10000\rhos\}$. An illustration of the situation is shown in Fig.~\ref{fig_cyl_kappa_variation_interface}, which compares contact line dynamics at three different $\kappa$ values. Numerical instabilities arise around $T = 25$ with $\kappa = \Delta t/10000\rhos$. Despite trying to dry out, the cylinder is unable to do so due to the combination of mass and force balance conditions. In the intermediate case of $\kappa = \Delta t/10\rhos$, numerical instabilities do not occur due to imperfectly imposed no-slip conditions. One can clearly see, however, the contact line's tendency to align with the cylinder surface (the 180 degree contact angle condition) in the intermediate case as well. The heave dynamics of the cylinder for the three $\kappa$ values are compared in Fig.~\ref{fig_heave_kappa_compare}. There is a very good match between all three curves and the results reported in the literature (except for the numerical instability in the lowest $\kappa$ case). 

 \begin{figure}
  \centering
  \subfigure[Relative residual when $\theta$ = 0]{
    \includegraphics[scale = 0.4]{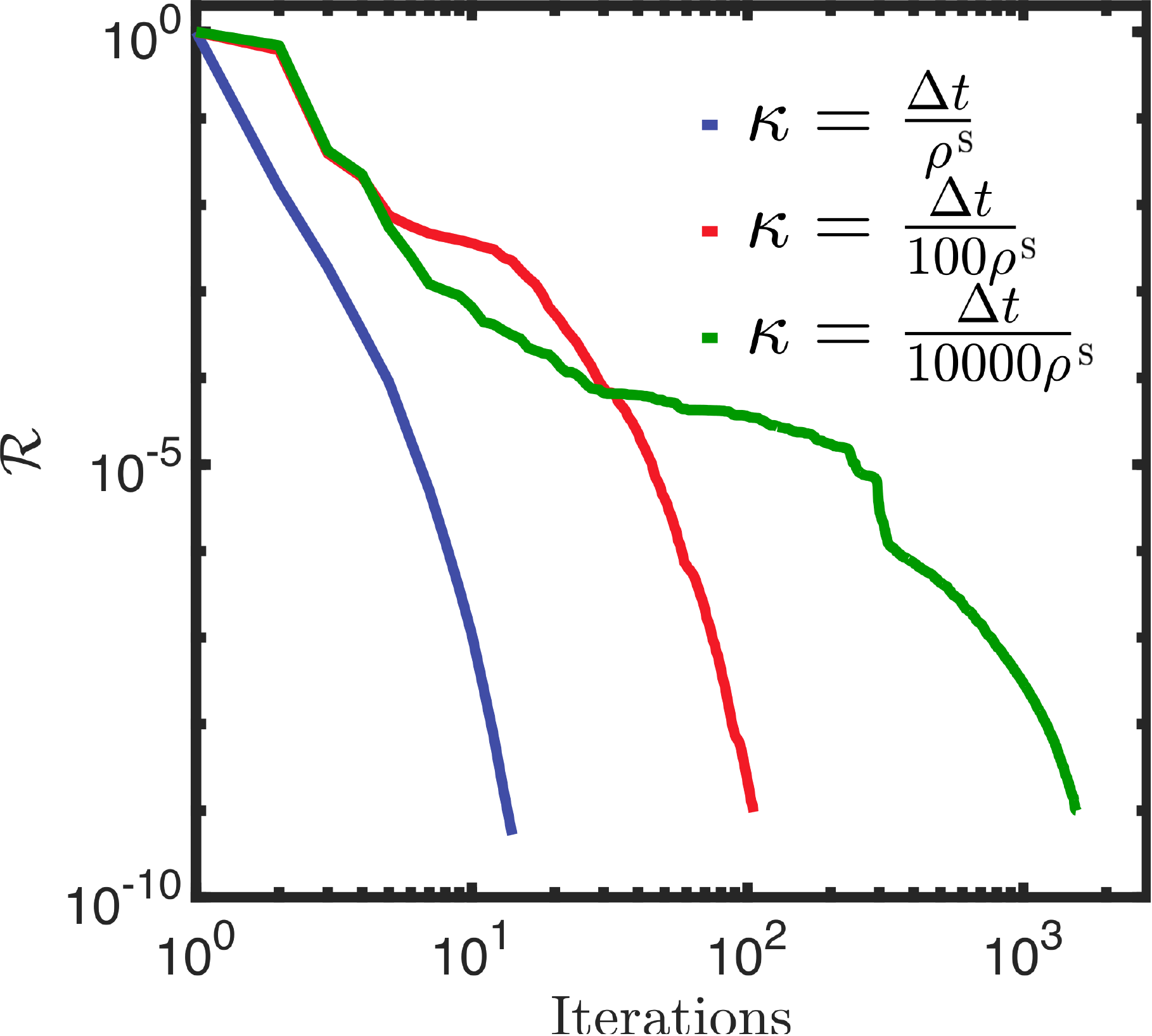} 
    \label{fig_residual_theta_0}
  }
   \subfigure[Relative residual when $\theta$ = 1]{
    \includegraphics[scale = 0.48]{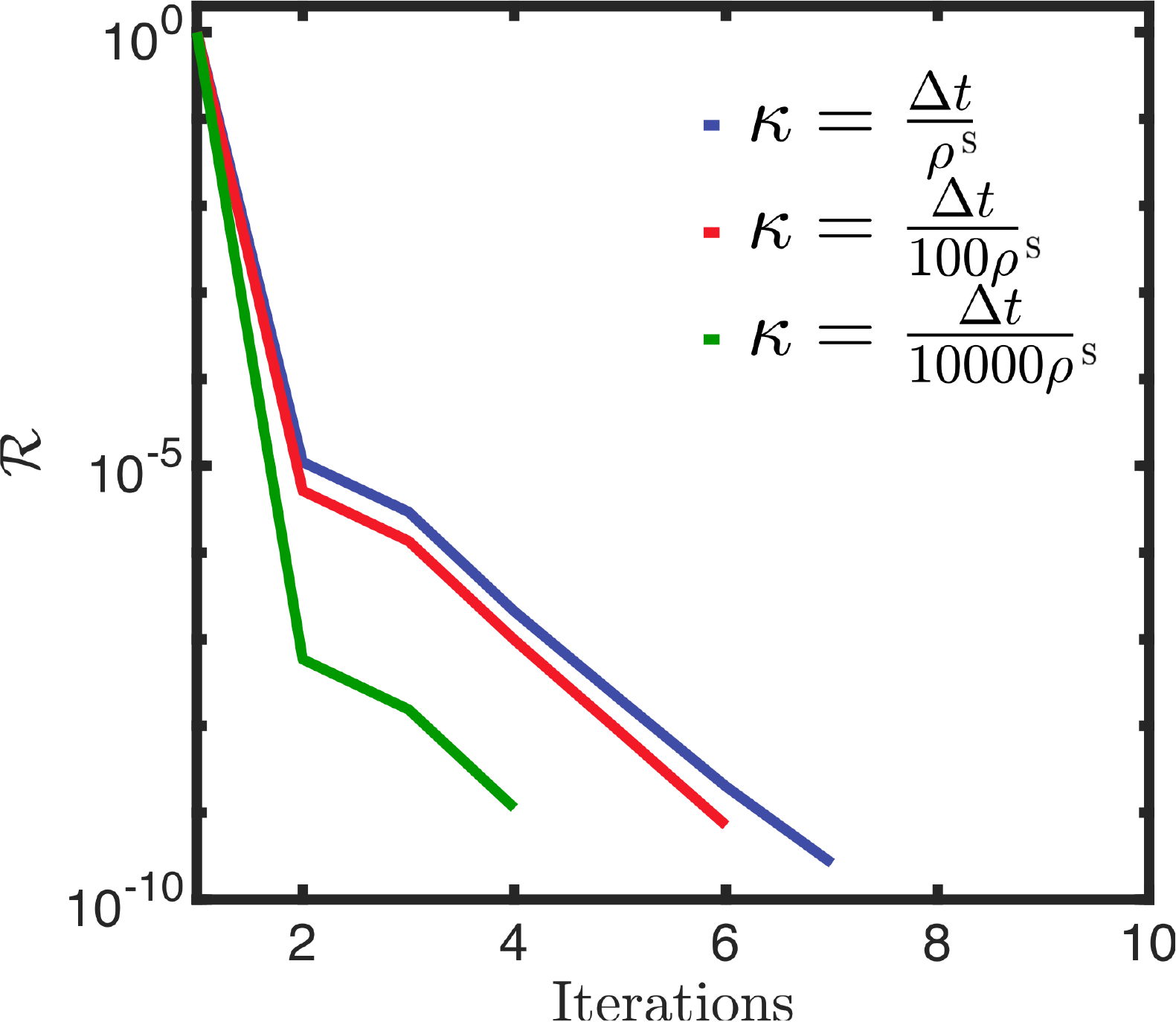}
    \label{fig_residual_theta_1}
  }
  \caption{Convergence rate of the FGMRES solver during the first time step of the simulation for the multiphase case of a cylinder heaving on the air-water interface.}
\label{fig_convergence_rate_cyl_decay}
\end{figure}

\begin{figure}[]
\centering
\includegraphics[scale = 0.4]{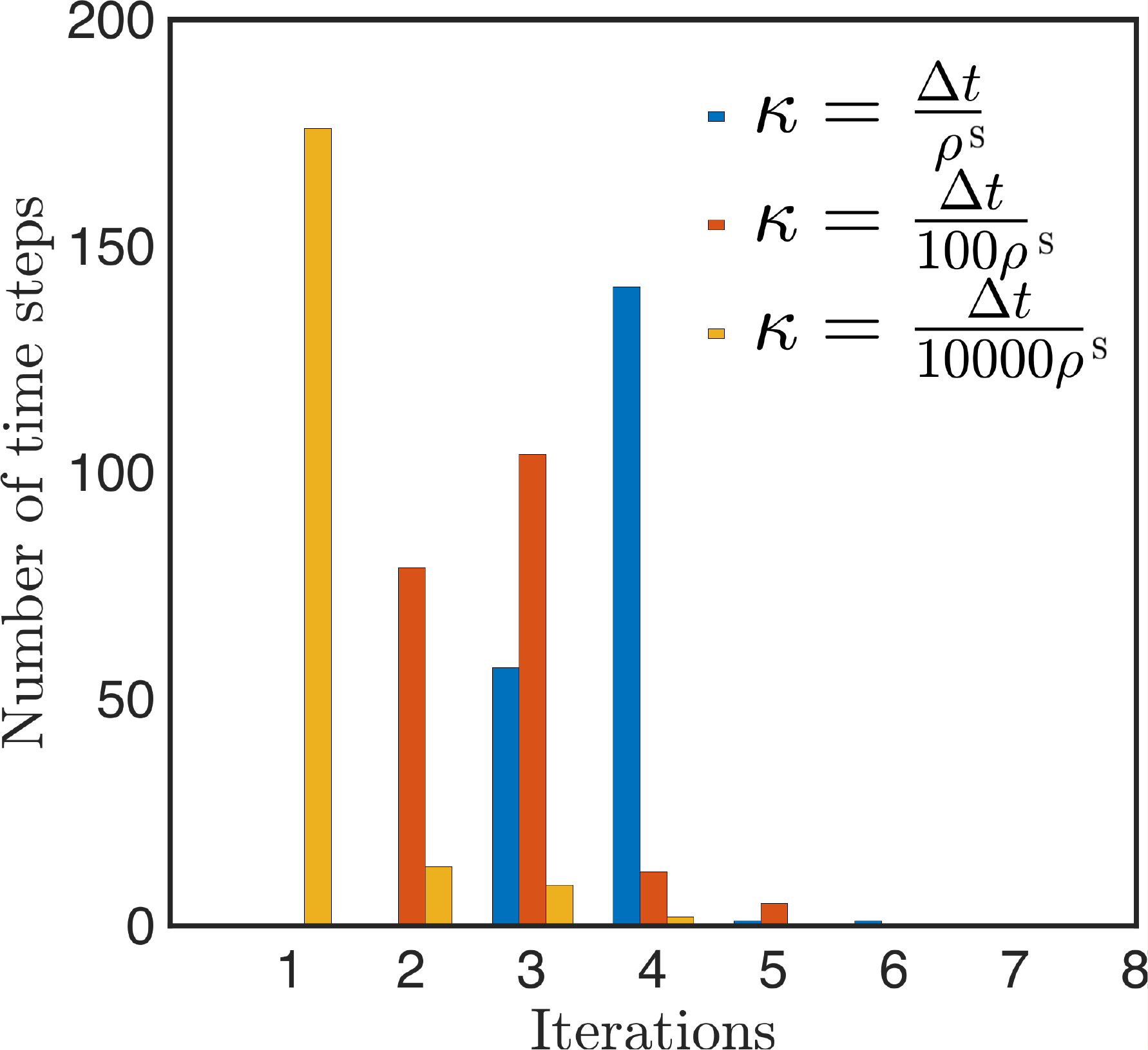}
  \caption{Performance of the preconditioned ($\theta$ = 1) FGMRES solver during the first 200 time steps of the simulations for the multiphase  case of cylinder heaving on the air-water interface. The average number of iterations required to converge for $\kappa = \Delta t / \rho^\textrm{S}$, $\kappa = \Delta t /100 \rho^\textrm{S}$, and $\kappa = \Delta t /10000 \rho^\textrm{S}$ are 4, 3 and 1, respectively. Here, only a single cycle of fixed-point iterations is employed.}
  \label{fig_stokes_iters}
\end{figure}  
 
Next, we examine the effect of $\kappa$ on the convergence rate of the preconditioned FGMRES solver by considering three different values of $\kappa =   \{ \Delta t/\rho^\textrm{S}, \, \Delta t/100 \rho^\textrm{S}, \, \Delta t/10000 \rho^\textrm{S}\}$. Here we are interested in testing the performance of the preconditioner at extremely low values of $\kappa$, despite the fact that very low values of $\kappa$ (which lead to the no-slip condition for the contact line) are not consistent with the multiphase FSI model \cite{huh1971hydrodynamic}. The solver performance is monitored only at the first time step for the reasons explained in the previous Sec.~\ref{sec_mms}. The tests are run with ($\theta = 1$) and without ($\theta = 0$) the Brinkman penalty term $\widetilde{\V{\chi}} / \kappa$ in the projection preconditioner; see Eq.~\ref{eq_rhochi_theta}. Fig.~\ref{fig_convergence_rate_cyl_decay} compares the convergence rate of the solver for the three $\kappa$ values. As observed in Fig.~\ref{fig_residual_theta_0}, without the penalty term, the number of iterations required to converge to a relative residual of $\mathcal{R} = 10^{-9}$ increases approximately by a factor of 10 for decreasing values of $\kappa$. With the new projection algorithm, the convergence of the solver remains robust and it convergences within 7 iterations as illustrated in  Fig.~\ref{fig_residual_theta_1}. Further, Fig.~\ref{fig_stokes_iters} shows the number of iterations to converge for the first 200 time steps for the three $\kappa$ values. Here, only a single cycle of fixed-point iterations is employed as the number of FGMRES iterations reduces (substantially) at iteration 2 and beyond. Similar to the results of the previous section, with decreasing $\kappa$ values, the average number of iterations to converge decreases. Based upon the results of this and the previous section, it can be concluded that the projection preconditioner is an effective and scalable strategy for both single and multiphase VP INS systems.

\subsubsection{Effect of $\ncells$}
We investigate how $\ncells$ values affect the solution accuracy, linear solver convergence rate, and mass/volume of conserved phases. We consider three different $\ncells$ values, $\ncells$ = 1, 2, and 3, to solve the heaving cylinder problem on a uniform $6000 \times 1200$ grid.  $\kappa=\Delta t/\rho^\textrm{S}$ for all cases considered in this subsection. The vertical displacement of the cylinder using different $\ncells$ values is plotted in Fig.~\ref{fig_heave_disp_cyl_decay}. As can be observed, there is no significant change in the numerical solution and it is similar to the prior IB simulations of Nangia et al.~\cite{Nangia2019WSI}.

\begin{figure}
\centering
\includegraphics[scale = 0.4]{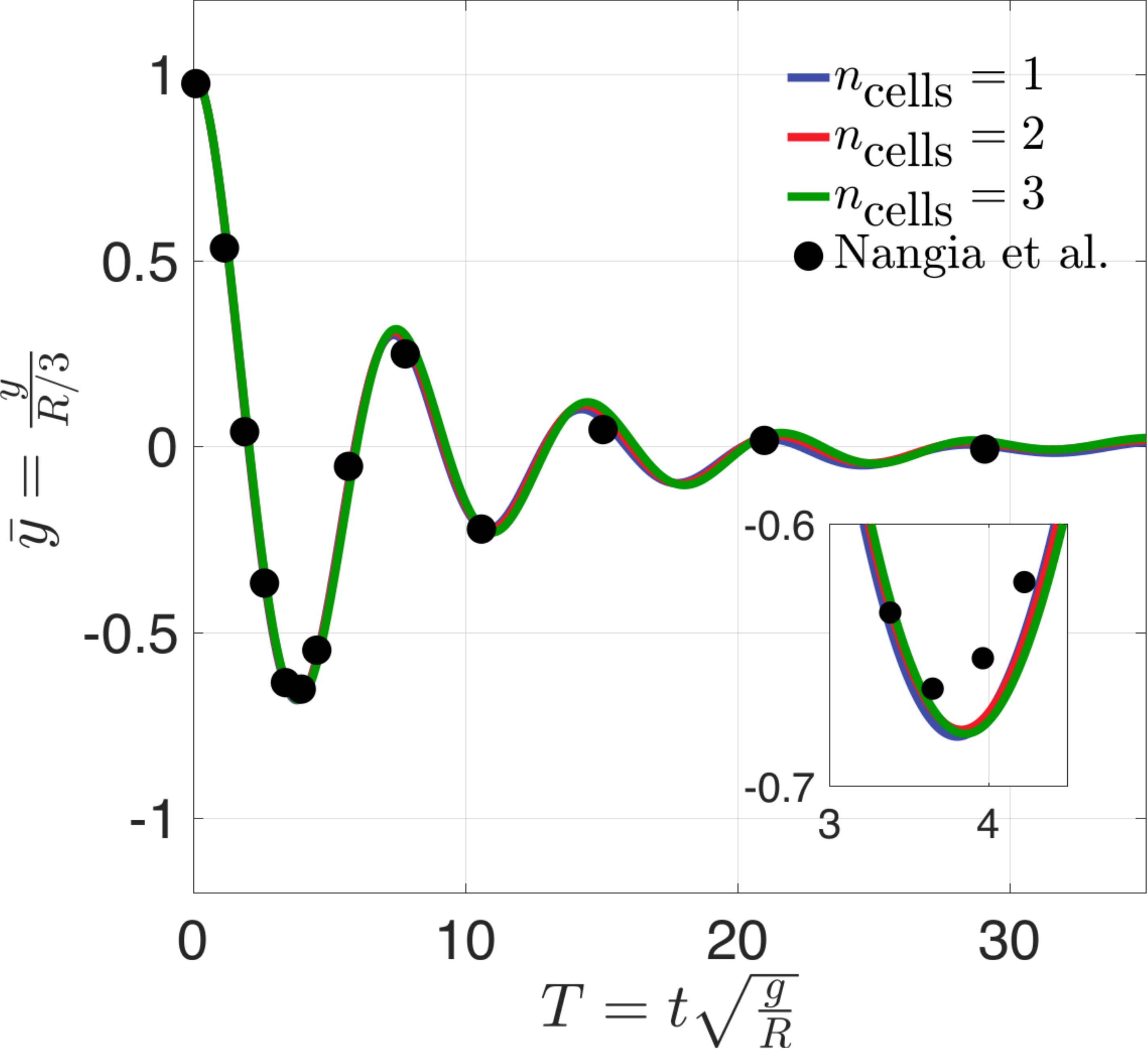}
  \caption{Temporal evolution of the non-dimensional vertical displacement $\bar{y}$ of the cylinder for different $\ncells$ values. The heave displacement of the cylinder is compared against the prior immersed boundary (IB) simulations of  Nangia et al.~\cite{Nangia2019WSI}}
  \label{fig_heave_disp_cyl_decay}
\end{figure}

\begin{figure}
  \centering
  \subfigure[Relative residual when $\theta = 0$]{
    \includegraphics[scale = 0.08]{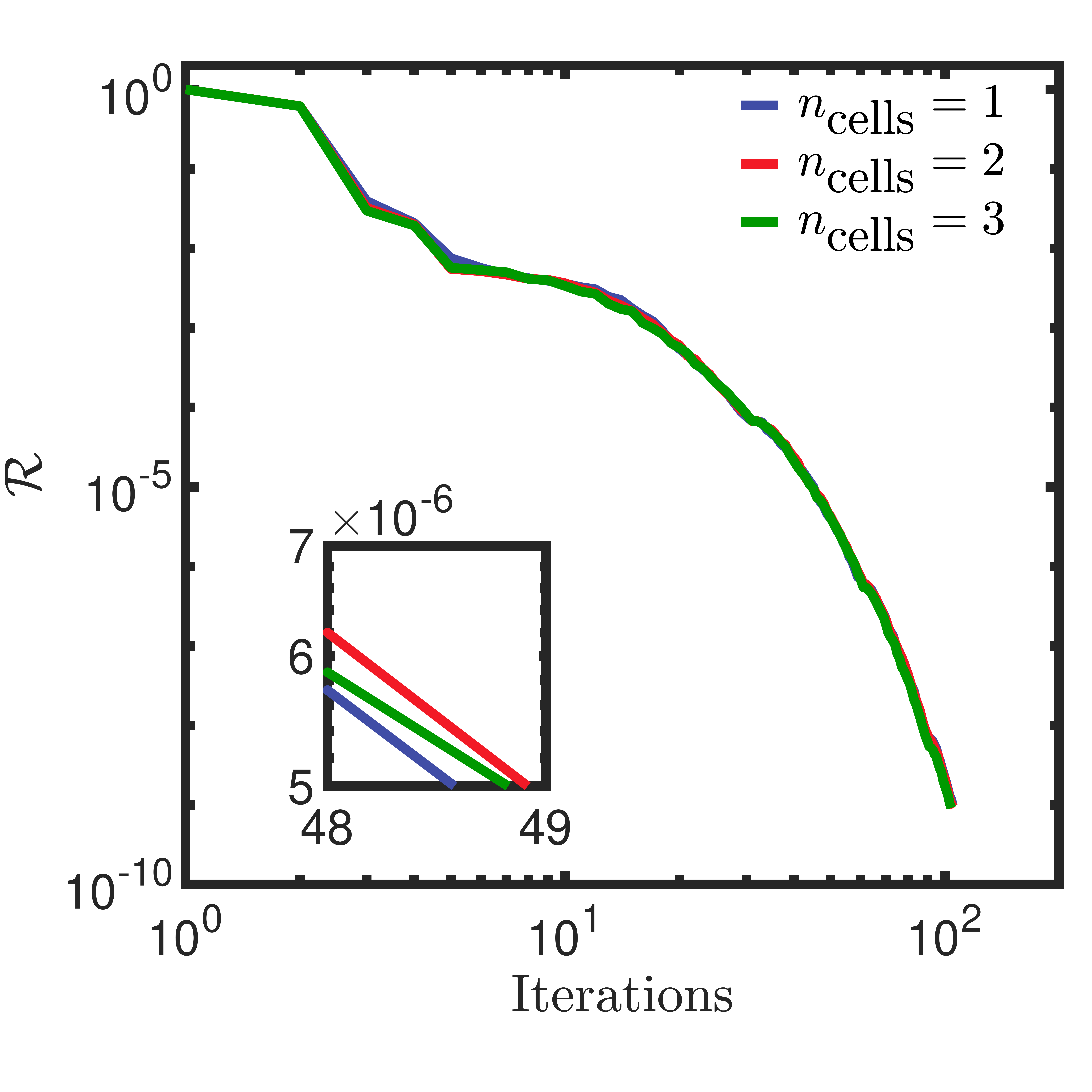} 
    \label{fig_heaving_cyl_error_wo_fix_ncells}
  }
   \subfigure[Relative residual when $\theta = 1$]{
    \includegraphics[scale = 0.08]{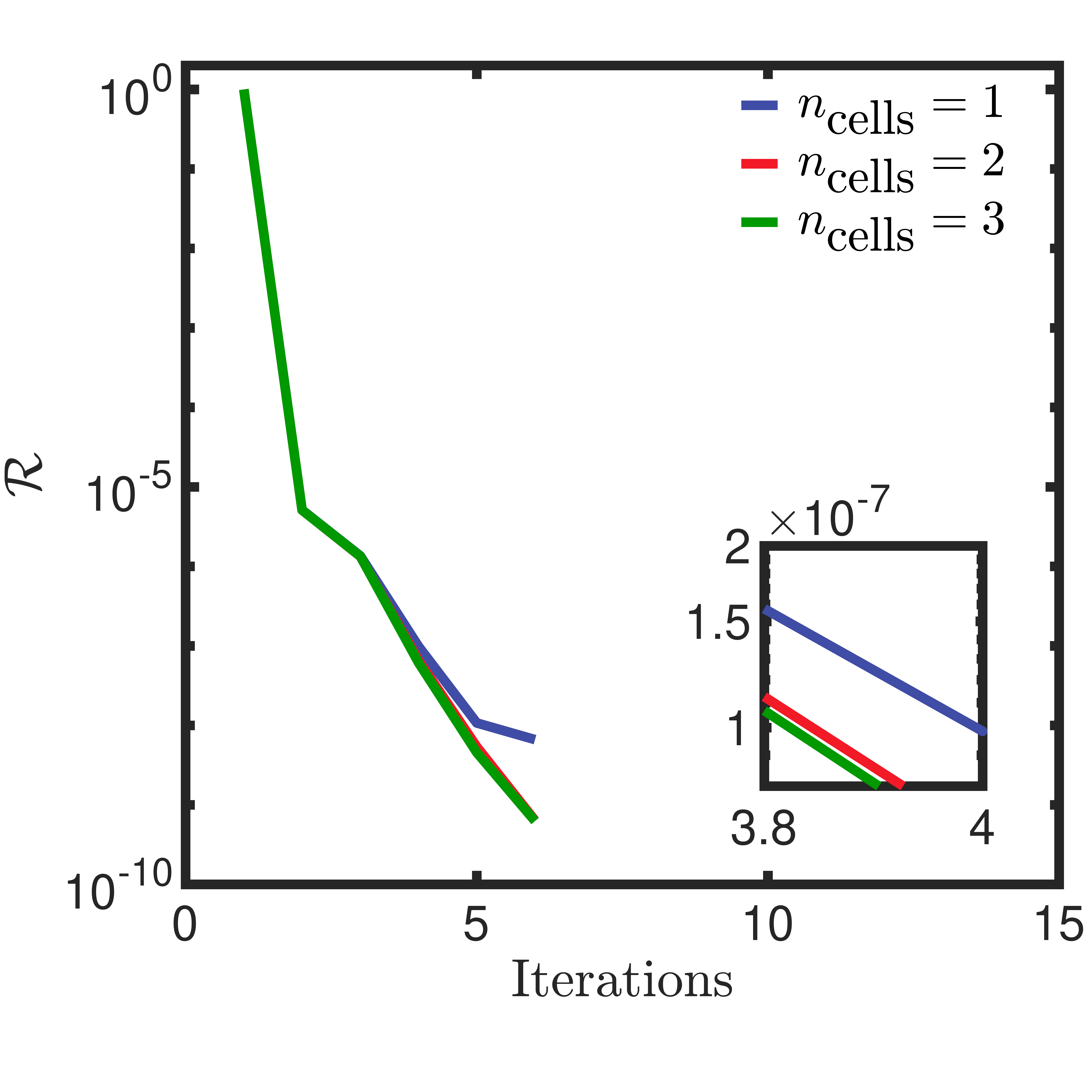}
    \label{fig_heaving_cyl_error_with_fix_ncells}
  }
  \caption{Convergence rate of the preconditioned FGMRES solver during the first time step of the simulation \subref{fig_residual_without_brinkman_in_PPE} without and \subref{fig_residual_with_brinkman_in_PPE} with the Brinkman penalty term in the projection preconditioner. The grid size is $N = 256$.}
\label{fig_heaving_cylinder_error}
\end{figure}

Using different values of $\ncells$, we compare the convergence rate of the linear solver during the first time step. Fig.~\ref{fig_heaving_cyl_error_wo_fix_ncells} illustrates the linear solver's convergence rate with $\theta = 0$. When $\theta = 0$, the solver requires more iterations to converge. However, the convergence rates are not significantly affected by $\ncells$ values. We achieve a robust convergence rate when the Brinkman penalty is included in the projection preconditioner, i.e., when $\theta = 1$ as shown in Fig.~\ref{fig_heaving_cyl_error_with_fix_ncells}. There is no significant difference in convergence rates in this case either with different values of $\ncells$. 
\begin{figure}
  \centering
    \subfigure[Percentage change in liquid volume]{
    \includegraphics[scale = 0.08]{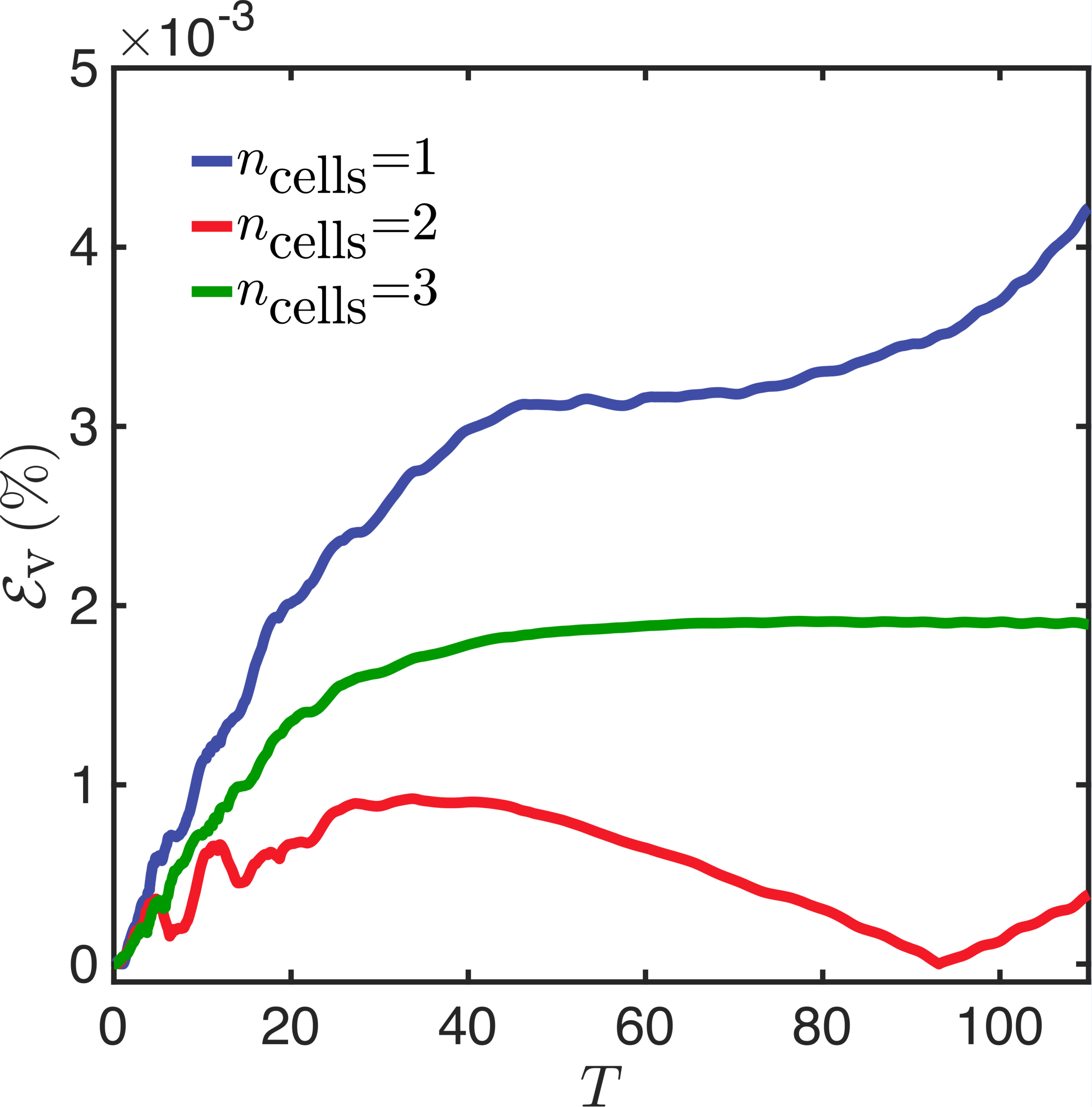} 
    \label{fig_heaving_cyl_liquid_colume}
  }
  \subfigure[Percentage change in gas volume]{
    \includegraphics[scale = 0.08]{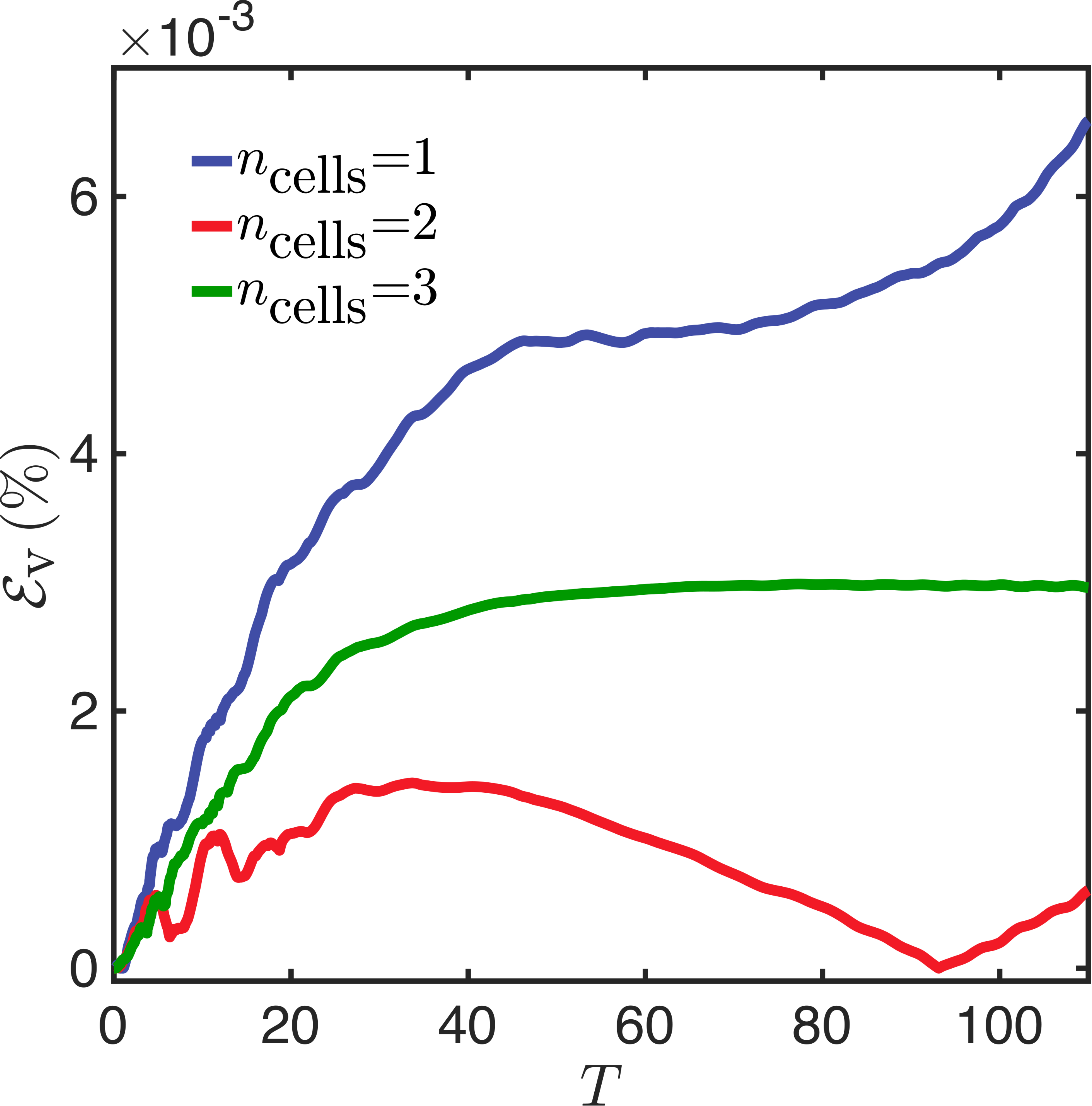} 
    \label{fig_heaving_cyl_gas_colume}
  }
 \caption{Percentage change $\mathcal{E}_\textrm{v}$ in the volume of liquid and gas over time for the heaving cylinder case with three $\ncells$ values.
}
\label{fig_heaving_cyl_ncells_volume_error}
\end{figure}

Finally, we compute the percentage volume change in liquid and gas over time as a function of $\ncells$, and is shown in Fig.~\ref{fig_heaving_cyl_ncells_volume_error}. A longer duration of simulation time $T$ is considered here. We consider a uniform grid of size $6000 \times 1200$ with a constant time step size of $\Delta t = 10^{-3}$ s. Again, $\ncells$ values do not significantly affect $\mathcal{E}_\textrm{v}$. The errors appear in the same order for all $\ncells$ values.

While increasing the $\ncells$ value does not affect the numerical solution quantitatively, it does make the solid-liquid and liquid-gas interfaces more diffuse. This produces  diffuse vortex structures as well. Therefore, we do not recommend using $\ncells > 2$ in the simulations.   

\section{Conclusions}
\label{sec_conclusions}
This paper presented a scalable preconditioner for the monolithic solution of the volume penalized single and multiphase incompressible Navier-Stokes equations. The same preconditioner can also be used to solve volume penalized low Mach Navier-Stokes equations monolithically. The preconditioner uses a projection algorithm that correctly accounts for the Brinkman penalty term in both pressure Poisson and velocity update equations. The accuracy and scalability of the solver under grid refinement were discussed. In contrast to prior experience, the solver convergence rate improves when the permeability parameter is decreased.  However, the penalized solution's error saturates at extremely low values of permeability. The inertial scale ($\kappa = \Delta t/\rho$) can be used as a reasonable starting point for permeability parameter, which can be gradually lowered until no further improvement in the penalized solution is observed. Our robust preconditioner allowed us to consider low $\kappa$ values for the multiphase FSI model. It was demonstrated that a dry contact line condition is imposed over the surface of the immersed body at low values of $\kappa$. In addition, low $\kappa$ values imply a no-slip condition, which leads to singularities in shear stress and pressure at the contact line~\cite{huh1971hydrodynamic}. In multiphase FSI simulations, low but not too small $\kappa$ values should be used. $\kappa = \Delta/\rhos$ is a robust choice. The robust preconditioner enabled us to select the right penalty value to arrest the solidified surface in the phase change problem. Additionally, we studied how $\kappa$ affected solution accuracy for manufactured solutions with our proposed preconditioner. We also showed that the use of the standard level set method leads to (an acceptable level of) volume loss of the conserved fluid phases. This loss, however, is not due to the Brinkman penalty. This can be further confirmed from the coupled Brinkman penalization and volume of fluid (VoF) simulations of Sharaborin et al.~\cite{Sharaborin2021}. These VoF simulations demonstrate mass/volume loss close to machine accuracy. Lastly,  we discussed the limitations of the multiphase FSI model used in this and other works~\cite{Zhang2010,Patel2018,Calderer2014,Sharaborin2021,bergmann2022numerical} in Sec.~\ref{sec_kappa_effect_wedge}, which are (i) fluid penetrating into the solid;  and (ii) inability to impose contact angle conditions. It was discussed that despite these limitations the model gives a converged FSI solution for a certain class of problems (e.g., in ocean engineering) that have length scales much larger than the capillary length scale and where the reservoir volume is much larger than the structure volume.  It remains a future endeavor to develop a VP/fictitious domain method that overcomes these two limitations while keeping the air-water interface smooth/continuous and satisfying the mass balance and momentum equations.   


\section*{Acknowledgements}
This research was supported by the Exascale Computing Project (17-SC-20-SC), a collaborative effort of the U.S. Department of Energy Office of Science and the National Nuclear Security Administration. We thank the Department of Energy’s Workforce Development of Teachers and Scientists as well as Workforce Development \& Education at Berkeley Lab programs that facilitated this work. R.T, K.K,  and A.P.S.B also acknowledge support from NSF awards OAC 1931368 and CBET CAREER 2234387.  SDSU's Fermi compute cluster was used to carry out the numerical simulations.  

\appendix
\renewcommand\thesection{\Alph{section}}

\section{Liquid fraction and enthalpy/temperature relations} \label{sec_enthalpy}

The specific enthalpy $\hen$ of solid, liquid, and mushy zones are defined in terms of their temperature $T$ as
\begin{align}
\hen = \begin{cases}
 \cps (T - T_r) ,&  T<\Tsol,\\ 
 \bar{C}(T - \Tsol) + \hsol + \varphi \frac{\displaystyle \rhol}{\displaystyle \rho}L,&\Tsol \le T \le \Tliq, \\ 
 \cpl(T-\Tliq)+\hliq,& T> \Tliq ,
\end{cases}
\label{eq_h_pcm}
\end{align}
and of the gas as
\begin{equation}
\hen = \cpg (T - T_r) .
\label{eq_h_gas}
\end{equation} 
Here, $\Tliq$ is the liquidus temperature at which initial solidification commences, and $\Tsol$ is the solidus temperature at which solidification concludes. In~Eq.~\eqref{eq_h_pcm}, $\hsol = \cps (\Tsol - T_r), \hliq = \bar{C}(\Tliq-\Tsol)+\hsol+L,  \text{ and  } \bar{C}=\frac{\cps+\cpl}{2}$. $\bar{C}$ is the specific heat of the mushy region, which is taken as an average of liquid  ($\cpl$) and solid ($\cps$) specific heats. Eqs.~(\ref{eq_h_pcm}) and (\ref{eq_h_gas}) imply that solid and gas enthalpies are zero at $T = T_r$. The numerical solution is not affected by this arbitrary choice of reference temperature $T_r$, and in the numerical simulations we set $T_r = T_m$ \footnote{We compared results of the simulations by considering $T_r$ = 0 and $T_m$, and they came out to be the same.}. We use a mixture model  to express density and specific enthalpy in terms of liquid fraction in the mushy region 
\begin{align}
\rho &= \varphi \rhol + (1-\varphi) \rhos   \label{eq_rho_mixture}, \\
\rho \hen &= \varphi \rhol \hliq + (1-\varphi) \rhos \hsol.   \label{eq_h_mixture}
\end{align}
Substituting $\hen$ from Eq.~\eqref{eq_h_pcm} and $\rho$ from Eq.~\eqref{eq_rho_mixture} into Eq.~\eqref{eq_h_mixture}, we obtain a $\varphi$-$T$ relation for the mushy region 
\begin{equation}
\varphi = \frac{\displaystyle \rho}{\displaystyle \rhol}\frac{\displaystyle T-\Tsol}{\displaystyle \Tliq-\Tsol}.
\label{eq_varphi_mixture}
\end{equation}
Knowing $\varphi$ in terms of $T$ (Eq.~\eqref{eq_varphi_mixture}) allows us to invert $\hen$-$T$ relations. The  temperature in the solid-liquid-mushy region
\begin{align}
T = \begin{cases}
 \frac{\displaystyle \hen}{\displaystyle \cps} + T_r, & \hen<\hsol,\\ 
  \Tsol + \frac{\displaystyle \hen-\hsol}{\displaystyle \hliq-\hsol}(\Tliq-\Tsol),&\hsol \le \hen \le \hliq, \\ 
 \Tliq + \frac{\displaystyle \hen -\hliq}{\displaystyle \cpl},& \hen > \hliq,
\end{cases}
\label{eq_T_pcm}
\end{align} 
and in the gas region
\begin{equation}
T =  \frac{\displaystyle \hen}{\displaystyle \cpg} + T_r
\label{eq_T_gas}
\end{equation}
can be written in terms of $\hen$. These $T$-$\hen$ relations are used in the Newton's iterations to solve the nonlinear enthalpy Eq.~\eqref{eq_enthalpy_lm} written in the main text. 
Similarly, substituting $\rho$ from Eq.~\eqref{eq_rho_mixture} into Eq.~\eqref{eq_h_mixture},  we get  a $\varphi$-$\hen$ relation 
\begin{align}
\varphi = \begin{cases}
 0,& \hen<\hsol,\\ 
  \frac{\displaystyle \rhos(\hsol-\hen)}{\displaystyle \hen(\rhol-\rhos)-\rhol \hliq + \rhos \hsol},&\hsol \le \hen \le \hliq, \\ 
 1,& \hen > \hliq.
\end{cases}
\label{eq_liquid_fraction}
\end{align}
Although arbitrary, $\varphi$ in the gas region is defined to be zero. 

Finally,  Eq.~\eqref{eq_liquid_fraction} allows us to define $\DDD{\varphi}{t}$ for the low Mach~Eq.~\eqref{eq_lm} as
\begin{align}
\DDD{\varphi}{t}=\begin{cases}
 0,& \hen <\hsol,\\ 
 \frac{\displaystyle -\rhos \rhol (\hsol-\hliq)}{\displaystyle (\hen(\rhol- \rhos)- \rhol \hliq + \rhos \hsol)^2}\displaystyle \DDD{\hen}{t},&\hsol \le \hen \le \hliq, \\ 
 0,& \hen > \hliq .
\end{cases}
\label{eq_dvarphi_dt}
\end{align}
The material derivative of $\hen$ in Eq.~\eqref{eq_dvarphi_dt} is obtained by expressing the enthalpy Eq.~\eqref{eq_enthalpy_lm} in non-conservative form as \[\DDD{h}{t}=\frac{1}{\rho} \left( \div\left({\kappa\grad{T}}\right) + Q_{\rm src} \right).\] It is clear from Eq.~\eqref{eq_dvarphi_dt} that $\DDD{\varphi}{t} \ne 0$ only in the mushy region where $\hsol \le \hen \le \hliq$ and $\Tsol \le T \le \Tliq$. Putting it together, the final form of the low Mach equation reads as
\begin{align}{}
& \div{\u} = \begin{cases} 
0, & \beta = 0 \; (\text {i.e., in the gas phase}), \\
0,& \hen <\hsol,\\ 
-\frac{ \rhos \rhol}{\rho^2}(\rhol-\rhos)\beta \frac{\displaystyle (\hliq-\hsol)}{\displaystyle \left(\hen(\rhol- \rhos)- \rhol \hliq +\rhos \hsol \right)^2}\displaystyle \left(\div {\kappa\grad{T}}  + Q_{\rm src} \right),&\hsol \le \hen \le \hliq, \\ 
0,& \hen>\hliq.\\ 
\end{cases} \label{eq_lowmach_discretized}
\end{align}

From Eq.~\eqref{eq_lowmach_discretized} it can be seen that velocity is non divergence-free only in the mushy region, but divergence-free elsewhere. In other words, in the absence of mushy regions, velocity is divergence-free. This can happen when a liquid phase has solidified completely or when a solid phase has melted completely. Our continuous and discrete formulations, therefore, guarantee that there will be no change in the volume of the system in the absence of phase change. It can also be seen from Eq.~\eqref{eq_lowmach_discretized} that when the densities of the solid and liquid phases match, there is no induced flow and the velocity is divergence-free.


\section{Formal derivation of the projection method} \label{sec_projection_derivation}

The saddle-point problem 
\begin{align}
\label{eq_saddle} 
\left[
\begin{array}{cc}
 \A & \G\\
 -\vD\cdot & \mathbf{0} \\
\end{array}
\right]
\left[
\begin{array}{c}
  \xu\\
  \xp \\
\end{array}
\right] & =
\left[
\begin{array}{c}
 \bu\\
 \bp \\
\end{array}
\right]
\end{align}
can formally be solved using the inverse of the Schur-complement  \[\S^{-1} = \left(-\vD \cdot {\A}^{-1} \G \right)^{-1}\] to obtain the exact pressure and velocity solutions 
\begin{subequations} 
\begin{equation}
\xp = -\S^{-1}(\vD \cdot \A^{-1} \bu + \bp), \label{eq_xp} 
\end{equation} 
\begin{equation}
\xu =  \A^{-1}(\bu - \G\xp) = \A^{-1}\bu + \A^{-1} \G \, \S^{-1}(\vD \cdot \A^{-1} \bu + \bp). \label{eq_xu}
\end{equation} 
\end{subequations} 
The projection method approximation to the velocity solution is obtained by approximating 
\begin{align*}
 \A^{-1} \G  \, \S^{-1} &=  \A^{-1} \G  \left(- \vD \cdot {\A}^{-1} \G \right)^{-1}  \\
                              &= \A^{-1}\G \left( - \vrhochi \vrhochi^{-1} \vD \cdot {\A}^{-1} \G \right)^{-1} \\
                              &\approx \A^{-1}\G \left( - (\vD \cdot \vrhochi^{-1} \G) \vrhochi  {\A}^{-1} \right)^{-1} =  \A^{-1}\G \left( -  \Lrhochi \vrhochi  {\A}^{-1} \right)^{-1}  \\
                              & = -\A^{-1}\G  \A  \vrhochi^{-1} \Lrhochi^{-1} \\
                              &\approx -\G \A^{-1} \A \vrhochi^{-1} \Lrhochi^{-1} \\
                              & = -\vrhochi^{-1} \G \Lrhochi^{-1}.
\end{align*} 
Here, we have commuted a few operators (which is valid only for constant-coefficient operators defined on periodic domains) to simplify $\A^{-1} \G  \, \S^{-1}$. This simplification allows to approximate the exact velocity solution $\xu$ in Eq.~\eqref{eq_xu} as
\begin{align*}
\xu &= \A^{-1}\bu + \A^{-1} \G \, \S^{-1}(\vD \cdot \A^{-1} \bu + \bp)  \\
      &\approx  \A^{-1}\bu  -\vrhochi^{-1} \G \, \Lrhochi^{-1} (\vD \cdot \A^{-1} \bu + \bp)  \\
      &=  \A^{-1}\bu  - \Delta t  \; \vrhochi^{-1} \G \, \left( \Lrhochi^{-1} \frac{1}{\Delta t} (\vD \cdot \A^{-1} \bu + \bp)  \right) \\
      & =  \widehat{\x}_{\u}   - \Delta t  \; \vrhochi^{-1} \G  \vvarphi. 
\end{align*}
Here,  $\vvarphi$ is an auxiliary pressure-like scalar field that satisfies the Poisson Eq.~\eqref{eq_frac_div}. Approximating the inverse of the Schur complement as (see Cai et al.~\cite{Cai2014})
\begin{align}
\S^{-1} = \left(-\vD \cdot {\A}^{-1} \G \right)^{-1}  \approx -\frac{1}{\Delta t} \Lrhochi^{-1} + \V{\mu} \label{eq_inv_s}
\end{align}
provides an approximate pressure solution $\xp$ 
\begin{align*}
 \xp &= -\S^{-1}(\vD \cdot \A^{-1} \bu + \bp) \\
        &\approx \left( \frac{1}{\Delta t}  \Lrhochi^{-1} - \V{\mu} \right) (\vD \cdot \A^{-1} \bu + \bp) \\
        & =   \Lrhochi^{-1}  \left( \frac{ \vD \cdot   \widehat{\x}_{\u} + \bp}{\Delta t} \right) - \V{\mu}(\vD \cdot   \widehat{\x}_{\u} + \bp) \\
        & = \vvarphi - \Delta t  \, \V{\mu} \, \Lrhochi \vvarphi.
 \end{align*}
Note that in this work we use the trapezoidal rule for time integrating the viscous Laplacian term. If instead backward Euler scheme is employed then
\begin{align}
\S^{-1} \approx -\frac{1}{\Delta t} \Lrhochi^{-1} + 2 \V{\mu} \label{eq_inv_s_backward_euler}
\end{align} 
is suggested based on the spectral analysis of the viscous operator (see Cai et al.~\cite{Cai2014}).

\section{Two-dimensional dam break problem} \label{sec_dam_break}

\begin{figure}
\centering
\includegraphics[scale = 0.12]{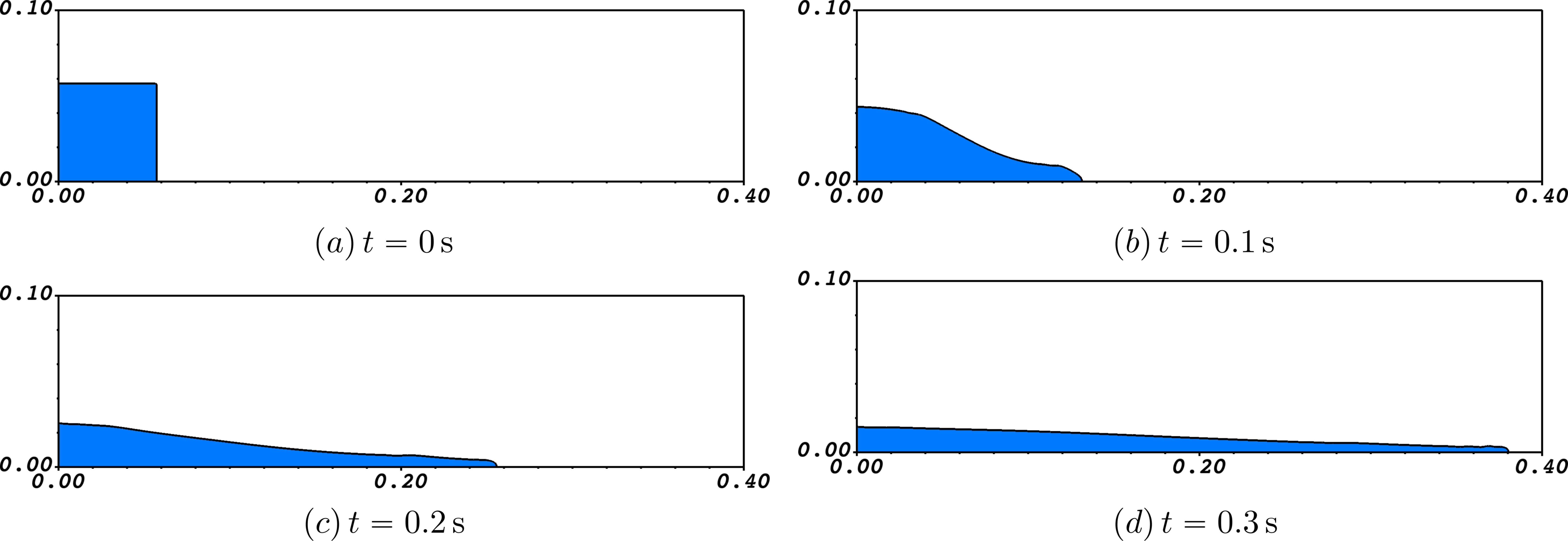}
  \caption{Evolution of the spreading water column in two dimensions considering a density ratio of $\rhol/\rhog = 815.66$ and a viscosity ratio of $\mul/\mug = 63.88$ between dense water and light air fluids.}
  \label{fig_dam_break}
\end{figure} 

\begin{figure}
  \centering
    \subfigure[]{
    \includegraphics[scale = 0.08]{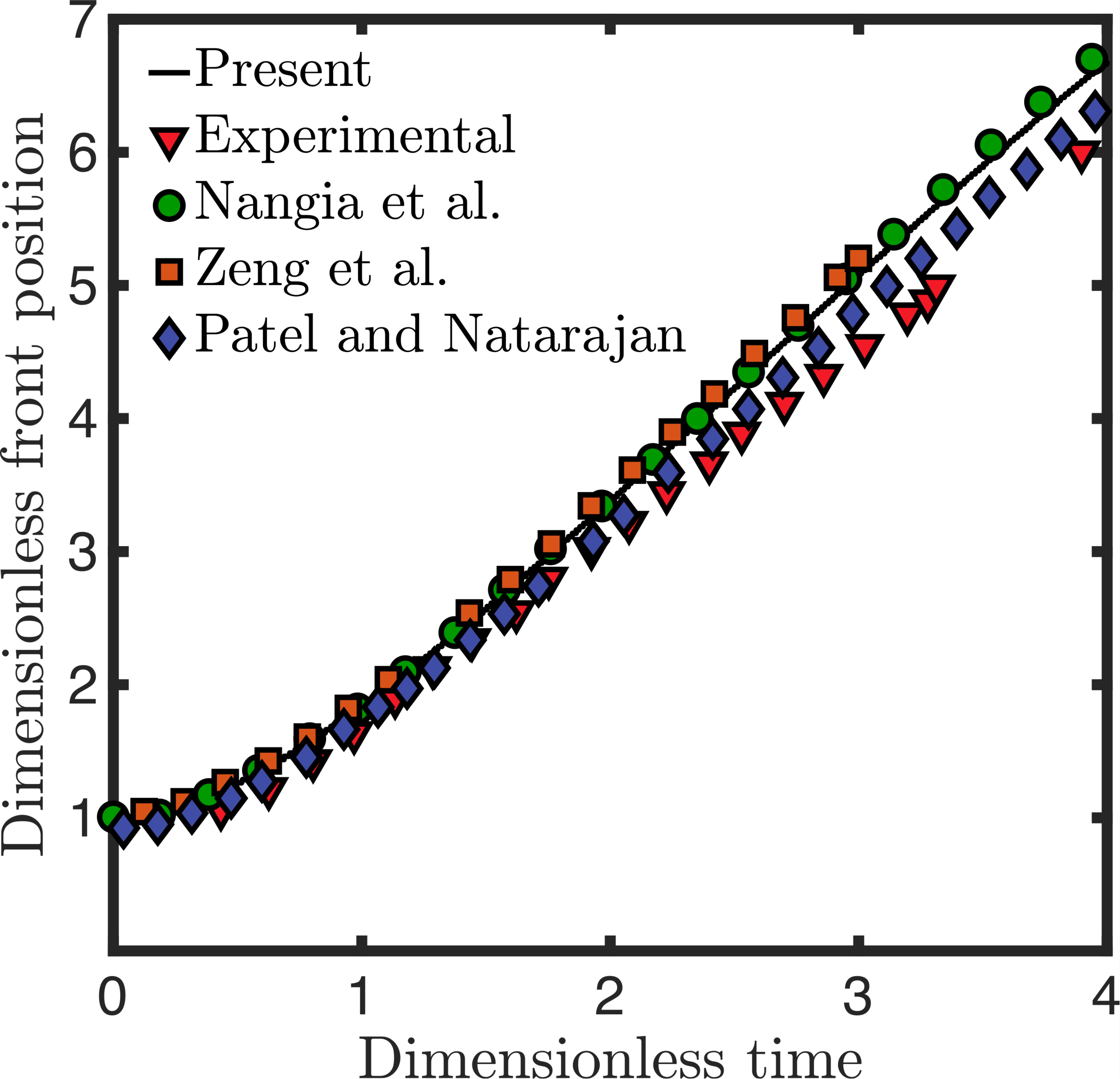} 
    \label{fig_wedge_liquid_volume}
  }
  \subfigure[]{
    \includegraphics[scale = 0.08]{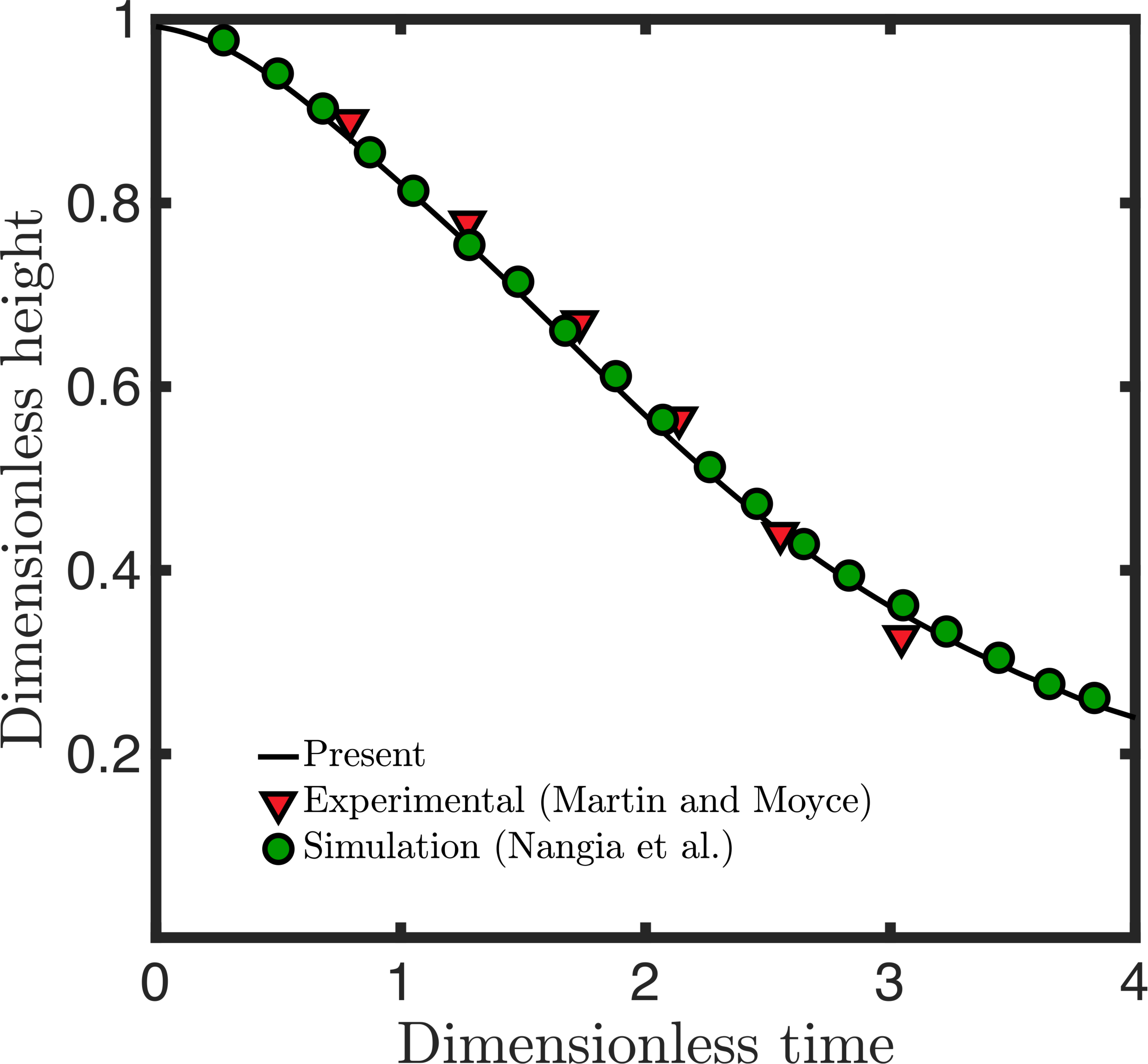} 
    \label{fig_wedge_gas_volume}
  }
 \caption{Comparison of the temporal evolution of dimensionless front position with prior studies: present work (black), experiments of Martin and Moyce (red); and numerical studies of Nangia et al. (green), Zeng et al. (magenta), and Patel and Natarajan (blue).
}
\label{fig_dam_break_dynamics}
\end{figure}

\begin{figure}
  \centering
    \subfigure[Percentage change in liquid volume]{
    \includegraphics[scale = 0.08]{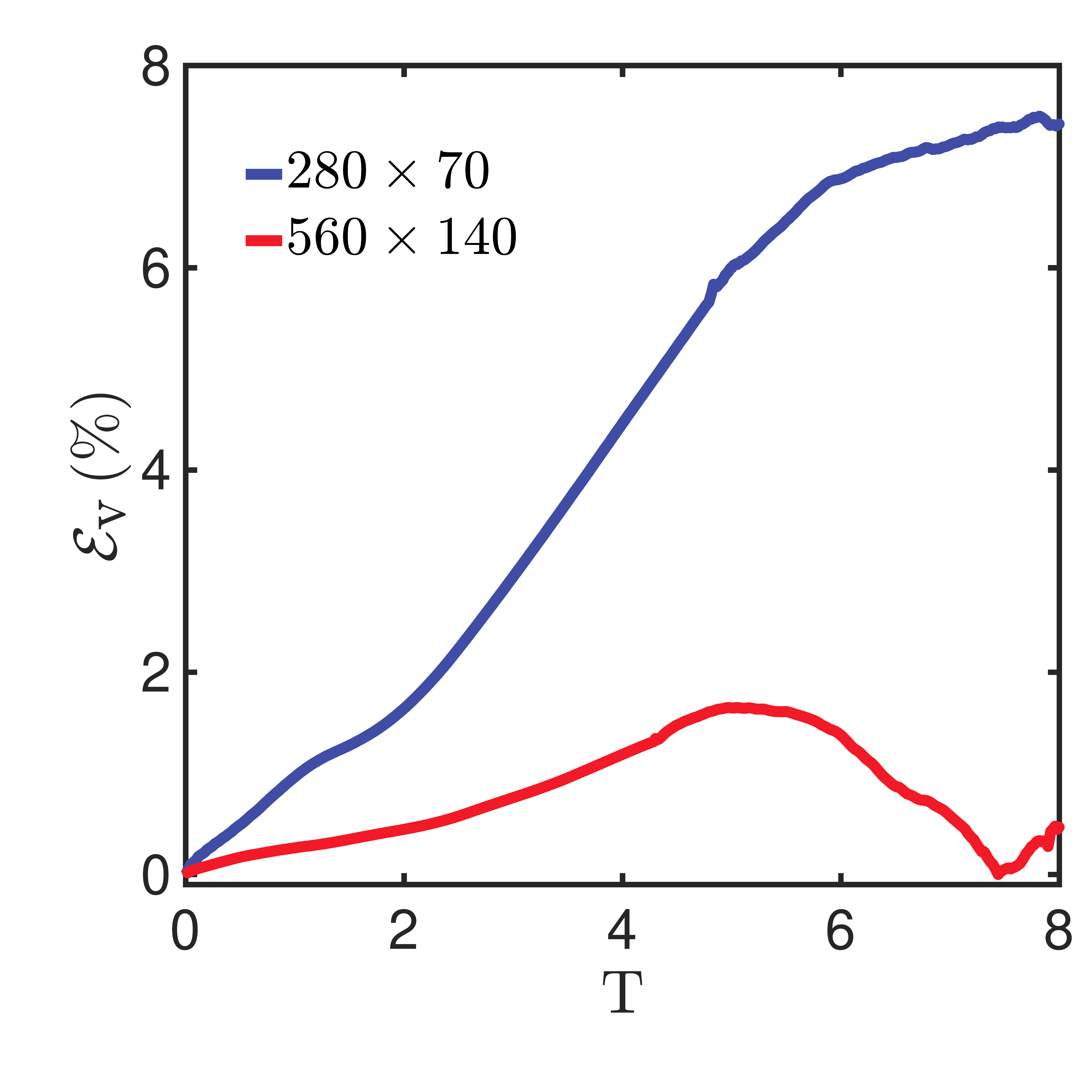} 
    \label{fig_wedge_liquid_volume}
  }
  \subfigure[Percentage change in gas volume]{
    \includegraphics[scale = 0.08]{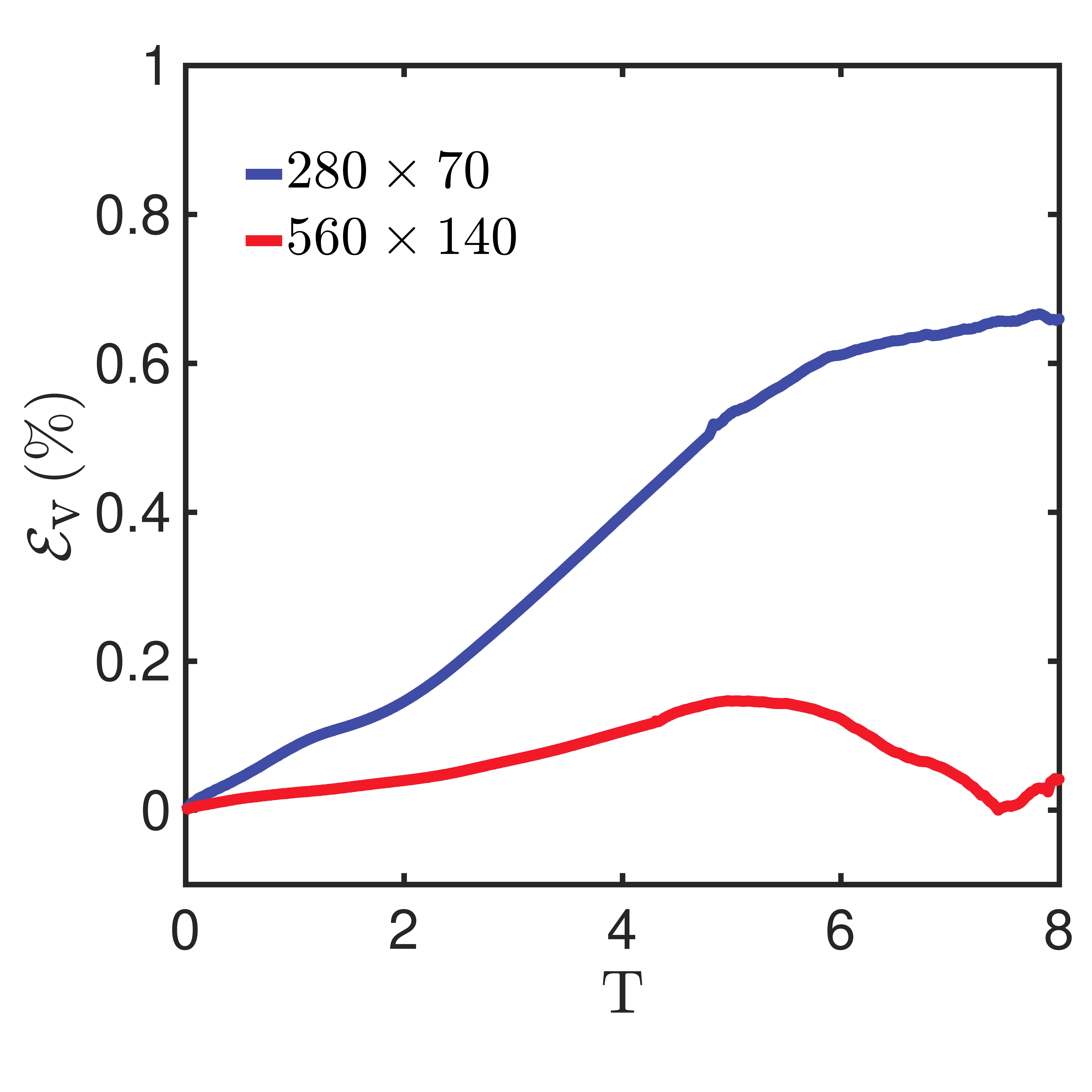} 
    \label{fig_wedge_gas_volume}
  }
 \caption{Percentage volume change $\mathcal{E}_\textrm{v}$ of liquid and gas over time for the two-dimensional dam break problem at different grid sizes. A uniform time step size of $\Delta t = 5\times10^{-5}$ s is used for coarser grid ($N = 70$) and it is halved for the finer grid ($N = 140$).
}
\label{fig_dam_break_volume_error}
\end{figure}

Using the two-dimensional dam break test case, we demonstrate two aspects of our multiphase solver: (1) consistent mass and momentum transport scheme that ensures numerical stability of high density ratio flows; and (2) the percentage change in the volume of conserved phases (liquid and gas) in the absence of rigid bodies. Initially, the water column occupies a square block of size $a=0.057$ m within a computational domain $\Omega \in [0,7a] \times [0,1.75a]$. The lower left corner of the water column aligns with the lower left corner of the computational domain; the remaining domain is occupied by air. A uniform grid of size $4N \times N$ is used for discretizing the computational domain. Two grid resolutions are considered $N = \{70, 140\}$. The domain has a no-slip velocity boundary condition on all sides.  Water and air have densities of  $\rhol =$ 1000 kg/m$^3$ and $\rhog =$ 1.226 kg/m$^3$, respectively, and viscosities of $\mul = 1.137\times10^{-3}$ Pa $\cdot$ s and $\mug = 1.78\times10^{-5}$ Pa $\cdot$ s, respectively. The surface tension coefficient between the air and water phases is taken to be $\gamma =0.0728$ N/m. The fluids are initially at rest. This problem has been studied numerically by Nangia et al.~\cite{Nangia2019MF}, Patel and Natarajan~\cite{Patel2017} and Zeng et al.~\cite{zeng2023consistent}, and experimentally by Martin and Moyce~\cite{Martin1952}.

Fig.~\ref{fig_dam_break} shows the evolution of the air-water interface using a $N=140$ grid at different time instants. Based on the results, we can conclude that our multiphase formulation provides a physical solution that remains stable over time.  Using prior experiments and numerical solutions, we compare the temporal evolution of the dimensionless front position (non-dimensionalized by $a$) with the numerical solution. The results of our study are in excellent agreement with those of previous studies. Also presented is the percentage volume change of liquid and gas as a function of non-dimensional time $T = t\sqrt{\frac{g}{a}}$ in Fig.~\ref{fig_dam_break_volume_error}.  As can be observed,  the percentage volume change is less than 1.5\% for the finer grid. This also agrees well with the previous level set-based simulations of  Zeng et al. \cite{zeng2023consistent}, who report a percentage volume change around $\mathcal{E}_{\rm v} \approx 2\%$ for this problem.

\section{Rayleigh-Taylor instability problem} \label{sec_RTI}

\begin{figure}
  \centering
  \subfigure[Interfacial dynamics]{
    \includegraphics[scale = 0.6]{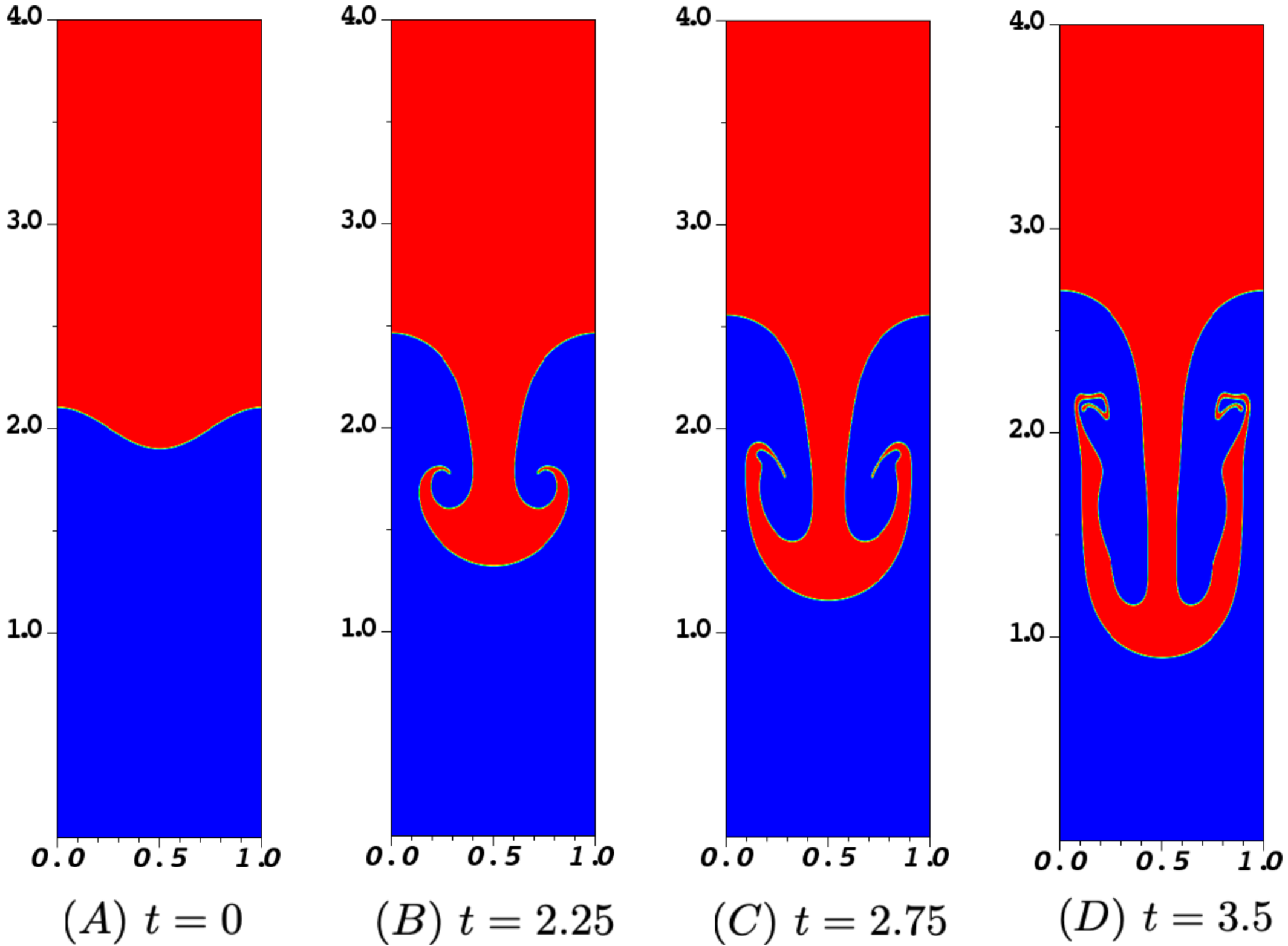} 
    \label{fig_RT2P}
  }
   \subfigure[Volume change of the top heavy fluid]{
    \includegraphics[scale = 0.35]{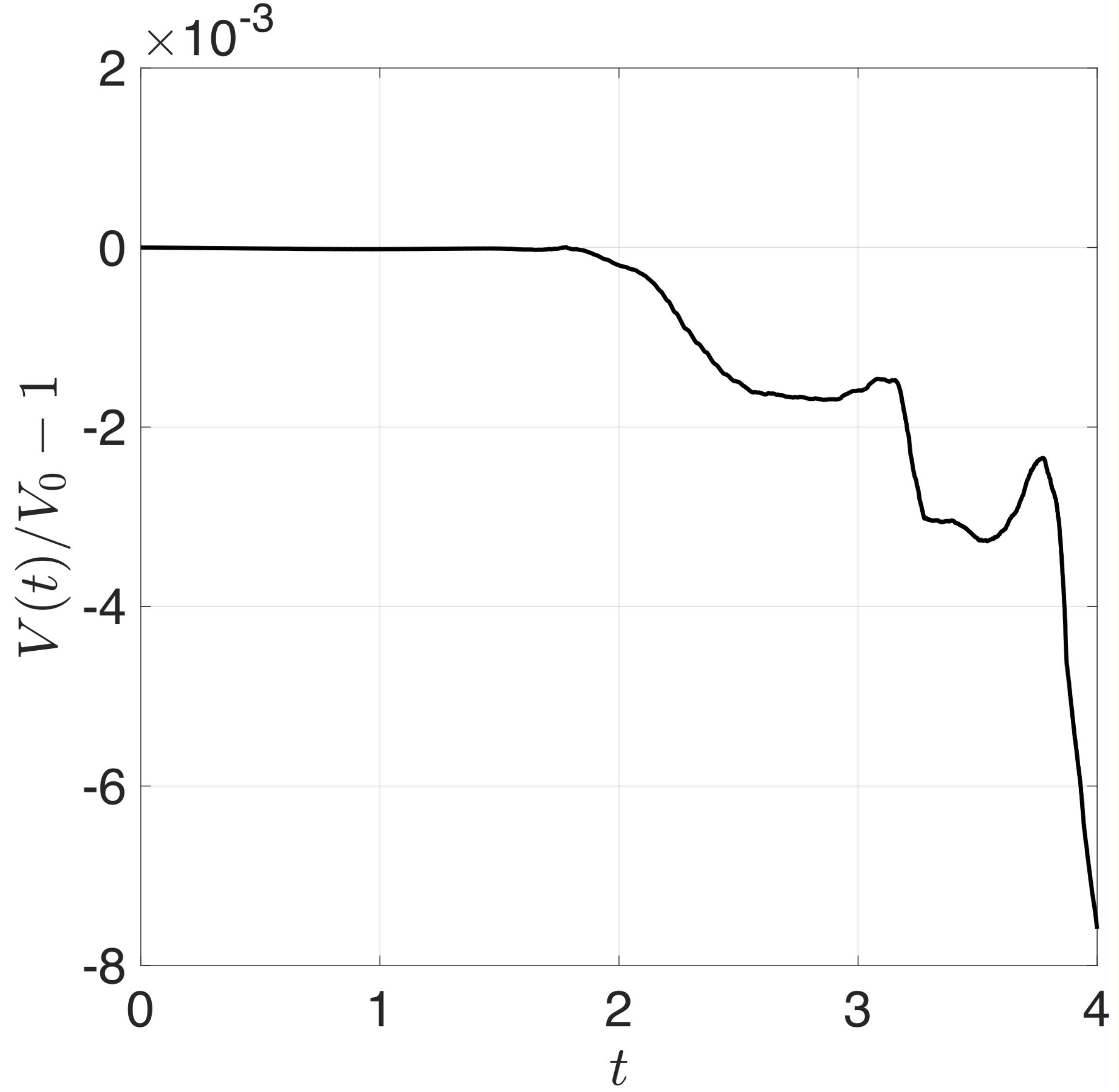}
    \label{fig_RT2PVol}
  }
  \caption{Two phase Rayleigh-Taylor instability problem at Atwood number $A = 0.5$ and Reynolds number $Re = 3000$. \subref{fig_RT2P} Temporal evolution of the top heavy (red color) and bottom light (blue color) fluids in the domain. \subref{fig_RT2PVol} Volume change of the heavy fluid over time. Here, time $t$ is non-dimensionalized by $d/\sqrt{gd}$.}
\label{fig_RT2P_interface_vol}
\end{figure}

\begin{figure}
  \centering
  \subfigure[Interfacial dynamics]{
    \includegraphics[scale = 0.6]{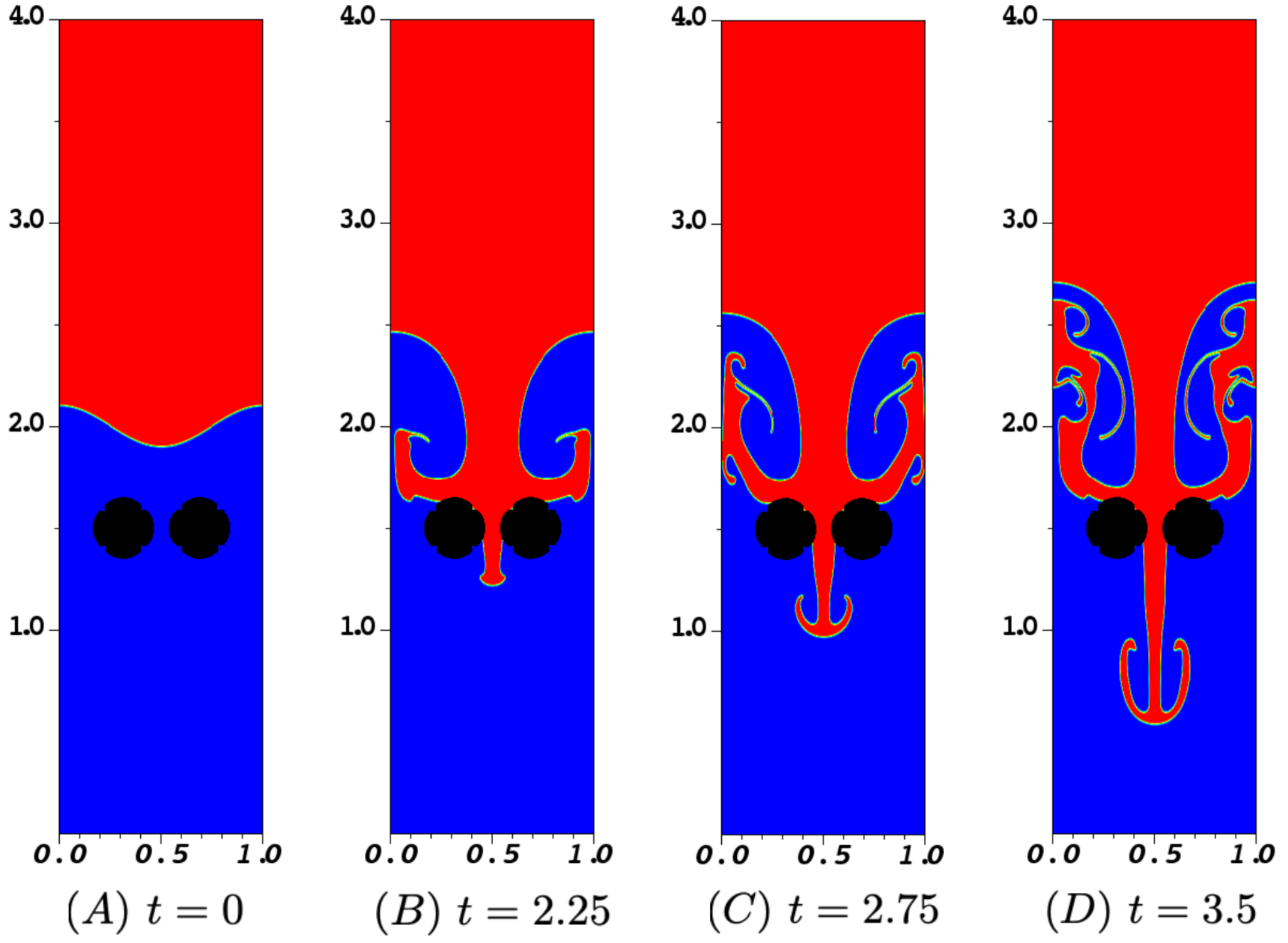} 
    \label{fig_RT3P}
  }
    \subfigure[Zoomed-in view near the obstacles at $t= 3.5$]{
    \includegraphics[scale = 0.2]{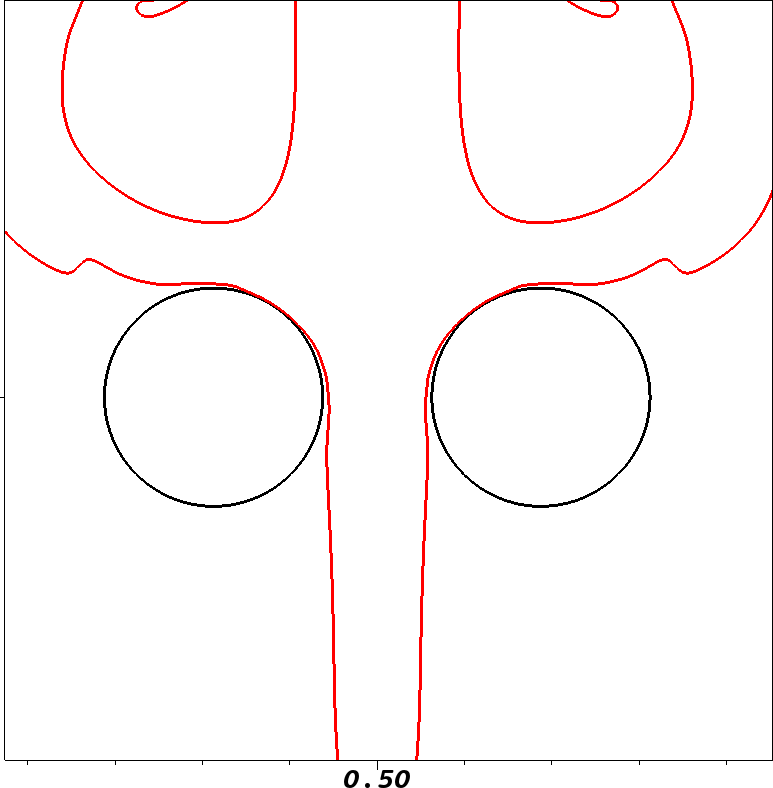} 
    \label{fig_RT3PContour}
  }
   \subfigure[Volume change of the top heavy fluid]{
    \includegraphics[scale = 0.3]{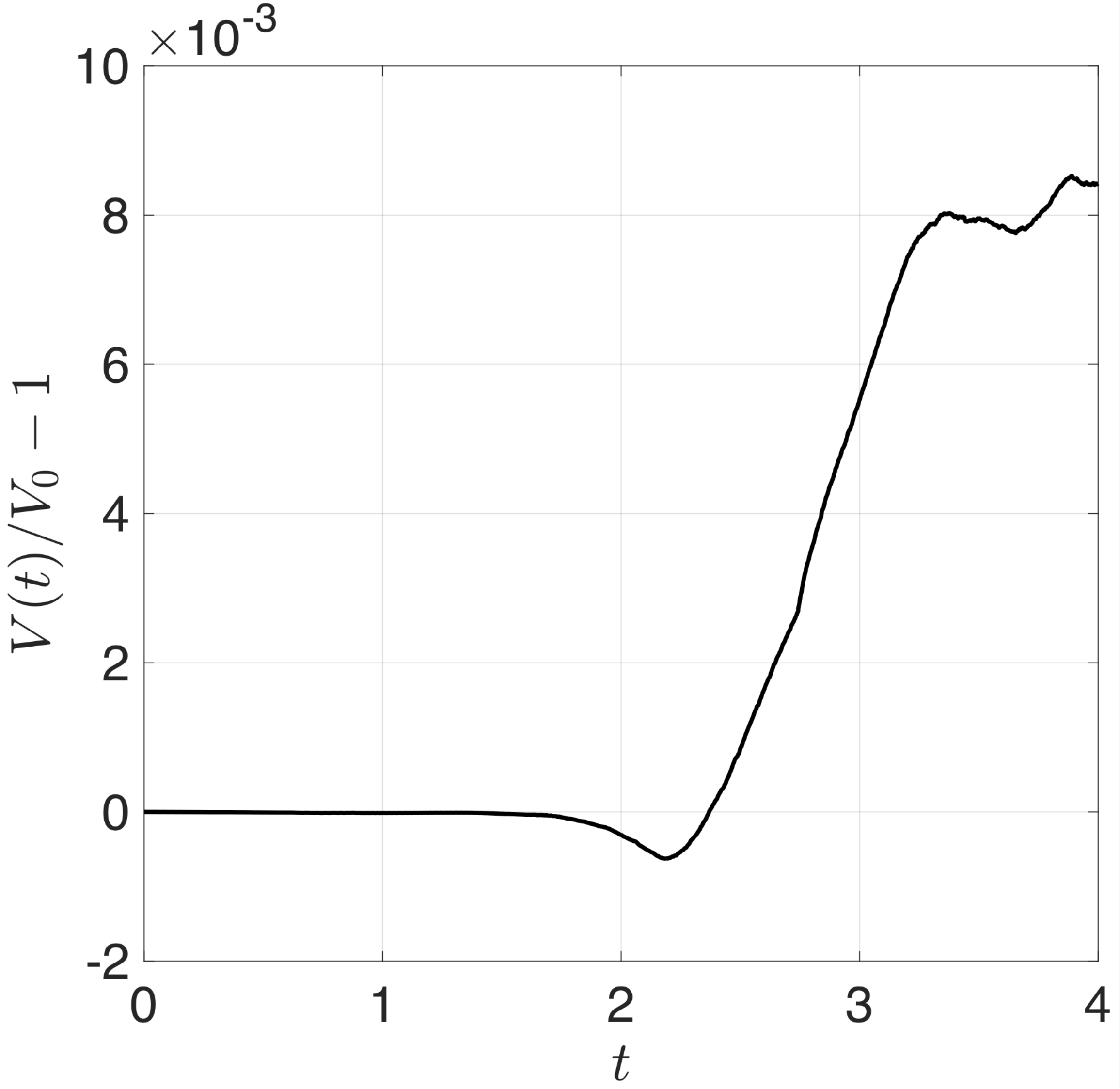}
    \label{fig_RT3PVol}
  }
  \caption{Three phase Rayleigh-Taylor instability problem at Atwood number $A = 0.5$ and Reynolds number $Re = 3000$. Two rigid cylinders (shown as black disks) are placed in the path of the falling heavy fluid. \subref{fig_RT3PContour} Zoomed-in view of the fluid-fluid interface (red color) and the outer surface of the cylinders (black color). \subref{fig_RT3P} Temporal evolution of the top heavy (red color) and bottom light (blue color) fluids in the domain. \subref{fig_RT3PVol} Volume change of the heavy fluid over time. Here, time $t$ is non-dimensionalized by $d/\sqrt{gd}$.}
\label{fig_RT3P_interface_vol}
\end{figure}

For the level set method, different discretization, advection, and time-stepping schemes could result in different mass/volume changes of the conserved phases. Several combinations of these schemes have been extensively tested by Solomenko et al.~\cite{solomenko2017mass} on a number of benchmarking two-phase problems, including the Rayleigh-Taylor instability (RTI) flow. It is important to note, however, that the authors did not include the subcell-fix method in their study as they mention that ``\textit{These methods were not tested here because we have found [their] implementation comparatively complex.}"  To complement the prior study, as well as to compare the performance of the subcell-fix method with the best performing method(s) reported in~\cite{solomenko2017mass}, we simulate the RTI case using the same grid resolution ($\Delta x = \Delta y = h = 1/192$) and problem  parameters as in Sec. 7 of Solomenko et al. Specifically, we consider the Atwood number  to be $A = (\rho_h - \rho_l)/(\rho_h + \rho_l)  = 0.5$, and the Reynolds number to be $Re = \rho_h d \sqrt{gd}/\mu = 3000$. Here, $\rho_h$ and $\rho_l$, denote density of heavier and lighter phase, respectively, $g$ is the acceleration to gravity (acting in the negative $y$ direction), and $\mu$ is the fluid viscosity, which is the same in both fluid phases. The initial interface between the top heavy  and bottom light fluid is considered to be a cosine function with a small amplitude. The initial signed distance is then expressed as $\sigma(x,y, t = 0) = d(2 + 0.1\cos(2\pi x/d)) - y$, in which $d$ is the domain width and $4d$ is the domain height.  Fig.~\ref{fig_RT2P} shows the two phase interface evolution at various non-dimensional time instants $t$ (normalized by time scale $d/\sqrt{gd}$), whereas Fig.~\ref{fig_RT2PVol} reports the change in the volume of the heavier fluid as a function of non-dimensional time. Solomenko et al. report volume changes till $t = 3.5$, and their best performing method (in terms of least amount of volume changes and parasitic oscillations at the interface) leads to a percentage volume change of 0.3\% (reported in Table 10 of~\cite{solomenko2017mass}). As can be seen in Fig.~\ref{fig_RT2P_interface_vol}, the subcell-fix method of Min also leads to similar volume change around $t = 3.5$, and does not cause parasitic oscillations at the evolving interface.

As a variation of the two phase RTI problem, and to examine the effect of the Brinkman penalty term on volume change of the phases, we introduce two cylinders of radius $R = 0.125 d$ in the path of the falling heavy fluid. Left and right cylinder centers are located at $(0.3125d, 1.5d)$ and $(0.6875d, 1.5d)$, respectively. These objects are held stationary throughout the simulation. Fig.~\ref{fig_RT3P} shows the perturbed dynamics of the fluid-fluid interface in the presence of rigid obstacles, which results in a more chaotic dynamics than the two phase RTI problem. A zoomed-in view of the fluid-fluid interface at $t = 3.5$ near the stationary cylinders is provided in Fig.~\ref{fig_RT3PContour}. The no-slip condition imposed by the Brinkman penalty term prevents the fluid-fluid interface from penetrating the rigid surfaces. Fig.~\ref{fig_RT3PVol} shows the volume change of the heavier fluid as a function of time. The percentage volume change for the heavier fluid at $t=4$ is approximately 0.8\%, which is similar to the two-phase RTI problem; see Fig.~\ref{fig_RT2PVol}. We can conclude from this example that the Brinkman penalization term does not result in additional mass/volume changes (in the conserved phases) beyond what is expected from the standard level set method.

\newpage
\section{Bibliography}
\begin{flushleft}
 \bibliography{Brinkman_solver_biblography.bib}

\begin{thebibliography}{10}
\expandafter\ifx\csname url\endcsname\relax
  \def\url#1{\texttt{#1}}\fi
\expandafter\ifx\csname urlprefix\endcsname\relax\def\urlprefix{URL }\fi
\expandafter\ifx\csname href\endcsname\relax
  \def\href#1#2{#2} \def\path#1{#1}\fi

\bibitem{glowinski1994fictitious}
R.~Glowinski, T.-W. Pan, J.~Periaux, {A fictitious domain method for Dirichlet
  problem and applications}, Computer Methods in Applied Mechanics and
  Engineering 111~(3-4) (1994) 283--303.

\bibitem{glowinski1999distributed}
R.~Glowinski, T.-W. Pan, T.~I. Hesla, D.~D. Joseph, {A distributed Lagrange
  multiplier/fictitious domain method for particulate flows}, International
  Journal of Multiphase Flow 25~(5) (1999) 755--794.

\bibitem{Patankar2000}
N.~A. Patankar, P.~Singh, D.~D. Joseph, R.~Glowinski, T.-W. Pan, A new
  formulation of the distributed lagrange multiplier/fictitious domain method
  for particulate flows, International Journal of Multiphase Flow 26~(9) (2000)
  1509--1524.

\bibitem{Arquis1984}
E.~Arquis, J.~P. Caltagirone, Sur les conditions hydrodynamiques au voisinage
  d’une interface milieu fluide-milieu poreux: Applicationa la convection
  naturelle, Comptes Rendus de l'AcadÃ©mie des Sciences - Series IIB 299
  (1984) 1--4.

\bibitem{Angot1999}
P.~Angot, C.-H. Bruneau, P.~Fabrie, A penalization method to take into account
  obstacles in incompressible viscous flows, Numerische Mathematik 81~(4)
  (1999) 497--520.

\bibitem{Brinkman1949}
H.~C. Brinkman, A calculation of the viscous force exerted by a flowing fluid
  on a dense swarm of particles, Applied Science Research 1 (1949) 27.

\bibitem{BhallaBP2019}
A.~P.~S. Bhalla, N.~Nangia, P.~Dafnakis, G.~Bracco, G.~Mattiazzo, {Simulating
  water-entry/exit problems using Eulerian-Lagrangian and fully-Eulerian
  fictitious domain methods within the open-source IBAMR library}, Applied
  Ocean Research 94 (2020) 101932.

\bibitem{Rossinelli2010}
D.~Rossinelli, M.~Bergdorf, G.-H. Cottet, P.~Koumoutsakos, {GPU} accelerated
  simulations of bluff body flows using vortex particle methods, Journal of
  Computational Physics 229~(9) (2010) 3316--3333.

\bibitem{Thirumalaisamy2021}
R.~Thirumalaisamy, N.~Nangia, A.~P.~S. Bhalla, {Critique on ``Volume
  penalization for inhomogeneous Neumann boundary conditions modeling scalar
  flux in complicated geometry"}, Journal of Computational Physics 433 (2021)
  110163.

\bibitem{Thirumalaisamy2022}
R.~Thirumalaisamy, N.~A. Patankar, A.~P.~S. Bhalla, Handling {N}eumann and
  {R}obin boundary conditions in a fictitious domain volume penalization
  framework, Journal of Computational Physics 448 (2022) 110726.

\bibitem{Kou2022}
J.~Kou, S.~Joshi, A.~Hurtado-de\mbox{-}Mendoza, K.~Puri, C.~Hirsch, E.~Ferrer,
  Immersed boundary method for high-order flux reconstruction based on volume
  penalization, Journal of Computational Physics 448 (2022) 110721.

\bibitem{Carman1937}
P.~C. Carman, Fluid flow through granular beds, Transactions of the Institute
  of Chemical Engineers 15 (1937) 150--166.

\bibitem{Voller1987}
V.~R. Voller, C.~Prakash, A fixed grid numerical modelling methodology for
  convection-diffusion mushy region phase-change problems, International
  journal of heat and mass transfer 30~(8) (1987) 1709--1719.

\bibitem{Huang2022}
Z.~Huang, G.~Lin, A.~M. Ardekani, A consistent and conservative phase-field
  model for thermo-gas-liquid-solid flows including liquid-solid phase change,
  Journal of Computational Physics 449 (2022) 110795.

\bibitem{Gazzola2011b}
M.~Gazzola, P.~Chatelain, W.~M. Van~Rees, P.~Koumoutsakos, Simulations of
  single and multiple swimmers with non-divergence free deforming geometries,
  Journal of Computational Physics 230~(19) (2011) 7093--7114.

\bibitem{Bergmann2011}
M.~Bergmann, A.~Iollo, Modeling and simulation of fish-like swimming, Journal
  of Computational Physics 230~(2) (2011) 329--348.

\bibitem{engels2015numerical}
T.~Engels, D.~Kolomenskiy, K.~Schneider, J.~Sesterhenn, Numerical simulation of
  fluid--structure interaction with the volume penalization method, Journal of
  Computational Physics 281 (2015) 96--115.

\bibitem{Khedkar2020}
K.~Khedkar, N.~Nangia, R.~Thirumalaisamy, A.~P.~S. Bhalla, {The inertial sea
  wave energy converter (ISWEC) technology: Device-physics, multiphase modeling
  and simulations}, Ocean Engineering 229 (2021) 108879.

\bibitem{khedkar2022model}
K.~Khedkar, A.~P.~S. Bhalla,
  \href{https://www.sciencedirect.com/science/article/pii/S0029801822012471}{{A
  model predictive control (MPC)-integrated multiphase immersed boundary (IB)
  framework for simulating wave energy converters (WECs)}}, Ocean Engineering
  260 (2022) 111908.
\newblock \href {https://doi.org/10.1016/j.oceaneng.2022.111908}
  {\path{doi:10.1016/j.oceaneng.2022.111908}}.
\newline\urlprefix\url{https://www.sciencedirect.com/science/article/pii/S0029801822012471}

\bibitem{Sharaborin2021}
E.~L. Sharaborin, O.~A. Rogozin, A.~R. Kasimov, The coupled volume of fluid and
  {B}rinkman penalization methods for simulation of incompressible multiphase
  flows, Fluids 6~(9) (2021) 334.

\bibitem{bergmann2022numerical}
M.~Bergmann, Numerical modeling of a self-propelled dolphin jump out of water,
  Bioinspiration \& Biomimetics 17~(6) (2022) 065010.

\bibitem{xie2020three}
Z.~Xie, T.~Stoesser, A three-dimensional cartesian cut-cell/volume-of-fluid
  method for two-phase flows with moving bodies, Journal of Computational
  Physics 416 (2020) 109536.

\bibitem{van2023two}
M.~van~der Eijk, P.~Wellens, Two-phase free-surface flow interaction with
  moving bodies using a consistent, momentum preserving method, Journal of
  Computational Physics 474 (2023) 111796.

\bibitem{Kolomenskiy2009}
D.~Kolomenskiy, K.~Schneider, A fourier spectral method for the
  {N}avier--{S}tokes equations with volume penalization for moving solid
  obstacles, Journal of Computational Physics 228~(16) (2009) 5687--5709.

\bibitem{Sakurai2019}
T.~Sakurai, K.~Yoshimatsu, N.~Okamoto, K.~Schneider, Volume penalization for
  inhomogeneous {N}eumann boundary conditions modeling scalar flux in
  complicated geometry, Journal of Computational Physics 390 (2019) 452--469.

\bibitem{Gazzola2011}
M.~Gazzola, O.~V. Vasilyev, P.~Koumoutsakos, Shape optimization for drag
  reduction in linked bodies using evolution strategies, Computers \&
  Structures 89~(11-12) (2011) 1224--1231.

\bibitem{Beaugendre2018}
H.~Beaugendre, F.~Morency, Penalization of the {S}palart--{A}llmaras turbulence
  model without and with a wall function: Methodology for a vortex in cell
  scheme, Computers \& Fluids 170 (2018) 313--323.

\bibitem{Griffith2009}
B.~E. Griffith, {An accurate and efficient method for the incompressible
  Navier-Stokes equations using the projection method as a preconditioner},
  Journal of Computational Physics 228~(20) (2009) 7565--7595.

\bibitem{Cai2014}
M.~Cai, A.~Nonaka, J.~B. Bell, B.~E. Griffith, A.~Donev, Efficient
  variable-coefficient finite-volume stokes solvers, Communications in
  Computational Physics 16~(5) (2014) 1263--1297.

\bibitem{Nangia2019MF}
N.~Nangia, B.~E. Griffith, N.~A. Patankar, A.~P.~S. Bhalla, {A robust
  incompressible Navier-Stokes solver for high density ratio multiphase flows},
  Journal of Computational Physics 390 (2019) 548--594.

\bibitem{ahlkrona2021cut}
J.~Ahlkrona, D.~Elfverson, A cut finite element method for non-newtonian free
  surface flows in 2d-application to glacier modelling, Journal of
  Computational Physics: X 11 (2021) 100090.

\bibitem{lofgren2022increasing}
A.~L{\"o}fgren, J.~Ahlkrona, C.~Helanow, Increasing stable time-step sizes of
  the free-surface problem arising in ice-sheet simulations, Journal of
  Computational Physics: X 16 (2022) 100114.

\bibitem{thirumalaisamy2023low}
R.~Thirumalaisamy, A.~P.~S. Bhalla, A low mach enthalpy method to model
  non-isothermal gas-liquid-solid flows with melting and solidification, arXiv
  preprint arXiv:2301.06256 (2023).

\bibitem{voller1991eral}
V.~R. Voller, C.~Swaminathan, {ERAL Source-based method for solidification
  phase change}, Numerical Heat Transfer, Part B Fundamentals 19~(2) (1991)
  175--189.

\bibitem{Kallemov16}
B.~Kallemov, A.~Bhalla, B.~Griffith, A.~Donev, An immersed boundary method for
  rigid bodies, Communications in Applied Mathematics and Computational Science
  11~(1) (2016) 79--141.

\bibitem{Usabiaga17}
F.~Balboa~Usabiaga, B.~Kallemov, B.~Delmotte, A.~Bhalla, B.~Griffith, A.~Donev,
  Hydrodynamics of suspensions of passive and active rigid particles: a rigid
  multiblob approach, Communications in Applied Mathematics and Computational
  Science 11~(2) (2017) 217--296.

\bibitem{Osher1988}
S.~Osher, J.~A. Sethian, Fronts propagating with curvature-dependent speed:
  algorithms based on hamilton-jacobi formulations, Journal of Computational
  Physics 79~(1) (1988) 12--49.

\bibitem{Sussman1994}
M.~Sussman, P.~Smereka, S.~Osher, A level set approach for computing solutions
  to incompressible two-phase flow, Journal of Computational Physics 114~(1)
  (1994) 146--159.

\bibitem{Zhang2019}
C.~Zhang, C.~Wu, K.~Nandakumar, Effective geometric algorithms for immersed
  boundary method using signed distance field, Journal of Fluids Engineering
  141~(6) (2019).

\bibitem{Harlow1965}
F.~H. Harlow, J.~E. Welch, Numerical calculation of time-dependent viscous
  incompressible flow of fluid with free surface, The physics of fluids 8~(12)
  (1965) 2182--2189.

\bibitem{Guermond2006}
J.-L. Guermond, P.~Minev, J.~Shen, An overview of projection methods for
  incompressible flows, Computer methods in applied mechanics and engineering
  195~(44-47) (2006) 6011--6045.

\bibitem{patankar2018numerical}
S.~V. Patankar, Numerical heat transfer and fluid flow, CRC press, 2018.

\bibitem{li2001immersed}
Z.~Li, M.-C. Lai, {The immersed interface method for the Navier--Stokes
  equations with singular forces}, Journal of Computational Physics 171~(2)
  (2001) 822--842.

\bibitem{kolahdouz2021sharp}
E.~M. Kolahdouz, A.~P.~S. Bhalla, L.~N. Scotten, B.~A. Craven, B.~E. Griffith,
  {A sharp interface Lagrangian-Eulerian method for rigid-body fluid-structure
  interaction}, Journal of computational physics 443 (2021) 110442.

\bibitem{gibou2002second}
F.~Gibou, R.~P. Fedkiw, L.-T. Cheng, M.~Kang, A second-order-accurate symmetric
  discretization of the poisson equation on irregular domains, Journal of
  Computational Physics 176~(1) (2002) 205--227.

\bibitem{Rider2007}
W.~J. Rider, J.~A. Greenough, J.~R. Kamm, Accurate monotonicity-and
  extrema-preserving methods through adaptive nonlinear hybridizations, Journal
  of Computational Physics 225~(2) (2007) 1827--1848.

\bibitem{Shu1998}
C.-W. Shu, Essentially non-oscillatory and weighted essentially non-oscillatory
  schemes for hyperbolic conservation laws, in: Advanced numerical
  approximation of nonlinear hyperbolic equations, Springer, 1998, pp.
  325--432.

\bibitem{Russo2000}
G.~Russo, P.~Smereka, A remark on computing distance functions, Journal of
  Computational Physics 163~(1) (2000) 51--67.

\bibitem{Min2010}
C.~Min, On reinitializing level set functions, Journal of Computational Physics
  229~(8) (2010) 2764--2772.

\bibitem{howard2021conservative}
A.~A. Howard, A.~M. Tartakovsky, {A conservative level set method for N-phase
  flows with a free-energy-based surface tension model}, Journal of
  Computational Physics 426 (2021) 109955.

\bibitem{Saad93}
Y.~Saad, A flexible inner-outer preconditioned {GMRES} algorithm,
  SIAM~J~Sci~Comput 14~(2) (1993) 461--469.

\bibitem{Brown2001}
D.~L. Brown, R.~Cortez, M.~L. Minion, {Accurate projection methods for the
  incompressible Navier--Stokes equations}, Journal of Computational Physics
  168~(2) (2001) 464--499.

\bibitem{Mccormick1986}
S.~McCormick, J.~Thomas, {The fast adaptive composite grid (FAC) method for
  elliptic equations}, Mathematics of Computation 46~(174) (1986) 439--456.

\bibitem{IBAMR-web-page}
{IBAMR}: {A}n adaptive and distributed-memory parallel implementation of the
  immersed boundary method, \url{https://github.com/IBAMR/IBAMR}.

\bibitem{HornungKohn02}
R.~D. Hornung, S.~R. Kohn, Managing application complexity in the {SAMRAI}
  object-oriented framework, Concurrency~Comput~Pract~Ex 14~(5) (2002)
  347--368.

\bibitem{samrai-web-page}
{SAMRAI}: {S}tructured {A}daptive {M}esh {R}efinement {A}pplication
  {I}nfrastructure, \url{http://www.llnl.gov/CASC/SAMRAI}.

\bibitem{petsc-efficient}
S.~Balay, W.~D. Gropp, L.~C. McInnes, B.~F. Smith, Efficient management of
  parallelism in object oriented numerical software libraries, in: E.~Arge,
  A.~M. Bruaset, H.~P. Langtangen (Eds.), Modern Software Tools in Scientific
  Computing, Birkh{\"{a}}user Press, 1997, pp. 163--202.

\bibitem{petsc-user-ref}
S.~Balay, S.~Abhyankar, M.~F. Adams, J.~Brown, P.~Brune, K.~Buschelman,
  L.~Dalcin, V.~Eijkhout, W.~D. Gropp, D.~Kaushik, M.~G. Knepley, L.~C.
  McInnes, K.~Rupp, B.~F. Smith, S.~Zampini, H.~Zhang,
  \href{http://www.mcs.anl.gov/petsc}{{PETS}c users manual}, Tech. Rep.
  ANL-95/11 - Revision 3.6, Argonne National Laboratory (2015).
\newline\urlprefix\url{http://www.mcs.anl.gov/petsc}

\bibitem{petsc-web-page}
S.~Balay, S.~Abhyankar, M.~F. Adams, J.~Brown, P.~Brune, K.~Buschelman,
  L.~Dalcin, V.~Eijkhout, W.~D. Gropp, D.~Kaushik, M.~G. Knepley, L.~C.
  McInnes, K.~Rupp, B.~F. Smith, S.~Zampini, H.~Zhang,
  \href{http://www.mcs.anl.gov/petsc}{{PETS}c {W}eb page},
  \url{http://www.mcs.anl.gov/petsc} (2015).
\newline\urlprefix\url{http://www.mcs.anl.gov/petsc}

\bibitem{Pathak16}
A.~Pathak, M.~Raessi, {A 3D, fully Eulerian, VOF-based solver to study the
  interaction between two fluids and moving rigid bodies using the fictitious
  domain method}, Journal of Computational Physics 311 (2016) 87--113.

\bibitem{Yettou2006}
E.-M. Yettou, A.~Desrochers, Y.~Champoux, Experimental study on the water
  impact of a symmetrical wedge, Fluid Dynamics Research 38~(1) (2006) 47.

\bibitem{Hamed2015}
A.~Hamed, Y.~Jin, L.~Chamorro, On the transient dynamics of the wake and
  trajectory of free falling cones with various apex angles, Experiments in
  Fluids 56 (2015).

\bibitem{Peskin02}
C.~S. Peskin, The immersed boundary method, Acta~Numer 11 (2002) 479--517.

\bibitem{Griffith2012vol}
B.~E. Griffith, On the volume conservation of the immersed boundary method,
  Communications in Computational Physics 12~(2) (2012) 401--432.

\bibitem{bale2021one}
R.~Bale, A.~P.~S. Bhalla, B.~E. Griffith, M.~Tsubokura, A one-sided direct
  forcing immersed boundary method using moving least squares, Journal of
  Computational Physics 440 (2021) 110359.

\bibitem{cheron2023hybrid}
V.~Ch{\'e}ron, F.~Evrard, B.~van Wachem, A hybrid immersed boundary method for
  dense particle-laden flows, Computers \& Fluids 259 (2023) 105892.

\bibitem{huh1971hydrodynamic}
C.~Huh, L.~E. Scriven, Hydrodynamic model of steady movement of a
  solid/liquid/fluid contact line, Journal of colloid and interface science
  35~(1) (1971) 85--101.

\bibitem{Zhang2010}
Y.~Zhang, Q.~Zou, D.~Greaves, D.~Reeve, A.~Hunt-Raby, D.~Graham, P.~James,
  X.~Lv, A level set immersed boundary method for water entry and exit, Comm.
  Comput. Phys 8~(2) (2010) 265--288.

\bibitem{Patel2018}
J.~K. Patel, G.~Natarajan, Diffuse interface immersed boundary method for
  multi-fluid flows with arbitrarily moving rigid bodies, Journal of
  Computational Physics 360 (2018) 202--228.

\bibitem{Calderer2014}
A.~Calderer, S.~Kang, F.~Sotiropoulos, Level set immersed boundary method for
  coupled simulation of air/water interaction with complex floating structures,
  Journal of Computational Physics 277 (2014) 201--227.

\bibitem{sanders2011new}
J.~Sanders, J.~E. Dolbow, P.~J. Mucha, T.~A. Laursen, A new method for
  simulating rigid body motion in incompressible two-phase flow, International
  Journal for Numerical Methods in Fluids 67~(6) (2011) 713--732.

\bibitem{Dafnakis2020}
P.~Dafnakis, A.~P.~S. Bhalla, S.~A. Sirigu, M.~Bonfanti, G.~Bracco,
  G.~Mattiazzo, {Comparison of wave--structure interaction dynamics of a
  submerged cylindrical point absorber with three degrees of freedom using
  potential flow and computational fluid dynamics models}, Physics of Fluids
  32~(9) (2020) 093307.

\bibitem{Nangia2019WSI}
N.~Nangia, N.~A. Patankar, A.~P.~S. Bhalla, {A DLM immersed boundary method
  based wave-structure interaction solver for high density ratio multiphase
  flows}, Journal of Computational Physics 398 (2019) 108804.

\bibitem{Patel2017}
J.~K. Patel, G.~Natarajan, A novel consistent and well-balanced algorithm for
  simulations of multiphase flows on unstructured grids, Journal of
  Computational Physics 350 (2017) 207--236.

\bibitem{zeng2023consistent}
Y.~Zeng, H.~Liu, Q.~Gao, A.~Almgren, A.~P.~S. Bhalla, L.~Shen, A consistent
  adaptive level set framework for incompressible two-phase flows with high
  density ratios and high reynolds numbers, Journal of Computational Physics
  478 (2023) 111971.

\bibitem{Martin1952}
J.~C. Martin, W.~J. Moyce, Penney, Part iv. an experimental study of the
  collapse of liquid columns on a rigid horizontal plane, Phil. Trans. R. Soc.
  Lond. A 244~(882) (1952) 312--324.

\bibitem{solomenko2017mass}
Z.~Solomenko, P.~D. Spelt, L.~O. Naraigh, P.~Alix, {Mass conservation and
  reduction of parasitic interfacial waves in level-set methods for the
  numerical simulation of two-phase flows: A comparative study}, International
  Journal of Multiphase Flow 95 (2017) 235--256.

\end{thebibliography}
\end{flushleft}

\end{document}